\documentclass[12pt]{article}
\pdfoutput=1
\usepackage[top=1in, bottom=1in, left=1in, right=1in]{geometry}
\usepackage{latexsym}
\usepackage{amssymb, amsfonts, amsmath}
\usepackage{indentfirst}
\usepackage{bbm}
\usepackage{amssymb}
\usepackage{verbatim}
\usepackage{amsmath, amsthm, amssymb}
\usepackage{mathrsfs}
\usepackage{amsfonts}
\usepackage{dsfont}
\usepackage{csquotes}
\usepackage{caption}
\usepackage{hyperref}
\usepackage{url}
\usepackage{xcolor}
\usepackage{xpatch}
\usepackage{multirow}
\usepackage{diagbox}
\usepackage{booktabs}
\usepackage[english]{babel}

\usepackage{graphicx} 
\usepackage[export]{adjustbox}
\usepackage{float}
\graphicspath{ {images/} }

\usepackage[style=phys, articletitle=true, biblabel=brackets, backend=bibtex,
chaptertitle=false, pageranges=false, eprint=true,%
natbib=true, maxbibnames=10, giveninits=true, sorting=none]{biblatex} 

\DeclareFieldFormat[article]{journaltitle}{\mkbibitalic{#1\isdot}} 
\makeatletter
\newrobustcmd{\mkbibfixedbrackets}[1]{%
	\begingroup
	\blx@blxinit
	\blx@setsfcodes
	\bibleftbracket#1\bibrightbracket
	\endgroup}

\renewbibmacro*{doi+eprint+url}{%
	\iftoggle{bbx:doi}
	{\printfield{doi}}
	{}%
	\ifboolexpr{test {\iffieldequalstr{eprinttype}{arxiv}} or test {\iffieldequalstr{eprinttype}{arXiv}}}
	{\setunit{\addspace}\newblock}
	{\newunit\newblock}%
	\iftoggle{bbx:eprint}
	{\usebibmacro{eprint}}
	{}%
	\newunit\newblock
	\iftoggle{bbx:url}
	{\usebibmacro{url+urldate}}
	{}}

\xpatchbibdriver{online}
{\newunit\newblock
	\iftoggle{bbx:eprint}
	{\usebibmacro{eprint}}
	{}}
{\ifboolexpr{test {\iffieldequalstr{eprinttype}{arxiv}} or test {\iffieldequalstr{eprinttype}{arXiv}}}
	{\setunit{\addspace}\newblock}
	{\newunit\newblock}%
	\iftoggle{bbx:eprint}
	{\usebibmacro{eprint}}
	{}}
{}{}

\DeclareFieldFormat{eprint:arxiv}{%
	\mkbibbrackets{%
		\ifhyperref
		{\href{http://arxiv.org/\abx@arxivpath/#1}{%
				arXiv\addcolon
				\nolinkurl{#1}%
				\iffieldundef{eprintclass}
				{}
				{\addspace\UrlFont{\mkbibfixedbrackets{\thefield{eprintclass}}}}}}
		{arXiv\addcolon
			\nolinkurl{#1}%
			\iffieldundef{eprintclass}
			{}
			{\addspace\UrlFont{\mkbibfixedbrackets{\thefield{eprintclass}}}}}}}
\makeatother

\parskip=\medskipamount

\DeclareMathAlphabet{\mathbbmsl}{U}{bbm}{m}{sl}
\DeclareMathAlphabet\mathbfcal{OMS}{cmsy}{b}{n}

\parskip=\medskipamount

\newcommand {\cA}{{\cal A}}
\newcommand {\cB}{{\cal B}}

\newcommand {\cD}{{\cal D}}

\newcommand {\cG}{{\cal G}}
\newcommand {\cH}{{\cal H}}
\newcommand {\cI}{{\cal I}}

\newcommand {\cN}{{\cal N}}
\newcommand {\cO}{{\cal O}}

\newcommand {\cQ}{{\cal Q}}

\def\a{\alpha}

\def\b{\beta}

\def\d{\delta}
\def\e{\epsilon}

\def\g{\gamma}
\def\G{\Gamma}

\def\l{\lambda}

\def\o{\omega}

\def\q{\theta}

\def\s{\sigma}

\def\x{\xi}

\def\D{\Delta}
\def\F{\Phi}
\def\J{\Psi}

\def\P{\Pi}
\def\Q{\Theta}

\def\ri{{\rm i}}

\newcommand{\ad}{{\dot{\alpha}}}                           
\newcommand{\bd}{{\dot{\beta}}}                            
\newcommand{\ve}{\varepsilon}                            

\newcommand{\pa}{\partial}                           
\newcommand{\hf}{\frac12}

\newcommand{\be}{\begin{equation}}
\newcommand{\ee}{\end{equation}}
\newcommand{\bea}{\begin{eqnarray}}
\newcommand{\eea}{\end{eqnarray}}
\newcommand{\non}{\nonumber}

\def\double #1{#1{\hbox{\kern-2pt $#1$}}}

\newcommand{\gd}{{\dot\g}}
\newcommand{\dd}{{\dot\d}}

\newif\ifdtup

\def\fn3{{{X}_{3}, {\Q}_{3}, \bar {\Q}_{3}}}
\def\fxq{{{X}, {\Q}, \bar {\Q}}}
\def\corr1{{\langle
J_{\a(r_1) \ad(r_1)} (z_1) \, J'{}_{\b(r_2) \bd(r_2)}(z_2)\,  J''{}_{\g(r_3) \gd(r_3)}(z_3)
\rangle}}

\newcommand{\bsubeq}{\begin{subequations}}
\newcommand{\esubeq}{\end{subequations}}

\newcommand{\lb}{{\bar{\l}}}

\numberwithin{equation}{section}

\newcommand{\sU}{\mathsf{U}}

\numberwithin{equation}{section}

\begin{document}

\begin{titlepage}
\begin{flushright}
July, 2024 \\
\end{flushright}
\vspace{5mm}

\begin{center}
{\Large \bf 
Three-point functions of higher-spin supercurrents in 4D ${\cN}=1$ SCFT: general formalism for arbitrary superspins}
\\ 
\end{center}

\begin{center}

{\bf
Evgeny I. Buchbinder${}^{a}$,
Jessica Hutomo${}^{b}$, and
Benjamin J. Stone${}^{a}$
} \\
\vspace{5mm}

\footnotesize{
${}^{a}$
{\it Department of Physics M013, The University of Western Australia\\
35 Stirling Highway, Crawley W.A. 6009, Australia
}
 \\~\\
${}^{b}$
{\it 
INFN, Sezione di Padova, Via Marzolo 8, 35131 Padova, Italy}
}
\vspace{2mm}
~\\
\texttt{evgeny.buchbinder@uwa.edu.au, jessica.hutomo@pd.infn.it, \\ benjamin.stone@uwa.edu.au}\\
\vspace{2mm}

\end{center}

\begin{abstract}
\baselineskip=14pt
We analyse the general structure of the three-point functions involving conserved higher-spin ``vector-like" supercurrents $J_{s}(z) := J_{\a(s) \ad(s)}(z)$ in
four-dimensional $\cN=1$ superconformal field theory. Using the constraints of superconformal symmetry and superfield conservation equations, we utilise a computational approach to analyse the general structure of the three-point function $\langle J^{}_{s_{1}}(z_{1}) \, J'_{s_{2}}(z_{2}) \, J''_{s_{3}}(z_{3}) \rangle$ and provide a general classification of the results. We demonstrate that the three-point function is fixed up to $2 \min (s_{i}) + 2$ independent conserved structures, which we propose to hold for arbitrary superspins. In addition, we show that the conserved structures can be classified as parity-even or parity-odd in superspace based on their transformation properties under superinversion.
\end{abstract}
\vspace{5mm}


\vfill
\end{titlepage}

\newpage
\renewcommand{\thefootnote}{\arabic{footnote}}
\setcounter{footnote}{0}

\tableofcontents{}
\vspace{1cm}
\bigskip\hrule

\allowdisplaybreaks


\section{Introduction}


Understanding how conformal symmetry constrains the correlation functions of primary operators remains an important problem in conformal field theory (see refs.~\cite{Polyakov:1970xd, Schreier:1971um, Migdal:1971xh, Migdal:1971fof, Ferrara:1972cq, Ferrara:1973yt, Koller:1974ut, Mack:1976pa, Stanev:1988ft, Fradkin:1978pp} for early results). An essential class of operators in conformal field theory are conserved currents; they encode information about the symmetries of a given CFT and correspond to the lowest weight states of the conformal algebra, with their scaling dimension saturating 
the unitarity bound. The study of their correlation functions is therefore of fundamental importance.
The systematic approach to study correlation functions of conserved currents was developed in~\cite{Osborn:1993cr, Erdmenger:1996yc}, and was later extended to superconformal field theories in various dimensions \cite{Park:1997bq, Osborn:1998qu, Park:1998nra, Park:1999cw, Park:1999pd, Kuzenko:1999pi, Nizami:2013tpa, Buchbinder:2015qsa, Buchbinder:2015wia, Kuzenko:2016cmf, Buchbinder:2021gwu, Buchbinder:2021izb, Buchbinder:2021kjk, Buchbinder:2021qlb, Buchbinder:2023fqv, Buchbinder:2023ndg, Jain:2022izp}.\footnote{Some novel approaches to the construction of correlation functions of conserved currents carry out the calculations in momentum space 
using methods such as spinor-helicity variables, see e.g.~\cite{Jain:2020puw, Jain:2020rmw, Jain:2021gwa, Jain:2021vrv, Jain:2021wyn, Isono:2019ihz, Bautista:2019qxj}. There are also embedding space formalisms~\cite{Elkhidir:2014woa,Weinberg:2010fx, Weinberg:2012mz, Costa:2011dw, Costa:2011mg, Costa:2014rya, Fortin:2020des} (see \cite{Goldberger:2011yp, Goldberger:2012xb, Fitzpatrick:2014oza,Khandker:2014mpa} for supersymmetric extensions).} The most studied examples of conserved currents in (super)conformal field theory are the energy-momentum tensor and vector currents; their most general 
three-point functions were determined in~\cite{Osborn:1993cr,Erdmenger:1996yc}. For three-point functions of conserved higher-spin ``vector-like" currents $J_{\a(s) \ad(s)}$, in four dimensions their general structure was analysed by Stanev~\cite{Stanev:2012nq} (see also Zhiboedov~\cite{Zhiboedov:2012bm} and the recent work by Karapetyan, Manvelyan \& Mkrtchyan \cite{Karapetyan:2023zdu} for results in general dimensions). However, more recently an auxiliary spinor formalism was utilised in \cite{Buchbinder:2023coi} to provide a complete classification of the three-point functions of conserved currents $J_{\a(i) \ad(j)}$ of Lorentz type $(i/2, j/2)$ with $i,j \geq 1$ in 4D CFT. There it was shown that the three-point function of higher-spin currents, $\langle J^{}_{\a(s_1) \ad(s_1)}(x_{1}) \, J'_{\b(s_2) \bd(s_2)}(x_{2}) \, J''_{\g(s_3) \gd(s_3)}(x_{3}) \rangle$ with $s_{i} \geq 1$, is fixed up to $2\min(s_{i}) +1$ independent structures, where, based on their transformation properties under inversion, $\min(s_{i}) +1$ structures are classified as parity-even and $\min(s_{i})$ structures are classified as parity-odd.

Superconformal field theories can also possess higher-spin conserved currents, and in general the bosonic and fermionic currents are assembled into supermultiplets. For example, the energy-momentum tensor
together with the fermionic supersymmetry current belong to the supercurrent 
multiplet \cite{Ferrara:1974pz} (see also \cite{Gates:1983nr, Magro:2001aj, Komargodski:2010rb, Kuzenko:2010am, Kuzenko:2010ni}). In the case of $\cN=1$ supersymmetry, there are three types of conformal current multiplets depending on the corresponding superfield Lorentz type $(i/2,j/2)$. These multiplets were explicitly described in flat \cite{Ceresole:1999zs} and curved \cite{Kuzenko:2019tys} backgrounds.\footnote{The structure of higher-spin current multiplets, including non-conformal ones, have been studied extensively in the past. For instance, explicit realisations in terms of free scalar and vector multiplets are known for $J_{\a(s) \ad(s)}$, in Minkowski \cite{Kuzenko:2017ujh, Buchbinder:2017nuc, Hutomo:2017phh, Hutomo:2017nce, Koutrolikos:2017qkx, Buchbinder:2018wwg, Buchbinder:2018gle, Gates:2019cnl}, AdS \cite{Buchbinder:2018nkp, Hutomo:2020wca} and conformally-flat backgrounds \cite{Kuzenko:2022hdv}.}
\begin{itemize} 
\item Given two positive integers $i$ and $j$, the 
conformal current multiplet $J_{\a(i) \ad(j)}:= J_{\a_1 \dots \a_i\ad_1 \dots \ad_j}
=J_{(\a_1 \dots \a_i) (\ad_1 \dots \ad_j)}$ obeys the conservation equations
\bea
D^\b J_{\a(i-1) \b \ad(j)} = 0~, \qquad \bar{D}^\bd J_{\a(i) \ad(j-1) \bd} = 0 ~.
\eea
This superfield has dimension $\big(2+ \hf (i+j) \big)$ and its $\sU(1)_R$ charge is 
equal to $\frac 13 (i -j) $, see \cite{Kuzenko:2019tys} for the technical details. 
When $i=j=s$, the supercurrent $ J_{\a(s) \ad(s)} $ is neutral with respect to 
the $\text{R}$-symmetry group $\sU(1)_{\text{R}}$, and so we may
restrict $ J_{\a(s) \ad(s)} $ to be real, i.e. $\bar{J}_{\a(s) \ad(s)} = J_{\a(s) \ad(s)}$. The $s=1$ case corresponds to the ordinary 
conformal supercurrent \cite{Ferrara:1974pz}, while the case $s>1$ was first discussed 
in \cite{Howe:1981qj}.
\item If $i>0$ and $j=0$, the conformal current multiplet $S_{\a(i)}$ obeys the 
the conservation equation
\bea
D^{\b}S_{\b\a_1 \dots \a_{i-1}} = 0\,,  \qquad  \bar D^2 S_{\a(i)} = 0~.
\label{zh02}
\eea
The case $i=1$ was first considered in \cite{Kuzenko:1999pi}, where it was shown that the spinor supercurrent 
$S_\a$ naturally originates from the reduction of the conformal $\cN=2$ supercurrent \cite{Sohnius:1978pk}
to $\cN=1$ superspace. 

\item Finally, the $i=j=0$ case corresponds to the flavour current multiplet \cite{Ferrara:1974ac}, 
$L=\bar L$, constrained by 
\bea
		D^2 L = 0~, \qquad \bar{D}^2 L = 0 ~.
\eea
\end{itemize}
Let us comment on some recent progress in our understanding of three-point correlation functions involving the above current multiplets. In the context of 4D $\cN=1$ SCFT, the study of three-point functions of conserved higher-spin supercurrents was only initiated three years ago in \cite{Buchbinder:2021izb,Buchbinder:2021kjk}, where a complete classification of the three-point functions of conserved fermionic higher-spin supercurrents $S_{\a(i)}, \bar{S}_{\ad(i)}$ was provided. Later, in \cite{Buchbinder:2022kmj}, the general structure of several new mixed three-point functions of $J_{\a(s) \ad(s)}$ with the supercurrent and flavour current multiplets (such as $\langle J_{\a(s_1) \ad(s_1)} J_{\b \bd} J_{\g \gd} \rangle$ and $\langle J_{\a(s_1) \ad(s_1)} J_{\b(s_2) \bd(s_2)} L \rangle$) were found in an explicit form for arbitrary superspins $s_1, s_2$. However, despite this progress it has remained an open problem to determine the general structure of $\langle J_{\a(s_1) \ad(s_1)}(z_{1}) \, J_{\b(s_2) \bd(s_2)}(z_{2}) \, J_{\g(s_3) \gd(s_3)}(z_{3}) \rangle$ for $s_{i} > 1$, and to classify the independent conserved structures based on their parity/superinversion properties. 

In this paper we develop a general formalism to study the three-point correlation function of higher-spin supercurrent multiplets $J_{\a(i) \ad(j)}$ corresponding to Lorentz type $(i/2, j/2)$ with $i, j \geq 0$. To undertake the analysis we augment the formalism developed by Park and Osborn \cite{Park:1997bq, Park:1999pd, Osborn:1998qu} with auxiliary spinors, similar to the approach of \cite{Buchbinder:2023coi} (see also \cite{Buchbinder:2022mys, Buchbinder:2023fqv, Buchbinder:2022kmj}). The advantages of our approach are two-fold; first, by using the formalism of Park and Osborn the three-point function is completely determined by a constrained tensor $\cH_{\bar{\cA}_{1} \bar{\cA}_{2} \cA_{3}}(X,\Q,\bar{\Q})$ which is a function of the superconformally covariant three-point building blocks $X_{\a \ad}$, $\Q^{\a}$, $\bar{\Q}_{\ad}$ (the dependence on the superspace points is implicit). Second, by introducing auxiliary spinors the tensor structure of the three-point function may be encoded in a polynomial which is a product of some simple monomial structures. This simplifies both the analysis and the resulting structure of the three-point functions of higher-spin supercurrents. 

Although the formalism is completely general and can be applied to supercurrent multiplets $J_{\a(i) \ad(j)}$ with $i,j \geq 0$, the main goal of this work is to determine the general structure of three-point functions of real ``vector-like" supercurrent multiplets $J_{\a(s) \ad(s)}$ of superspin $s\geq1$. For these cases one must impose conservation conditions on all three superspace points, a reality condition, and any symmetries under permutations of superspace points. After imposing the aforementioned constraints the three-point function of real supercurrent multiplets is fixed up to finitely many independent structures, and similar to 4D CFT it is possible to classify them as parity-even or parity-odd in superspace based on their transformation properties under superinversion.\footnote{In 4D CFT the three-point functions of conserved currents are fixed up to finitely many independent structures which may be classified as parity-even or parity-odd based on their transformation properties under inversion. See \cite{Buchbinder:2023coi} for details.} Based on our analysis we propose that the three-point function $\langle J^{}_{\a(s_1) \ad(s_1)}(z_{1}) \, J'_{\b(s_2) \bd(s_2)}(z_{2}) \, J''_{\g(s_3) \gd(s_3)}(z_{3}) \rangle$ is fixed up to $2 \min(s_{i}) + 2$ independent structures, where $\min(s_{i}) + 1$ are parity-even in superspace, and $\min(s_{i}) + 1$ are parity-odd in superspace. More precisely we found that the three-point function is fixed to the following form:
\begin{align}
	\langle  J^{}_{s_{1}}(z_{1}) \, J'_{s_{2}}(z_{2}) \, J''_{s_{3}}(z_{3}) \rangle &= \sum_{i=1}^{\min(s_{1}, s_{2}, s_{3}) + 1} a_{i} \, \langle  J^{}_{s_{1}}(z_{1}) \, J'_{s_{2}}(z_{2}) \,  J''_{s_{3}}(z_{3}) \rangle_{i}^{E} \\
    & \hspace{10mm} + \sum_{i=1}^{\min(s_{1}, s_{2}, s_{3}) + 1} \text{i} b_{i} \, \langle  J^{}_{s_{1}}(z_{1}) \, J'_{s_{2}}(z_{2}) \, J''_{s_{3}}(z_{3}) \rangle_{i}^{O} \, , \non
\end{align}
where $\langle  J^{}_{s_{1}}(z_{1}) \, J'_{s_{2}}(z_{2}) \,  J''_{s_{3}}(z_{3}) \rangle_{i}^{E/O}$ are structures which are classified as parity-even/odd under superinversion, and $a_{i}$ and $b_{i}$ are real coefficients. We demonstrate that the superfield analysis is completely consistent with the component analysis presented in \cite{Buchbinder:2023coi} after bar-projection, and that the parity-even sector of certain component correlators residing in $\langle J^{}_{\a(s_1) \ad(s_1)}(z_{1}) \, J'_{\b(s_2) \bd(s_2)}(z_{2}) \, J''_{\g(s_3) \gd(s_3)}(z_{3}) \rangle$ is further constrained by $\cN=1$ superconformal symmetry compared to their non-supersymmetric counterparts.

The content of this paper is organised as follows. In Section \ref{section2} we review the properties of the conformal building blocks used to construct correlation functions 
of primary operators in 4D $\cN=1$ SCFT. In Section \ref{section3} we develop the formalism to construct three-point functions of superconformal primary operators of the form $J_{\a(i) \ad(j)}(z)$, 
where we demonstrate how to impose all constraints arising from conservation equations, reality conditions and point-switch symmetries. In Subsection \ref{section3.1} we show how to classify the structures in the three-point function based on their transformation properties under superinversion. In Subsection \ref{GeneratingFunctionFormalism} we introduce an index-free auxiliary spinor formalism based on \cite{Buchbinder:2023coi} (see also \cite{Buchbinder:2022mys,Buchbinder:2023fqv, Buchbinder:2023ndg, Buchbinder:2022kmj})
to construct the supersymmetric three-point functions, and we outline the pertinent aspects of our computational approach. Next, in Section \ref{section4} we demonstrate 
our formalism by analysing the three-point functions involving the supercurrent and flavour current multiplets, where we reproduce the results previously found in~\cite{Osborn:1998qu,Buchbinder:2021izb,Buchbinder:2021kjk,Buchbinder:2022kmj}. We then expand the analysis to three-point functions of 
higher-spin ``vector-like" supercurrents $J_{\a(s) \ad(s)}$ and provide a classification of the results, which we summarise in Subsection \ref{section4.2}. Finally, in Subsection \ref{section4.3} we comment on the superspace reduction of three-point functions of higher-spin supercurrents and the classification of parity-even and parity-odd structures, where we compare our results with \cite{Buchbinder:2023coi}. The Appendix \ref{AppA} is devoted to useful identities. In Appendix \ref{AppD}, we discuss the implications of pseudo-covariance under superinversion for the structure of three-point functions. In Appendix \ref{AppB}, as a consistency check we provide further examples of three-point functions involving higher-spin spinor current multiplets to compare against the results in \cite{Buchbinder:2021kjk}. In Appendix \ref{AppC} we 
provide some examples of the three-point functions of higher-spin supercurrents to demonstrate the pattern in the number of independent structures.

\newpage


\section{Superconformal building blocks} \label{section2}


This section contains a concise summary of two- and three-point superconformal building blocks in four-dimensional $\cN=1$ superspace, which are essential to the construction of three-point functions of primary superfields. The superconformal structures were introduced in \cite{Park:1997bq,Osborn:1998qu}, and later generalised to arbitrary ${\cN}$ in \cite{Park:1999pd} (see also \cite{Kuzenko:1999pi} for a review). Our notation and conventions are those of \cite{Buchbinder:1998qv}.


\subsection{Two-point structures}


We denote the $\cN=1$ Minkowski superspace by ${\mathbb M}^{4|4}$. It is parametrised by coordinates $z^A = (x^a, \q^\a, {\bar \q}_\ad)$, where $a = 0,1, 2, 3;~ \a, \ad = 1, 2 $. Let $z_1$ and $z_2$ be two different points in superspace. 
All building blocks for the two- and three-point correlation functions are composed of the two-point structures:
\bsubeq
\bea
x^a_{2\bar{1}} &=& -x^{a}_{\bar{1} 2} = x^{a}_{2+} - x^a_{1-} - 2 \ri \,\q_{2}\s^{a} {\bar \q}_{1}~,\\
\q_{12} &=& \q_1 - \q_2~, \qquad {\bar \q}_{12} = {\bar \q}_1 - {\bar \q}_2~,
\eea 
\esubeq
which are invariants of the $Q$-supersymmetry transformations. Here $x^a_{\pm}= x^a \pm \ri \theta \sigma^a {\bar \theta}$.
In spinor notation, we write
\bsubeq 
\bea
{x}_{\bar{1} 2}{}^{\ad \a} &=& (\tilde{\s}_a)^{\ad \a}x^{a}_{\bar{1} 2}~, \\
{x}_{2 \bar{1} \,\a \ad} &=& (\s_a)_{\a \ad} x^a_{2 \bar{1} }=
-(\s_a)_{\a \ad} \, x^a_{\bar{1} 2}= - \ve_{\a \b} \ve_{\ad \bd} {x}_{\bar{1} 2}{}^{\bd \b}~, \label{eps-x}\\
({x}_{2 \bar{1}})_{\a \ad}^{\dagger} &=& {x}_{\bar{2} 1 \, \a \ad} = - {x}_{1 \bar{2} \, \a \ad}~. \label{hc-x}
\eea
\esubeq
Note that ${x}^{}_{\bar{1} 2}{}^{\ad \a} x_{2 \bar{1} \, \a \bd} = x_{\bar{1} 2}{}^{2}\, \d^{\ad}{}_{\bd}$ and ${x}_{2 \bar{1} \, \a \ad} \,x_{\bar{1} 2}{}^{\ad \b} = x_{\bar{1} 2}{}^{2}\, \d_{\a}{}^{\b}$.
Hence, it follows that 
%
\bea
({x}_{\bar{1} 2}^{-1})_{\a \ad} = -\frac{({x}_{\bar{1} 2})_{\a \ad}}{x_{\bar{1} 2}{}^{2}} = \frac{({x}_{2 \bar{1}})_{\a \ad}}{x_{\bar{1} 2}{}^{2}}~.
\eea
The notation `${x}_{2 \bar{1}}$' means that it is antichiral with respect to $z_1$ and chiral with respect to $z_2$. 
That is,
\be 
D_{(1)  \a} \,{x}_{2 \bar{1}\, \b \bd}=0\,, \qquad \bar D_{(2) \ad} \,{x}_{2 \bar{1} \, \b \bd}=0~,
\label{E1}
\ee
where $D_{(i) \a}$ and $\bar D_{(i) \ad}$ are the superspace covariant spinor derivatives acting on the superspace point $z_i$. They are defined as follows:
\bea
D_{\a} = \frac{\pa}{\pa \q^{\a}}+ \ri (\s^a)_{\a \ad} \bar{\q}^{\ad} \frac{\pa}{\pa x^a}~, \qquad \bar D_{\ad} = -\frac{\pa}{\pa \bar \q^{\ad}} - \ri {\q}^{\a} (\s^a)_{\a \ad}  \frac{\pa}{\pa x^a}~.
\eea
Likewise, the $Q$-supersymmetry generators are defined as
\bea
Q_{\a} =  \ri \frac{\pa}{\pa \q^{\a}}+(\s^a)_{\a \ad} \bar{\q}^{\ad} \frac{\pa}{\pa x^a}~, \qquad 
\bar Q_{\ad} = -\ri \frac{\pa}{\pa \bar \q^{\ad}} - {\q}^{\a} (\s^a)_{\a \ad}  \frac{\pa}{\pa x^a}~,
\eea
and thus we have that
\bea
\{ D_{\a}, Q_{\b} \} = \{ D_{\a}, \bar Q_{\bd} \} = \{ \bar{D}_{\ad}, Q_{\b} \}= \{ \bar D_{\ad}, \bar Q_{\bd} \} = 0~.
\eea
Indeed, it can be checked that ${x}_{2 \bar{1}}$ is annihilated by the supercharge operators,
\bea
Q_{(1)\a} {x}_{2 \bar{1}\, \b \bd} = \bar{Q}_{(1)\ad} {x}_{2 \bar{1}\, \b \bd} = Q_{(2)\a} {x}_{2 \bar{1}\, \b \bd} = \bar{Q}_{(2)\ad} {x}_{2 \bar{1}\, \b \bd} = 0~.
\eea
We also note several useful differential identities:
\bsubeq \label{2ptids}
\bea
&&D_{(1) \a} (x_{\bar{2} 1})^{\bd \b} = 4 \ri \d_{\a}{}^{\b} \bar{\q}_{12}^{\bd}~, \qquad \bar{D}_{(1) \ad}\, (x_{\bar{1} 2})^{\bd \b} = 4 \ri \d_{\ad}{}^{\bd} \q_{12}^{\b}~, \\
&&D_{(1) \a} \bigg(\frac{1}{x_{\bar{2} 1}{}^{2}} \bigg) = - \frac{4 \ri}{x_{\bar{2} 1}{}^2}  ({x}_{\bar{2} 1}{}^{-1})_{\a \bd} \bar{\q}_{12}^{\bd}~, \quad \bar D_{(1) \ad} \bigg(\frac{1}{x_{\bar{1} 2}{}^{2}} \bigg) = - \frac{4 \ri}{x_{\bar{1} 2}{}^2}  ({x}_{\bar{1} 2}{}^{-1})_{\a \ad} {\q}_{12}^{\a}~.~~~~~
\eea
\esubeq
Here and throughout, we assume that the superspace points are not coincident, $z_1 \neq z_2$. 

It is convenient to define the normalised two-point functions \cite{Osborn:1998qu},
\bea
\hat{x}_{1\bar{2} \,\a \ad} = \frac{x_{1\bar{2} \a \ad}}{(x_{1 \bar{2}}{}^2)^{1/2}}~, \qquad  
\hat{x}_{\bar{1} 2}{}^{\ad \a} = \frac{{x}_{\bar{1} 2}{}^{\ad \a}}{(x_{\bar{1} 2}{}^2)^{1/2}}~,
\label{I-def}
\eea 
which have the properties
\bea
\hat{x}^{}_{\bar{1} 2}{}^{\ad \a} 
\hat{x}^{}_{2 \bar{1} \,\b \ad} = \d^{\a}_{\b}~, \qquad 
\hat{x}^{}_{\bar{1} 2}{}^{\ad \a} \hat{x}_{2 \bar{1} \,\a \bd} = \d^{\ad}_{\bd}~.
\label{I-inv}
\eea
In what follows, for any two-point function we note $x_{\bar{2}1}{}^{k} = x_{1\bar{2}}{}^{k} = (x_{1\bar{2}}{}^{2})^{k/2} = (x_{\bar{2}1}{}^{2})^{k/2}$. 

We now construct an operator analogous to the conformal inversion tensor which acts on the space of symmetric traceless tensors of arbitrary rank. Specifically, we define
\bea
\cI_{\a(k)\ad(k)} (x_{1 \bar{2}}) &=& 
\hat{x}_{1 \bar{2} (\a_1 (\ad_1} \dots \hat{x}_{1 \bar{2} \a_k) \ad_k)}~,
\label{I-hs}
\eea
along with its inverse
\bea
\bar{\cI}^{\ad{(k)} \a(k)} ({x}_{\bar{2} 1 }) &=& 
\hat{x}_{\bar{2}1 }{}^{(\ad_1 (\a_1} \dots \hat{x}_{\bar{2}1 }{}^{\ad_k) \a_k)} = (-1)^{k} \bar{\cI}^{\ad{(k)} \a(k)} ({x}_{1 \bar{2}  }) \, .
\label{I-hs-up}
\eea
Due to the properties \eqref{eps-x} and \eqref{hc-x}, the spinor indices may be raised and lowered as follows:
\bsubeq \label{2pt inversion raise and lower}
\bea
\cI_{\a(k)}{}^{\ad(k)} (x_{1 \bar{2}}) &=& 
\ve^{\ad_1 \bd_1} \dots \ve^{\ad_k \bd_k} \cI_{\a(k)\bd(k)} (x_{1 \bar{2}})~, \\
&=& \ve_{\a_1 \b_1} \dots \ve_{\a_k \b_k} \bar{\cI}^{\ad{(k)} \b(k)} ({x}_{1\bar{2}})~.
\eea
\esubeq
As such, we also have
\bea
\bar \cI_{\ad(k)}{}^{\a(k)} (x_{\bar{1} 2}) = \Big(\cI_{\a(k)}{}^{\ad(k)} (x_{1 \bar{2}})\Big)^{\dagger}~,
\eea
and from \eqref{I-inv} we may derive the following identities for products of inversion tensors:
\bsubeq \label{2pt inversion properties}
\bea
\cI_{\a(k)}{}^{\ad(k)} (x_{1 \bar{2}}) \,
\bar{\cI}_{\ad(k)}{}^{\b(k)} (x_{1\bar{2} }) \, &=& \d_{(\a_1}^{(\b_1} \dots \d_{\a_k)}^{\b_k)} = \d_{\a(k)}{}^{\b(k)} \, , \\
\bar{\cI}_{\ad(k)}{}^{\a(k)} (x_{\bar{1} 2}) \,
\cI_{\a(k)}{}^{\bd(k)} (x_{\bar{1}2}) \, &=& \d_{(\ad_1}^{(\bd_1} \dots \d_{\ad_k)}^{\bd_k)} = \d_{\ad(k)}{}^{\bd(k)}~.
\eea
\esubeq
Since all superconformal transformations may be generated by combining inversions with
ordinary supersymmetry, the operators \eqref{I-hs} and \eqref{I-hs-up} play a crucial role in constructing correlation functions of primary superfields.

In accordance with \cite{Osborn:1998qu}, one can construct the vector representation of the inversion tensor in terms of the spinor two-point functions \eqref{I-def} as follows:
\bsubeq \label{vect-2pt}
\bea
&&I_{ab} (x_{1 \bar 2}, x_{\bar{1} 2}) = \bar{I}_{ba}(x_{\bar{2} 1}, x_{2 \bar{1}})=
\hf {\rm tr}\,(\tilde{\s}_a \,\hat{x}_{1 \bar{2}} \,\tilde{\s}_b\, \hat{x}_{2 \bar{1}})~,\\
&&I_{ac} (x_{1 \bar 2}, x_{\bar{1} 2}) \, \bar{I}^{cb}(x_{\bar{2} 1}, x_{2 \bar{1}}) = \d_{a}{}^{b}~.
\eea
\esubeq
In the purely bosonic case, $I_{ab} (x_{1 \bar 2}, x_{\bar{1} 2})$ reduces to the conformal inversion tensor \cite{Osborn:1993cr,Osborn:1998qu}
\bea
I_{ab} (x_{12})= \eta_{a b} - \frac{2}{x_{12}^2} x_{12 a}{} x_{12 b}~,
\eea
which played a pivotal role in studying conformal invariance in arbitrary dimensions \cite{Osborn:1993cr}.

Consider a tensor superfield $\cO^{\cA}(z)$ transforming in an irreducible representation of the Lorentz group with respect to the index $\cA$. Here $\cA$ denotes a collection of totally symmetric undotted and dotted spinor indices on which the Lorentz generators act, i.e. $\cA = \{ \a(i), \ad(j) \}$. Such a superfield is called \textit{primary} if its infinitesimal superconformal transformation law reads
\be
\begin{aligned}
\d \cO^\cA(z) &= - \x \, \cO^\cA (z) 
+ (\hat{\o}^{\a \b} (z) M_{\a \b}+ 
\hat{ \bar{\o}}^{\dot{\a} \dot{\b}} (z) 
\bar{M}_{\dot{\a} \dot{\b}} )^\cA{}_\cB\,
\cO^\cB (z) \\
& - 2\left( q\, \s(z) + \bar{q}\, \bar{\s} (z) \right) 
\cO^\cA (z)~.
\end{aligned}
\ee
In the above, $\xi$ is the superconformal Killing vector,
\bea
\xi = \bar{\xi} = \xi^a (z) \pa_a + \xi^{\a} (z) D_{\a}+ \bar{\xi}_{\ad} (z) \bar{D}^{\ad}~.
\eea
The superfield parameters $\hat{\o}^{\a \b} (z), ~\s(z)$ correspond to the `local' Lorentz and scale transformations: they are expressed in terms of $\xi^{A} = (\xi^a, \xi^{\a}, \bar \xi_{\ad})$, see \cite{Osborn:1998qu} for details. Note that the superspin `$s$' of the superfield $\cO^{\cA}(z)$ is $s = \frac{1}{2}(i + j)$. The conformal weights $q = \frac{i}{2} + 1$ and $\bar q = \frac{j}{2} + 1$ are such that $\D = q+\bar q = s + 2$ is the scale dimension, while $q-\bar q$ is proportional to the $\rm U(1)_R$ charge of the superfield $\cO^{\cA}$.

Following the general formalism of \cite{Osborn:1998qu,Park:1997bq,Park:1999pd},
the two-point function of a primary superfield
$\cO^{\cA}$ with its
conjugate $\bar{\cO}^{\bar{\cB}}$ is given by
\be
\langle \cO_{\cA} (z_1)\, \bar{\cO}^{\bar{\cB}} (z_2)\rangle
~=~ C_{\cO}\;\frac{ 
\mathfrak{I}_{\cA}{}^{\bar{\cB}} ({{x}_{1 \bar{2}}}, {{x}_{\bar{1} 2}}) }
{  (x_{1\bar{2}}{}^2)^q (x_{\bar{1}2}{}^2)^{\bar q} }~,
\label{2pt-gen}
\ee
where $ C_{\cO}$ is an overall normalisation constant and $\mathfrak{I}_{\cA}{}^{\bar{\cB}}$ is an appropriate representation of the inversion tensor. For a superfield representation $\cA = \{ \a(i), \ad(j) \}$ (and hence, $\bar{\cA} = \{ \a(j), \ad(i) \}$), where $\a(i), \ad(j)$ are totally symmetric sets of indices, we introduce the following compact notation:
\begin{subequations} \label{2pt shorthand}
\begin{align}
    \mathfrak{I}_{\cA}{}^{\bar{\cB}} (x_{1 \bar{2}}, x_{\bar{1} 2}) &\equiv {\cI}_{\a(i)}{}^{\bd(i)} (x_{1 \bar{2}}) \, \bar{\cI}_{\ad(j)}{}^{\b(j)} (x_{\bar{1}2}) \, ,  \label{2pt shorthand A} \\
    \bar{\mathfrak{I}}_{\bar{\cA}}{}^{\cB} (x_{\bar{1} 2}, x_{1 \bar{2}}) &\equiv \bar{\cI}_{\ad(i)}{}^{\b(i)} (x_{\bar{1}2}) \, {\cI}_{\a(j)}{}^{\bd(j)} (x_{1 \bar{2}}) \, .  \label{2pt shorthand B}
\end{align}
\end{subequations}
These objects are conjugate to each other
\begin{align}
    \bar{\mathfrak{I}}_{\bar{\cA}}{}^{\cB} (x_{\bar{1} 2}, x_{1 \bar{2}}) = \Big(\mathfrak{I}_{\cA}{}^{\bar{\cB}} (x_{1 \bar{2}}, x_{\bar{1} 2})\Big)^{\dagger}~,
\end{align}
and by virtue of the properties \eqref{2pt inversion properties} satisfy
\begin{subequations}
\begin{align}
    \mathfrak{I}_{\cA}{}^{\bar{\cB}} (x_{1 \bar{2}}, x_{\bar{1} 2}) \, \bar{\mathfrak{I}}_{\bar{\cB}}{}^{\cA'} (x_{1\bar{2}}, x_{\bar{1}2 }) &= \d_{\cA}{}^{\cA'} \equiv \d_{\a(i)}{}^{\a'(i)} \d_{\ad(j)}{}^{\ad'(j)} \, , \\
    \bar{\mathfrak{I}}_{\bar{\cA}}{}^{\cB} (x_{\bar{1} 2}, x_{1 \bar{2}}) \, \mathfrak{I}_{\cB}{}^{\bar{\cA}'} (x_{\bar{1}2}, x_{1\bar{2}})&= \d_{\bar{\cA}}{}^{\bar{\cA}'} \equiv \d_{\ad(i)}{}^{\ad'(i)} \d_{\a(j)}{}^{\a'(j)}\, .
\end{align}
\end{subequations}
Note that for self-conjugate representations, $\cA = \bar{\cA}$, $\cB = \bar{\cB}$. Hence, the objects \eqref{2pt shorthand A}, \eqref{2pt shorthand B} become equivalent and satisfy $\mathfrak{I}_{\cA}{}^{\cB}(x, \bar{x}) = \bar{\mathfrak{I}}_{\cA}{}^{\cB}(\bar{x}, x)$, where $x_{\a \ad}$ is an arbitrary (vector) two-point function. 

In this paper, we are interested in the conserved higher-spin supercurrent multiplet $J_{\a(s) \ad(s)}$. By using the general formula \eqref{2pt-gen} and the compact notation \eqref{2pt shorthand} with $\cA = \{ \a(s), \ad(s) \}$, $\cB = \{ \b(s), \bd(s) \}$, its two-point function is fixed by the superconformal symmetry to the form:
\bea
\langle J_{\a(s) \ad(s)} (z_1) \, J^{\bd(s) \b(s) } (z_2)\rangle
= C_{J} \frac{\cI_{\a(s)}{}^{\bd(s)} (x_{1 \bar{2}})\, \bar{\cI}_{\ad(s)}{}^{\b(s)} (x_{\bar{1} 2})}{( x_{ 1 \bar{2}}{}^2 \, x_{\bar{1} 2}{}^2 )^\frac{s+2}{2}}~,
\eea
for some real constant $C_J$. Here we have used the fact that for $J_{\a(s) \ad(s)}$, the conformal weights are $q=\bar q= \frac{s+2}{2}$.

\subsection{Three-point structures}


Given three superspace points $z_1, z_2$ and $z_3$, we have the following three-point structures ${Z}_i = ({X}_i^a, {\Q}_i^{\a}, {\bar \Q}_{i}^{\ad})$, where $i=1,2,3$ (see \cite{Park:1997bq, Osborn:1998qu} for details). The structure $Z_1$ is defined as follows:
\bsubeq \label{Z1}
\bea
{X}_{1 \a \ad} &=& (x_{\bar{2} 1}^{-1})_{\a \bd} \, x_{\bar{2} 3}^{\bd \b} \,(x_{\bar{1} 3}^{-1})_{\b \ad}~,\\
{{\Q}}_{1\a} &=& \ri \Big(({x}^{-1}_{\bar{2} 1})_{\a \bd} \bar{\q}_{12}^{\bd} -  ({x}^{-1}_{\bar{3} 1})_{\a \bd} \bar{\q}_{13}^{\bd} \Big)~, \\
{\bar{{\Q}}}_{1 \ad} &=& \ri \Big({\q}_{12}^{\a} ({x}^{-1}_{\bar{1} 2})_{\a \ad} \ - {\q}_{13}^{\a} ({x}^{-1}_{\bar{1} 3})_{\a \ad} \Big)~.
\eea
\esubeq
The objects $Z_2$ and $Z_3$ are obtained through cyclic permutations of superspace points; hence, it suffices to study the properties of \eqref{Z1}. 
As implied by \eqref{I-inv}, here we have
\bsubeq \label{Three-point bb properties}
\begin{align}
    X_1^{\ad \a} X_{1 \a \bd} = -X_1^2 \,\d^{\ad}{}_{\bd}  \hspace{5mm} \implies \hspace{5mm} (X_1^{-1})_{\a \ad} = -\frac{X_{1 \a \ad}}{X_1^2} \, ,
\end{align}
\vspace{-8mm}
\begin{align}
    X_1^2 = -\hf X_1^{\ad \a} X_{1\,\a \ad}= \frac{x_{\bar{2} 3}{}^2}{x_{\bar{2} 1}{}^2 x_{\bar{1} 3}{}^2}~.
\end{align}
\esubeq
Let us also define 
\bea
\bar{{X}}_{1 \a \ad} = ({X}_1^{\dagger})_{\a \ad} &=& {(x}_{\bar{3}1}^{-1})_{\a \bd} \,{x}_{2\bar{3} }^{\bd \b} \,({x}_{\bar{1} 2}^{-1})_{\b \ad}~.
\eea
Similar relations hold for $\bar{{X}}_{2}, \bar{{X}}_{3}$.
We now list several properties of ${Z}$'s which will be useful later:
\bsubeq \label{XbarX}
\bea
&&{\bar X}_{\a \ad} = {X}_{\a \ad} -4 \ri {\Q}_{\a} {\bar \Q}_{\ad}~, \\
&&\Q_{\a} \Q_{\b}= \hf \ve_{\a \b} \Q^{\g}\Q_{\g} = \hf \ve_{\a \b} \Q^2 ~, \hspace{5mm} \bar{\Q}_{\ad} \bar{\Q}_{\bd} = -\hf \ve_{\ad \bd} \bar{\Q}_{\gd} \bar{\Q}^{\gd}= -\hf \ve_{\ad \bd} \bar{\Q}^2~,\\
&& \frac{1}{\bar{X}^{2k}} = 
\frac{1}{{X}^{2k+2}} \Big(
X^2 -4\ri k\, \Q^{\a} \bar{\Q}^{\ad} X_{\a \ad} - 4k(k-1) \Q^2 \bar{\Q}^2 \Big)~,
\eea
\esubeq
where, throughout the paper, we adopt the notation $X^k \equiv (X^2)^{k/2}$.
In particular, we see that ${\bar X}$ is not an independent variable for it can be expressed in terms of 
${X}, {\Q}, {\bar \Q}$.
Variables ${Z}$ with different labels are related to each other via the identities
\bsubeq \label{Z13}
\bea
{x}^{\ad \b}_{\bar{1} 3} \,{X}_{3 \b \bd} \,{x}_{\bar{3} 1}^{\bd \a} &=& \frac{\bar{X}^{\ad \a}_1}{\bar{X}_1^2}~, \hspace{1.5cm} {x}^{\ad \b}_{\bar{1} 3} \,\bar{X}_{3 \b \bd} \,{x}_{\bar{3} 1}^{\bd \a} =  \frac{X_1^{\ad\a}}{X_1^2}~, \\
\frac{x_{1\bar{3} }{}^2}{x_{\bar{1} 3}{}^2} \,{x}_{\bar{1} 3}^{\ad \a} \,{{\Q}}_{3 \a} &=&
\frac{X_1^{\ad\a}}{X_1^2} {\Q}_{1 \a}~, \qquad \frac{x_{\bar{1} 3}{}^2}{x_{1\bar{3} }{}^2}  \,{\bar {\Q}}_{3\ad} \,{x}_{\bar{3} 1}^{\ad \a} =
-{\bar{{\Q}}}_{1\ad} \frac{\bar{X}^{\ad \a}_1}{\bar{X}_1^2}~.
\eea
\esubeq
Analogous to the two-point structures, it is also convenient to define the normalised three-point building blocks $\hat{X}_{\a \ad}, \hat{\Q}_{\a}, \hat{\bar{\Q}}_{\ad}$:
\begin{align}
    \hat{X}_{\a \ad} = \frac{X_{\a \ad}}{(X^{2})^{1/2}}~, \hspace{8mm} \hat{\Theta}_{\a} = \frac{\Q_{\a}}{X^{1/2}}, \hspace{8mm} \hat{\bar{\Theta}}_{\ad} = \frac{\bar{\Q}_{\ad}}{X^{1/2}} \, .
\end{align}
We then construct the following higher-spin operators:
\begin{subequations}
    \bea
    \cI_{\a(k) \ad(k)} (X) &= \hat{X}_{(\a_1 (\ad_1} \dots \hat{X}_{\a_k) \ad_k)}~, \\
    \bar{{\cI}}^{\ad(k) \a(k)}(X) &= \hat{X}^{(\ad_1 (\a_1} \dots \hat{X}^{\ad_k) \a_k)}~.
    \eea
\end{subequations}
The normalised, higher-spin operators for $\bar{X}$ can also be defined in a similar way:
\begin{subequations}
    \bea
    \cI_{\a(k) \ad(k)} (\bar{X}) &= \hat{\bar{X}}_{(\a_1 (\ad_1} \dots \hat{\bar X}_{\a_k) \ad_k)}~, \\
    \bar{{\cI}}^{\ad(k) \a(k)}(\bar{X}) &= \hat{\bar{X}}^{(\ad_1 (\a_1} \dots \hat{\bar{X}}^{\ad_k) \a_k)}~.
    \eea
\end{subequations}
Due to the properties \eqref{Three-point bb properties} (see also \eqref{2pt inversion raise and lower}), the spinor indices may be raised and lowered as follows:
\bsubeq
\bea
\cI_{\a(k)}{}^{\ad(k)} (\bar{X}) &=& 
\ve^{\ad_1 \bd_1} \dots \ve^{\ad_k \bd_k} \cI_{\a(k)\bd(k)} (\bar{X})~, \\
&=&  \ve_{\a_1 \b_1} \dots \ve_{\a_k \b_k} \bar{\cI}^{\ad{(k)} \b(k)} (\bar{X})~.
\eea
\esubeq
As such, we have
\bea
\bar{\cI}_{\ad(k)}{}^{\a(k)} (X) = \Big(\cI_{\a(k)}{}^{\ad(k)} (\bar{X}) \Big)^{\dagger}~.
\eea
The object $\bar{\cI}_{\ad(k)}{}^{\a(k)} (X)$ acts on $X_{\a \ad}$ and $\Q_{\a}$, while $\cI_{\a(k)}{}^{\ad(k)} (\bar{X})$ acts on $X_{\a \ad}$ and $\bar{\Q}_{\ad}$. In addition, we have the fundamental identities
\bsubeq
\bea
&&\cI_{\a(k)}{}^{\ad(k)}(X) \, \bar{\cI}_{\ad{(k)}}{}^{\b(k)} (X) = \d_{(\a_1}^{(\b_1} \dots \d_{\a_k)}^{\b_k)} = \d_{\a(k)}{}^{\b(k)} ~,\\
&&\bar{\cI}_{\ad(k)}{}^{\a(k)}(X) \, \cI_{\a{(k)}}{}^{\bd(k)} (X) = \d_{(\ad_1}^{(\bd_1} \dots \d_{\ad_k)}^{\bd_k)} = \d_{\ad(k)}{}^{\bd(k)}~,
\eea
\esubeq
with analogous identities holding for inversion tensors involving $\bar{X}$. For totally symmetric and traceless indices $\cA = \{ \a(i), \ad(j) \}$ (and hence, $\bar{\cA} = \{ \a(j), \ad(i) \}$) it is convenient to introduce the following compact notation:
\begin{subequations} \label{3pt shorthand}
\begin{align}
    \mathfrak{I}_{\cA}{}^{\bar{\cB}} (\bar{X}, X) &\equiv {\cI}_{\a(i)}{}^{\bd(i)} (\bar{X}) \, \bar{\cI}_{\ad(j)}{}^{\b(j)} (X) \, , \label{3pt shorthand A} \\
    \bar{\mathfrak{I}}_{\bar{\cA}}{}^{\cB} (X, \bar{X}) &\equiv \bar{\cI}_{\ad(i)}{}^{\b(i)} (X) \, {\cI}_{\a(j)}{}^{\bd(j)} (\bar{X}) \, . \label{3pt shorthand B}
\end{align} 
\end{subequations}
These objects are conjugate to each other
\begin{align}
    \bar{\mathfrak{I}}_{\bar{\cA}}{}^{\cB} (X, \bar{X}) = \Big(\mathfrak{I}_{\cA}{}^{\bar{\cB}} (\bar{X}, X)\Big)^{\dagger}~,
\end{align}
and satisfy the fundamental relations
\begin{subequations}
\begin{align}
    \mathfrak{I}_{\cA}{}^{\bar{\cB}} (\bar{X}, X) \, \bar{\mathfrak{I}}_{\bar{\cB}}{}^{\cA'} (\bar{X}, X) &= \d_{\cA}{}^{\cA'} \equiv \d_{\a(i)}{}^{\a'(i)} \d_{\ad(j)}{}^{\ad'(j)} \, , \\
    \bar{\mathfrak{I}}_{\bar{\cA}}{}^{\cB} (X, \bar{X}) \, \mathfrak{I}_{\cB}{}^{\bar{\cA}'} (X, \bar{X})&= \d_{\bar{\cA}}{}^{\bar{\cA}'} \equiv \d_{\ad(i)}{}^{\ad'(i)} \d_{\a(j)}{}^{\a'(j)}\, .
\end{align}
\end{subequations}
Note that for self-conjugate representations, $\cA = \bar{\cA}$, $\cB = \bar{\cB}$, and the objects \eqref{3pt shorthand A}, \eqref{3pt shorthand B} are equivalent and satisfy $\mathfrak{I}_{\cA}{}^{\cB}(\bar{X}, X) = \bar{\mathfrak{I}}_{\cA}{}^{\cB}(X, \bar{X})$.

Various primary superfields, such as conserved current multiplets, are subject to certain differential constraints. For imposing such constraints, the action of covariant spinor derivatives on an arbitrary function $\cH( X_3, \Q_3, \bar{\Q}_3)$ can be simplified using the following differential identities \cite{Osborn:1998qu}:
\bsubeq \label{Cderivs}
\bea
D_{(1)\a} \,\cH(\fn3) &=& - \frac{\ri}{x_{\bar{1} 3}{}^{2}} (x_{1 \bar{3} })_{\a \ad} \bar{\cD}_{(3)}^{\ad} \cH(\fn3)~,\\
\bar{D}_{(1)\ad} \,\cH(\fn3) &=& \frac{\ri}{x_{1\bar{3} }{}^2}(x_{\bar{1} 3 })_{\a \ad} {\cD}_{(3)}^{\a} \cH(\fn3)~,\\
D_{(2)\a}\, \cH(\fn3) &=& \frac{\ri}{x_{\bar{2} 3}{}^2}(x_{2 \bar{3}})_{\a \ad} \bar{\cQ}_{(3)}^{\ad} \cH(\fn3)~, \label{Cderiv-2a}\\
\bar{D}_{(2) \ad}\, \cH(\fn3) &=& -\frac{\ri}{x_{2\bar{3} }{}^2}(x_{\bar{2}3})_{\a \ad} {\cQ}_{(3)}^{\a} \cH(\fn3)~, \label{Cderiv-2b}
\eea
\esubeq
where, for $( X_3, \Q_3, \bar{\Q}_3) \longrightarrow (\fxq)$;~$\cD_{(3)}, \bar{\cD}_{(3)} \longrightarrow \cD, \bar{\cD}$ and $\cQ_{(3)}, \bar{\cQ}_{(3)}\longrightarrow \cQ, \bar{\cQ}$, we define the conformally covariant 
operators \cite{Osborn:1998qu}
\be
\begin{aligned}
&\cD_{\bar A} = (\pa / \pa X^a, \cD_{\a}, \bar{\cD}^{\ad})~, \qquad \cQ_{\bar A} = (\pa / \pa X^a, {\cal Q}_{\a},
\bar{\cal Q}^{\ad})~, \\
&\cD_{\a} = \frac{\pa}{ \pa {\Q}^{\a} }
-2{\rm i}\,(\s^a)_{\a \ad} \bar{\Q}^\ad
\frac{\pa }{ \pa X^a }~, \qquad
\bar{\cD}^{\ad} = \frac{\pa}{ \pa \bar{\Q}_{\ad} }~, \\
&{\cal Q}_{\a}  =  \frac{\pa}{ \pa {\Q}^{\a} }~, \qquad
\bar{\cal Q}^{\ad} = \frac{\pa}{ \pa \bar{\Q}_{\ad} } 
+ 2{\rm i}\, {\Q}_{\a} (\tilde{\s}^a)^{\ad \a}
\frac{\pa}{ \pa X^a}~,  \\
& [\cD_{\bar A}  ,  \cQ_{\bar B} \} ~=~0 ~.
\end{aligned}
\ee
We can also derive these anti-commutation relations
\bea
\{ \cD^{\a}, \bar{\cD}^{\ad} \} = 2 \ri (\tilde{\s}^a)^{\ad \a} \frac{\pa}{\pa X^a}~, \qquad \{ {\cQ}^{\a}, \bar{\cQ}^{\ad} \} = -2 \ri\, (\tilde{\s}^a)^{\ad \a} \frac{\pa}{\pa X^a}~.
\eea

Let $\F^{\cA_1}, \J^{\cA_2}$ and $\Pi^{\cA_3}$ be primary superfields with conformal weights $(q_1, \bar{q}_1),(q_2, \bar{q}_2)$ and $(q_3, \bar{q}_3)$ respectively. Then, the three-point correlation function may be constructed using the general expression \cite{Osborn:1998qu, Park:1997bq,Park:1999pd}:
\bea \label{3ptgen}
&&\langle
\F_{\cA_1} (z_1) \, \J_{\cA_2}(z_2)\,  \Pi_{\cA_3}(z_3)
\rangle 
=\frac{ 
\mathfrak{I}_{\cA_1}{}^{\bar{\cA}'_1} ({{x}_{1 \bar{3}}}, {{x}_{\bar{1}3}})\,
\mathfrak{I}_{\cA_2}{}^{\bar{\cA}'_2} ({{x}_{2 \bar{3}}}, {{x}_{\bar{2}3}})\,
}
{ 
(x_{1\bar{3}}{}^2)^{q_1} (x_{\bar{1}3}{}^2)^{\bar{q}_1} 
(x_{2\bar{3}}{}^2)^{q_2} (x_{\bar{2}3}{}^2)^{\bar{q}_2} 
}
\cH_{\bar{\cA}'_1 \bar{\cA}'_2 \cA_3} (\fn3)~,~~~~~~~~
\eea 
where the functional form of the tensor $\cH_{\bar{\cA}_1 \bar{\cA_2} \cA_3}$ is constrained by $\cN=1$ superconformal symmetry as follows:
\begin{itemize}

\item[\textbf{(i)}] By construction, the superfield ansatz~\eqref{3ptgen} has the correct transformation properties at superspace points $z_1$ and $z_2$. However, to ensure that the three-point function has the correct scaling properties at $z_{3}$ we require that $\cH$ possesses the homogeneity property
\begin{subequations} \label{Homogeneity property}
    \begin{align}
    & \cH_{\bar{\cA}_1 \bar{\cA}_2 \cA_3} ( \l \bar{\l} {X},
    \l {\Q}, \bar{\l} \bar {\Q}) =
    \l^{2a} \bar{\l}^{2\bar{a}}
    \cH_{\bar{\cA}_{1} \bar{\cA}_{2} \cA_3}( {X}, {\Q}, \bar {\Q})~,
    \end{align}
\end{subequations}
where $a$ and $\bar{a}$ are determined by the requirements of invariance under combined scale and $\rm{U(1)}_R$-symmetry transformations \cite{Osborn:1998qu, Kuzenko:1999pi}. Invariance of the three-point function under R-symmetry requires
\be
a - \bar{a} = - \frac{1}{3} \sum_{i} ( q_i - \bar{q}_i) = 0 \, .
\label{r-sym-constraint}
\ee
Therefore $a = \bar{a}$ and we require that the total $\rm{U(1)}_R$-symmetry charge vanishes. On the other hand, invariance under scale transformations requires 
\begin{align}
    a+\bar{a} &= q_{3} + \bar{q}_{3} - q_{1} - \bar{q}_{1} - q_{2} - \bar{q}_{2} = \D_{3} - \D_{1} - \D_{2} \, ,
\end{align}
and, hence
\begin{align}
    a &= \bar{a} = \frac{1}{2}( \D_{3} - \D_{1} - \D_{2} ) \, .
\end{align}

\item[\textbf{(ii)}]If any of the superfields $\F$, $\J$ and $\P$ obey differential equations (e.g. conservation laws for conserved current multiplets), then $ \cH_{\bar{\cA}_{1} \bar{\cA}_{2} \cA_3}$
is constrained by certain differential equations too. The latter may be derived using \eqref{Cderivs}.

\item[\textbf{(iii)}] If any (or all) of
the superfields $\F$, $\J$ and $\P$ coincide,
then $ \cH_{\bar{\cA}_1 \bar{\cA}_{2} \cA_3}$
obeys additional constraints, the so-called ``point-switch symmetries". These are consequences of the symmetry
under permutations of superspace points. As an example,
\be
\langle \Phi_{{\cA_{1}}}(z_1) \Phi_{{\cA_{2}}}(z_2)
\P^{{\cA_{3}}}(z_3) \rangle =
(-1)^{\epsilon(\Phi)}
\langle \Phi_{{\cA_{2}}}(z_2) \Phi_{{\cA_{1}}}(z_1)
\P_{{\cA_{3}}}(z_3) \rangle~,
\ee
where $\epsilon(\Phi)$ denotes the Grassmann parity of $\Phi_{{\cA}}$. Note that under permutations of any two superspace points, the three-point building blocks transform as
\bsubeq
\bea
			{X}_{3 \, \a \ad} &\stackrel{1 \leftrightarrow 2}{\longrightarrow} - \bar{X}_{3 \, \a \ad} \, , \hspace{10mm} {\Q}_{3 \, \a} \stackrel{1 \leftrightarrow 2}{\longrightarrow} - {\Q}_{3 \, \a} \, , \label{pt12} \\[2mm]
			{X}_{3 \, \a \ad} &\stackrel{2 \leftrightarrow 3}{\longrightarrow} - \bar{X}_{2 \, \a \ad} \, , \hspace{10mm} {\Q}_{3 \, \a} \stackrel{2 \leftrightarrow 3}{\longrightarrow} - {\Q}_{2 \, \a} \, , \label{pt23} \\[2mm]
			{X}_{3 \, \a \ad} &\stackrel{1 \leftrightarrow 3}{\longrightarrow} - \bar{X}_{1 \, \a \ad} \, , \hspace{10mm} {\Q}_{3 \, \a} \stackrel{1 \leftrightarrow 3}{\longrightarrow} - {\Q}_{1 \, \a} \, . \label{pt13}
\eea
\esubeq
\end{itemize}
The above conditions fix the functional form of $ \cH_{\bar{\cA}_1 \bar{\cA}_2 \cA_3}$ (and, therefore, the three-point function under consideration) up to a few arbitrary constants.

A few comments are in order regarding the three-point functions of conserved current multiplets. It is worth pointing out that, depending on the exact way in which one constructs the general expression \eqref{3ptgen}, it can be impractical to impose conservation equations on one of the three superfields due to a lack of useful identities such as \eqref{Cderivs}. To illustrate this, let us go back to eq. \eqref{3ptgen}:
\bea 
&&\langle
\F_{\cA_1} (z_1) \, \J_{\cA_2}(z_2)\,  \Pi_{\cA_3}(z_3)
\rangle 
=\frac{ 
\mathfrak{I}_{\cA_1}{}^{\bar{\cA}'_1} ({{x}_{1 \bar{3}}}, {{x}_{\bar{1}3}})\,
\mathfrak{I}_{\cA_2}{}^{\bar{\cA}'_2} ({{x}_{2 \bar{3}}}, {{x}_{\bar{2}3}})\,
}
{ 
(x_{1\bar{3}}{}^2)^{q_1} (x_{\bar{1}3}{}^2)^{\bar{q}_1} 
(x_{2\bar{3}}{}^2)^{q_2} (x_{\bar{2}3}{}^2)^{\bar{q}_2}  
}
\cH_{\bar{\cA}'_1 \bar{\cA}'_2 \cA_3} (\fn3)~.~~~~~~~~
\label{3ptgen-new}
\eea 
All information about this correlation function is encoded in the tensor $\cH$, however, with this formulation of the three-point function it is difficult to impose conservation on $\P$ in a straightforward way. A way around this problem is to rearrange the fields in the three-point function so that $\P$ is at the first point:
\bea 
&&\langle
 \Pi_{\cA_3}(z_3) \, \J_{\cA_2}(z_2)\, \F_{\cA_1} (z_1) 
\rangle 
=\frac{ 
\mathfrak{I}_{\cA_3}{}^{\bar{\cA}'_3} ({{x}_{3 \bar{1}}}, {{x}_{\bar{3}1}})\,
\mathfrak{I}_{\cA_2}{}^{\bar{\cA}'_2} ({{x}_{2 \bar{1}}}, {{x}_{\bar{2}1}})\,
}
{ 
(x_{3\bar{1}}{}^2)^{q_3} (x_{\bar{3}1}{}^2)^{\bar{q}_3} 
(x_{2\bar{1}}{}^2)^{q_2} (x_{\bar{2}1}{}^2)^{\bar{q}_2} 
}
\tilde{\cH}_{\bar{\cA}'_3 \bar{\cA}'_2 \cA_1} (X_{1}, \Q_{1}, \bar{\Q}_{1})~.~~~~~~~~
\label{3ptgen-231}
\eea 
Here, all information about the correlator is now encoded in the tensor $\tilde{\cH}$, which is a completely different solution compared to $\cH$. 
Thus, we require a simple equation relating the tensors $\cH$ and $\tilde{\cH}$, which corresponds to different representations of the same correlation function.
Indeed, once $\tilde{\cH}$ is obtained, we can then easily impose conservation on $\Pi$ as if it were located at the ``first point'', with the aid of identities analogous 
to \eqref{Cderiv-2a} and \eqref{Cderiv-2b}. In Subsection \ref{section3.2} we will obtain an explicit formula relating $\tilde{\cH}$ and $\cH$, which allows us to easily impose conservation on $z_{3}$.

\section{Three-point functions: general formalism}
\label{section3}

First let us review the constraints which we must impose on the three-point function of higher-spin supercurrents of the form 
$J_{\a (i) \ad (j)}$. In this work we are primarily interested in the three-point functions of current multiplets with $i, j > 0$. Cases with $i = j = 0$, $i=0$, or $j=0$ correspond to special cases of three-point functions involving flavour currents or (higher-spin) spinor currents, which were analysed in \cite{Buchbinder:2021izb, Buchbinder:2021kjk, Buchbinder:2022kmj}.

The ansatz for the three-point function of three conserved supercurrents, consistent with the general formula \eqref{3ptgen} (after expanding out the compact notation), is given by
\bea \label{3HS-currents}
&& \langle J^{}_{\a(i_1) \ad(j_1)} (z_1) \, J'_{\b(i_2) \bd(j_2)}(z_2)\,  J''_{\g(i_3) \gd(j_3)}(z_3)
\rangle \non\\
&& \hspace{10mm} =\frac{\cI_{\a(i_1)}{}^{\ad'(i_1)} (x_{1 \bar{3}}) \,\bar{\cI}_{{\ad}(j_1)}{}^{\a'(j_1)} (x_{\bar{1} 3}) 
\, \cI_{\b(i_2)}{}^{\bd'(i_2)} (x_{2 \bar{3}}) 
\,\bar{\cI}_{{\bd}(j_2)}{}^{\b'(j_2)} (x_{\bar{2} 3})}{ (x_{1\bar{3} }{}^{2})^{q_{1}} (x_{\bar{1} 3}{}^{2})^{\bar{q}_{1}} (x_{2\bar{3}}{}^{2})^{q_{2}} (x_{\bar{2} 3}{}^{2})^{\bar{q}_{2}} } \non\\
&& \hspace{20mm} \times \cH_{\a'(j_1) \ad'(i_1) \b'(j_2) \bd'(i_2) \g(i_3) \gd(j_3)} (\fn3)~.~~~~~~~~
\eea 
We recall that $J_{\a(i) \ad(j)}$ is a primary superfield with conformal weight $(q, \bar{q}) = (\frac{i}{2}+1, \frac{j}{2}+1)$ and scaling dimension $\D = q+\bar{q}$. Assuming that all the operators are distinct, the three-point function \eqref{3HS-currents} is subject to the following constraints:
\begin{enumerate}
	\item[\textbf{(i)}] \textbf{Homogeneity and $\rm U(1)_R$-symmetry:} The tensor $\cH$ has the scaling property
	\begin{align}
	    &\cH_{\a(j_1) \ad(i_1) \b(j_2) \bd(i_2) \g(i_3) \gd(j_3)} (\l \bar{\l} X, \l \Q, \bar{\l} \bar{\Q}) \non \\
        &\hspace{30mm} = \l^{2a} \bar{\l}^{2 \bar{a}} \cH_{\a(j_1) \ad(i_1) \b(j_2) \bd(i_2) \g(i_3) \gd(j_3)} (\fxq)~,	
        \label{r-sym-H}
	\end{align}
    where $a = \bar{a} = \hf (\D_{3} - \D_{2} - \D_{1})$, in accordance with \eqref{r-sym-constraint}. This ensures that the three-point function does not carry a non-trivial $\rm U(1)_R$-symmetry charge and transforms correctly under scale transformations. The $\rm U(1)_R$ invariance also amounts to requiring $i_1+ i_2+ i_3 = j_1+j_2+ j_3$.
	\item[\textbf{(ii)}] \textbf{Conservation:} The conservation of $J_{\a(i) \ad(j)}$ at $z_{1}$ and $z_{2}$ (for $i,j>1$) imply
    \bsubeq
    \bea
    D^{\a_1}_{(1)} \langle J^{}_{\a(i_1) \ad(j_1)} (z_1) \, J'_{\b(i_2) \bd(j_2)}(z_2)\,  J''_{\g(i_3) \gd(j_3)}(z_3)
    \rangle &=& 0~, \\
    \bar D^{\ad_1}_{(1)} \langle J^{}_{\a(i_1) \ad(j_1)} (z_1) \, J'_{\b(i_2) \bd(j_2)}(z_2)\,  J''_{\g(i_3) \gd(j_3)}(z_3)
    \rangle &=& 0~, \\
    D^{\b_1}_{(2)} \langle J^{}_{\a(i_1) \ad(j_1)} (z_1) \, J'_{\b(i_2) \bd(j_2)}(z_2)\,  J''_{\g(i_3) \gd(j_3)}(z_3)
    \rangle &=& 0~, \\
    \bar D^{\bd_1}_{(2)} \langle J^{}_{\a(i_1) \ad(j_1)} (z_1) \, J'_{\b(i_2) \bd(j_2)}(z_2)\,  J''_{\g(i_3) \gd(j_3)}(z_3)
    \rangle &=& 0~.
    \eea
    \esubeq
    With the use of identities \eqref{Cderivs}, these requirements are translated to the following differential constraints on $\cH$:
    \bsubeq \label{ceq}
    \bea
    &&\bar{\cD}^{\dot{\d}} \cH_{\a(i_1) \dd\ad(j_1-1) \b(i_2) \bd(j_2) \g(i_3) \gd(j_3)} = 0~,
    \label{ceq1}\\
    &&{\cD}^{{\d}} \cH_{\d \a(i_1-1) \ad(j_1) \b(i_2) \bd(j_2) \g(i_3) \gd(j_3)} = 0~, \label{ceq2}\\
    &&\bar{\cQ}^{\dot{\d}} \cH_{\a(i_1) \ad(j_1) \b(i_2) \dd \bd(j_2-1) \g(i_3) \gd(j_3)} = 0~, \label{ceq3}\\
    &&\cQ^{\d} \cH_{\a(i_1) \ad(j_1) \d \b(i_2-1) \bd(j_2) \g(i_3) \gd(j_3)} = 0~.\label{ceq4}
    \eea
    \esubeq
    There are further constraints arising from the conservation at $z_3$:
    \bea
    D^{\g_1}_{(3)} \langle J^{}_{\a(i_1) \ad(j_1)} (z_1) \, J'_{\b(i_2) \bd(j_2)}(z_2)\,  J''_{\g(i_3) \gd(j_3)}(z_3)
    \rangle &=& 0~, \\
    \bar D^{\gd_1}_{(3)} \langle J^{}_{\a(i_1) \ad(j_1)} (z_1) \, J'_{\b(i_2) \bd(j_2)}(z_2)\,  J''_{\g(i_3) \gd(j_3)}(z_3)
    \rangle &=& 0~.	
    \eea
    However, these constraints are considerably more difficult to impose as there are no identities analogous to \eqref{Cderivs} which allow the spinor derivatives $D^{\a}_{(3)},\bar{D}^{\ad}_{(3)}$ to pass through the prefactor of \eqref{3HS-currents}. Hence, we employ the procedure outlined at the end of subsection \ref{section3.2}, where we utilise the reformulated ansatz eq.\,\eqref{3ptgen-231} which is described in terms of the tensor $\tilde{\cH}$. The details of this calculation will be outlined in the next subsection.

    \item[\textbf{(iii)}] \textbf{Reality:} In the case where the higher-spin supercurrents are ``vector-like" or bosonic, i.e. they are of the form $J_{\a(s) \ad(s)}$, we must impose a superfield reality condition on the three-point function, which leads to the following constraint on the tensor $\cH$:
    \bea \label{zh2}
    \cH_{\a(s_{1}) \ad(s_1) \b(s_2) \bd(s_2) \g(s_3) \gd(s_3)} (\fxq) = \bar{\cH}_{\a(s_1) \ad(s_1) \b(s_2) \bd(s_2) \g(s_3) \gd(s_3)} (\fxq),~~~~
    \eea
    where $\bar{\cH}(\fxq)$ is the conjugate of $\cH(\fxq)$. 

\end{enumerate}
The problem of computing higher-spin correlator \eqref{3HS-currents} is therefore reduced to determining the most general form of $\cH(\fxq)$ subject to the above constraints. 

In 4D $\cN=1$ SCFT, a useful observation regarding the general structure of three-point functions of conserved supercurrents is that they must have a vanishing total R-symmetry charge. As shown in \eqref{r-sym-H}, this condition implies that $\cH$ must be Grassmann even; hence, is an even function of $\Q$ and $\bar{\Q}$. We can write a general expansion for $\cH$ as follows:
\bea
\cH(\fxq) &=& F(X) + \Q^{\d} \bar{\Q}^{\dd} G_{\d \dot{\d}} (X) \non\\
&& + A^{(1)} (X) \,\Q^2 + A^{(2)} (X) \,\bar{\Q}^2 + A^{(3)} (X)\, \Q^2 \bar{\Q}^2~.
\eea
Now, conservation conditions \eqref{ceq1} and \eqref{ceq4} at $z_1$ and $z_2$ allow us to impose weaker constraints which read
\bea
\bar{\cD}^2 \cH(\fxq)=0~, \qquad
{\cQ}^2 \cH(\fxq)=0~.
\eea
The constraints above imply
\bea
A^{(1)} (X) = A^{(2)} (X)= A^{(3)} (X) = 0~,
\eea
hence, the general solution for $\cH(\fxq)$ can always be presented in the form \cite{Buchbinder:2022kmj}
\begin{align} \label{H-FG}
\cH_{\a(i_1) \ad(j_1) \b(i_2) \bd(j_2) \g(i_3) \gd(j_3)} (\fxq) &=F_{\a(i_1) \ad(j_1) \b(i_2) \bd(j_2)\g(i_3) \gd(j_3)} ({X}) \\
&\quad + \Q^{\d} \bar{\Q}^{\dd} G_{\d \dot{\d}, \a(i_1) \ad(j_1) \b(i_2) \bd(j_2) \g(i_3) \gd(j_3)} ({X}) \, . \non
\end{align}
The tensor $\cH (X, \Theta, \bar{\Theta})$ is constructed from totally symmetric and traceless combinations of the following tensor building blocks:
\begin{subequations} \label{building blocks}
    \begin{align}
    \ve_{\a \b}~, \hspace{8mm} \ve_{\ad \bd}~, \hspace{8mm} X_{\a \ad} \, , \hspace{8mm} \Theta_{\a} \, , \hspace{8mm} \bar{\Theta}_{\ad} \, , 
    \end{align}
    \vspace{-10mm}
    \begin{align}
    \hspace{10mm} (X \cdot \Theta)_{\ad} = X_{\a \ad} \Theta^{\a}~, \hspace{4mm} (X \cdot \bar{\Theta})_{\a} = X_{\a \ad} \bar{\Theta}^{\ad}~, \hspace{4mm} J = \Theta^{\a} \bar{\Theta}^{\ad} X_{\a \ad} \, .
\end{align}
\end{subequations}
The task is to construct a general solution for $\cH$ for supercurrent multiplets of arbitrary superspins using the building blocks \eqref{building blocks}. For this, it is advantageous to develop a supersymmetric extension of the generating function formalism proposed in the 4D CFT case \cite{Buchbinder:2023coi}, which encodes the tensor structure of $\cH$ in a polynomial. Before we develop the supersymmetric generating function formalism it is useful understand how superinversion acts on $\cH$ and to develop an effective method to impose conservation conditions at the third superspace point.

\subsection{Superinversion transformation}\label{section3.1}


Before we discuss how to impose conservation on the third point, it is useful to understand precisely how superinversion acts on the 
tensor $\cH_{\bar{\cA}_{1}\bar{\cA}_{2} \cA_{3}}(X,\Q,\bar{\Q})$. The action of superinversion on the three-point building blocks is defined similarly to its action on the superspace coordinates as in \cite{Buchbinder:1998qv} (see also \cite{Gates:1983nr}). Under the action of superinversion, the tensor $\cH(X,\Q,\bar{\Q})$ transforms into a tensor $\cH^{I}(X, \Q, \bar{\Q} )$ 
which we define as follows:
\begin{align} \label{InvertH2}
\cH^{I}_{\cA_{1} \cA_{2} \bar{\cA}_{3}}(X, \Q, \bar{\Q} ) = \mathfrak{I}_{\cA_{1}}{}^{\bar{\cA}'_{1}} (\bar{X}, X) \, 
\mathfrak{I}_{\cA_{2}}{}^{\bar{\cA}'_{2}} (\bar{X}, X) \, \bar{\mathfrak{I}}_{\bar{\cA}_{3}}{}^{\cA'_{3}} ( X, \bar{X}) \, \cH_{\bar{\cA}'_{1}\bar{\cA}'_{2} \cA'_{3}}(X,\Q,\bar{\Q}) \,, 
\end{align}
where $\cA_{1} = \{ \a(i_{1}), \ad(j_{1}) \}$, $\cA_{2} = \{ \b(i_{2}), \bd(j_{2}) \}$, $\cA_{3} = \{ \g(i_{3}), \gd(j_{3}) \}$. 
Note that a superinversion transformation transforms the representation $(\bar{\cA}_{1}, \bar{\cA}_{2}, \cA_{3})$ into its complex conjugate.
From eq.~\eqref{InvertH2} it follows that $X, \Q, \bar{\Q}$ transform under superinversion as
\begin{subequations} \label{inverted-Z3}
\begin{align}
    \bar{X}^I_{\a \ad} :=& \; \bar{\cI}_{\ad}{}^{\b}(X) \, {\cI}_{\a}{}^{\bd}(\bar{X}) \, X_{\b \bd}  & X^I_{\a \ad} :=& \; \bar{\cI}_{\ad}{}^{\b}(X) \, {\cI}_{\a}{}^{\bd}(\bar{X}) \, \bar{X}_{\b \bd} \\
    =& - \bigg(\frac{X^{2}}{\bar{X}^{2}} \bigg)^{\frac{1}{2}} \bar{X}_{ \a \ad} ~, & =& - \bigg(\frac{\bar{X}^{2}}{X^{2}} \bigg)^{\frac{1}{2}} X_{ \a \ad} \, , \non \\[2mm]
    \bar{\Q}_{\ad}^I :=& \; \bar{\cI}_{\ad}{}^{\a} (X) \, \Q_{\a} \, , & \Q_{\a}^I :=& \; \cI_{\a}{}^{\ad} (\bar{X}) \, \bar{\Q}_{\ad} \, .
\end{align}
\end{subequations}
Note that the definitions \eqref{inverted-Z3} are slightly different compared to \cite{Osborn:1998qu, Buchbinder:2021kjk, Buchbinder:2021izb, Buchbinder:2022kmj}, however, it is now more apparent that the action of 
superinversion is a type of pseudo-conjugation. The tensor $\cH$ is constructed from the building blocks \eqref{building blocks}, 
and by using the definitions for the inversion operators, in addition to \eqref{inverted-Z3}, it follows from eq.~\eqref{InvertH2} that
\begin{subequations}  \label{Z13-hat-transform}
\begin{align} 
     \bar{\cI}_{\ad}{}^{\a}(X) \, \bar{\cI}_{\bd}{}^{\b}(X) \, \ve_{ \a \b} &= -\ve_{ \ad \bd}~, & \cI_{\a}{}^{\ad}(\bar{X}) \, \cI_{\b}{}^{\bd}(\bar{X}) \, \ve_{ \ad \bd} &= - \ve_{ \a \b}~, \\[2mm]
     \bar{\cI}_{\ad}{}^{\b}(X) \, {\cI}_{\a}{}^{\bd}(\bar{X}) \, X_{ \b \bd} &= \bar{X}^I_{\a \ad}~, & \bar{\cI}_{\ad}{}^{\b}(X) \, {\cI}_{\a}{}^{\bd}(\bar{X}) \, \bar{X}_{ \b \bd} &= X^I_{\a \ad} \, , \\[2mm]
      \bar{\cI}_{\ad}{}^{\a} (X) \, \Q_{\a} &= \bar{\Q}_{\ad}^I\, , &   \cI_{\a}{}^{\ad} (\bar{X}) \, \bar{\Q}_{\ad} &= \Q_{\a}^I \, ,
\end{align}
\vspace{-9mm}
\begin{align} 
    \cI_{\a}{}^{\ad}(\bar{X}) (X \cdot \Q)_{\ad} &= - ( \bar{X}^{I} \cdot \bar{\Q}^{I})_{\a}~, & 
    \bar{\cI}_{\ad}{}^{\a} (X) \, (X \cdot \bar{\Q})_{\a} &= - (\bar{X}^{I} \cdot \Q^{I})_{\ad}~,
\end{align}
\vspace{-9mm}
\begin{align}
    J = \Q^{\a} \bar{\Q}^{\ad} X_{\ad \a} = \bar{\Q}^{I \ad} \Q^{I \a} \bar{X}^{I}_{\ad \a} \equiv \bar{J}^I~.
\end{align}
\end{subequations}
Hence, the action of superinversion is equivalent to the following replacement rules:
\begin{subequations} \label{inversion replacements}
\begin{align} \label{even-X}
\begin{split}  
    & \hspace{10mm} X_{\a \ad} \stackrel{\cI_{X,\bar{X}}}{\longrightarrow} \bar{X}^I_{\a \ad}~,  \hspace{10mm} \bar{X}_{\a \ad} \stackrel{\cI_{X,\bar{X}}}{\longrightarrow} X^I_{\a \ad}~, \\[1mm]
    &{\Q}_{\a} \stackrel{\cI_{X}}{\longrightarrow} \, {\bar{\Q}}^I_{\ad}~, \hspace{10mm}
    \bar{\Q}_{\ad} \stackrel{\cI_{\bar{X}}}{\longrightarrow} \Q^I_{\a}~, \hspace{10mm} J \stackrel{\cI_{X, \bar{X}}}{\longrightarrow}\bar{J}^{I}~,~~~~~~~ 
\end{split}
\end{align} 
\vspace{-5mm}
\begin{align} \label{odd-X}
\begin{split}
    & \hspace{25mm} \ve_{\a \b} \stackrel{\cI_{{X}}}{\longrightarrow} -\ve_{\ad \bd}~, \hspace{10mm}
    \ve_{\ad \bd} \stackrel{\cI_{\bar{X}}}{\longrightarrow} -\ve_{\a \b}~, \\[1mm]
    &(X \cdot \Q)_{\ad} \stackrel{\cI_{\bar{X}}}{\longrightarrow} - (\bar{X}^I \cdot \bar{\Q}^I)_{\a}~, \hspace{10mm}
    (X \cdot \bar{\Q})_{\a} \stackrel{\cI_{{X}}}{\longrightarrow} - (\bar{X}^I \cdot \Q^I)_{\ad}~. 
\end{split}
\end{align}
\end{subequations}
Note that the transformation properties of the objects above appear analogous to conjugation, except with an overall sign introduced for the objects \eqref{odd-X}. As a result, we obtain
\begin{align} \label{inversion formula}
    \cH^{I}_{\cA_{1} \cA_{2} \bar{\cA}_{3}}(X, \Q, \bar{\Q} ) = \cH_{\cA_{1} \cA_{2} \bar{\cA}_{3}}
    (\ve \rightarrow - \bar{\ve}, \bar{\ve} \rightarrow - \ve, X \rightarrow \bar{X}^{I}, \Q \rightarrow \bar{\Q}^{I}, \bar{\Q} \rightarrow \Q^{I} ) \, .
\end{align}
%

If the superfields in the three-point function are ``vector-like", then $\cA_{i} = \bar{\cA}_{i}$ and there exists a natural map from $\cH$ onto itself 
under superinversion, as $\cH$ and $\cH^{I}$ now belong to the same representation space. Hence, we can impose the condition
\begin{align} \label{Parity condition}
    \cH_{\cA_{1} \cA_{2} \cA_{3}}(X^{I}, \Q^{I}, \bar{\Q}^{I} ) = \pm \, \cH^{I}_{\cA_{1} \cA_{2} \cA_{3}}(X, \Q, \bar{\Q} ) \, .
\end{align}
Using eq.~\eqref{inversion formula} it can also be written as
\begin{align} \label{Parity conditionzh1}
    \cH_{\cA_{1} \cA_{2} \cA_{3}}(X, \Q, \bar{\Q}) = \pm \, 
\cH_{\cA_{1} \cA_{2} \cA_{3}}
    (\ve \rightarrow - \bar{\ve}, \bar{\ve} \rightarrow - \ve, X \rightarrow \bar{X}, \Q \rightarrow \bar{\Q}, \bar{\Q} \rightarrow \Q ) \, .
\end{align}
%
We will classify the tensors $\cH^{E}$ which satisfy \eqref{Parity condition} with ``$+$" as parity-even under superinversion, and the tensors $\cH^{O}$ which satisfy \eqref{Parity condition} 
with ``$-$" as parity-odd under superinversion. 

Since superinversion acts as a pseudo-conjugation, a consequence of the condition \eqref{Parity condition} is that $\cH$ can be decomposed as follows:
\begin{subequations}
\begin{align}
    \cH^{E}(X,\Q,\bar{\Q}) &= \sum^{N_{E}}_{i = 1} c^{E}_{i} \cH^{E}_{i}(X,\Q,\bar{\Q}) \, , &
    \cH^{O}(X,\Q,\bar{\Q}) &= \sum^{N_{O}}_{i = 1} c^{O}_{i} \cH^{O}_{i}(X,\Q,\bar{\Q}) \, , \\[2mm]
    \cH^{E}_{i}(X,\Q,\bar{\Q}) &= \bar{\cH}^{E}_{i}(X,\Q,\bar{\Q}) \, , & \cH^{O}_{i}(X,\Q,\bar{\Q}) &= - \bar{\cH}^{O}_{i}(X,\Q,\bar{\Q}) \, , 
\end{align}
\end{subequations}
where $c^{E}_{i}$, $c^{O}_{i}$ are complex parameters, and $\cH^{E}_{i}$, $\cH^{O}_{i}$ are linearly independent bases of tensor structures which are hermitian and 
anti-hermitian respectively. A basic proof of the result above is outlined in Appendix \ref{AppD}. 

If the superfields in the three-point functions are ``vector-like" and real, then imposing the reality condition \eqref{zh2} 
further implies $\text{Re}[c^{O}_{i}] = \text{Im}[c^{E}_{i}] = 0$, so that $\cH$ is of the form
\begin{align}
    \cH(X,\Q,\bar{\Q}) = \sum^{N_{E}}_{i} a_{i} \cH^{E}_{i}(X,\Q,\bar{\Q}) + \text{i} \sum^{N_{O}}_{i} b_{i} \cH^{O}_{i}(X,\Q,\bar{\Q}) \, .
    \label{zh1}
\end{align}
where $a_{i}$, $b_{i}$ are real parameters. This decomposition is  useful for studying three-point functions of conserved ``vector-like" supercurrents
and classifying the results in terms of contributions which are parity-even or parity-odd in superspace. In fact, all results presented in the later sections are of the form~\eqref{zh1}. 
This is a generic feature of three-point functions of real superfields which satisfy the superinversion pseudo-covariance condition \eqref{Parity condition}. 
Similar results were obtained for the 4D CFT case in \cite{Buchbinder:2023coi}.
Imposing conservation conditions and point-switch symmetries on $\cH$ results in relations among the $a_{i}$, and relations among the $b_{i}$. 


\subsection{\texorpdfstring{Conservation on $z_{3}$}{Conservation on z3} }\label{section3.2}


We now turn to analysing the conservation conditions on the third superspace point, $z_3$, for the general ansatz \eqref{3HS-currents}. We may reformulate the ansatz with $J''$ at the front as follows:
\begin{align} \label{HtildeAnsatz}
   \langle J''_{\cA_{3}}(z_3)\, J'_{\cA_{2}}(z_2)\,   J^{}_{\cA_{1}} (z_1) \rangle
    &= \frac{\mathfrak{I}_{\cA_{3}}{}^{\bar{\cA}_{3}'} (x_{3 \bar{1}}, x_{\bar{3} 1}) \,
    \mathfrak{I}_{\cA_{2}}{}^{\bar{\cA}_{2}'} (x_{2 \bar{1}}, x_{\bar{2} 1}) }{ (x_{3 \bar{1} }{}^{2})^{q_{3}} (x_{\bar{3} 1}{}^{2})^{\bar{q}_{3}} (x_{2\bar{1} }{}^{2})^{q_{2}} (x_{\bar{2} 1}{}^{2})^{\bar{q}_{2}} } \, \tilde{\cH}_{\bar{\cA}'_{3}\bar{\cA}'_{2} \cA_{1}} (X_1, \Q_1, \bar{\Q}_1)~.
\end{align}
For superfield representations $\cA_{1} = \{ \a(i_{1}), \ad(j_{1}) \}$, $\cA_{2} = \{ \b(i_{2}), \bd(j_{2}) \}$, $\cA_{3} = \{ \g(i_{3}), \gd(j_{3}) \}$, the result~\eqref{HtildeAnsatz} 
is equivalent to the following after expanding out the compact notation \eqref{3pt shorthand}
\begin{align}
    \langle J''_{\g(i_3) \gd(j_3)}(z_3)\, & J'_{\b(i_2) \bd(j_2)}(z_2)\,   J^{}_{\a(i_1) \ad(j_1)} (z_1) \rangle \non\\
    &= \frac{\cI_{\g(i_3)}{}^{\gd'(i_3)} (x_{3 \bar{1}}) \,
    \bar{\cI}_{\gd(j_3)}{}^{\g'(j_3)}(x_{\bar{3} 1}) \, \cI_{\b(i_2)}{}^{\bd'(i_2)} (x_{2 \bar{1}}) \,\bar{\cI}_{\bd(j_2)}{}^{\b'(j_2)} (x_{ \bar{2} 1}) }{ (x_{3\bar{1} }{}^{2})^{q_{3}} (x_{\bar{3} 1}{}^{2})^{\bar{q}_{3}} (x_{2\bar{1}}{}^{2})^{q_{2}} (x_{\bar{2} 1}{}^{2})^{\bar{q}_{2}} }\non\\
    &\qquad \times \tilde{\cH}_{\g'(j_3) \gd'(i_3) \b'(j_2) \bd'(i_2) \a(i_1) \ad(j_1)} (X_1, \Q_1, \bar{\Q}_1)~.
\end{align}
In what follows, we will proceed with the calculation using the compact notation. 

We now equate the reformulated ansatz above to \eqref{3HS-currents}; after some manipulations we obtain the following equation relating the representations $\tilde{\cH}$ and $\cH$:
\begin{align} \label{e0}
\begin{split}
    \tilde{\cH}_{\bar{\cA}_{3}\bar{\cA}_{2}\cA_{1}} (X_1, \Q_1, \bar{\Q}_1)
    &= (-1)^{i_{2} + j_{2} + i_{3} + j_{3}}\frac{(x_{3\bar{1} }{}^{2})^{q_{3}} (x_{\bar{3} 1}{}^{2})^{\bar{q}_{3}} (x_{2 \bar{1} }{}^{2})^{q_{2}} (x_{\bar{2} 1}{}^{2})^{\bar{q}_{2}}}{ (x_{1\bar{3} }{}^{2})^{q_{1}} (x_{\bar{1} 3}{}^{2})^{\bar{q}_{1}} (x_{2 \bar{3}}{}^{2})^{q_{2}} (x_{\bar{2} 3}{}^{2})^{\bar{q}_{2}}} \\
    & \times \mathfrak{I}_{\cA_{1}}{}^{\bar{\cA}'_{1}}(x_{1\bar{3}}, x_{\bar{1}3}) \, \bar{\mathfrak{I}}_{\bar{\cA}_{2}}{}^{\cB}(x_{\bar{1}2}, x_{1\bar{2}}) \, \mathfrak{I}_{\cB}{}^{\bar{\cA}'_{2}}(x_{2\bar{3}}, x_{\bar{2}3}) \\
    & \times \bar{\mathfrak{I}}_{\bar{\cA}_{3}}{}^{\cA'_{3}}(x_{\bar{1}3}, x_{1\bar{3}}) \, 
    \cH_{\bar{\cA}'_{1} \bar{\cA}'_{2} \cA'_{3}}(\fn3)~. ~~~~ 
\end{split}
\end{align}
By making use of the relations
\bsubeq
\begin{align}
    \bar{\cI}_{\ad(k)}{}^{\g(k)} (x_{ \bar{1} 2}) \, \cI_{\g(k)}{}^{\bd(k)}(x_{{2} \bar{3}})
    &= \bar{\cI}_{\ad(k)}{}^{\g(k)}(\bar{X}_{1}) \, \cI_{\g(k)}{}^{\bd(k)} (x_{1 \bar{3}})~,\\
    \cI_{\a(k)}{}^{\gd(k)}(x_{1\bar{2}}) \, \bar{\cI}_{\gd(k)}{}^{\b(k)} (x_{\bar{2} 3})
    &= \cI_{\a(k)}{}^{\gd(k)}(X_1) \, \bar{\cI}_{\gd(k)}{}^{\b(k)} (x_{\bar{1} {3}})~, \\[2mm]
    \implies \hspace{5mm} \bar{\mathfrak{I}}_{\bar{\cA}}{}^{\cB}( x_{\bar{1}2}, x_{1\bar{2}}) \, \mathfrak{I}_{\cB}{}^{\bar{\cA}'} (x_{2 \bar{3}}, x_{\bar{2} 3})
    &= \bar{\mathfrak{I}}_{\bar{\cA}}{}^{\cB}( \bar{X}_{1}, X_{1} ) \, \mathfrak{I}_{\cB}{}^{\bar{\cA}'} ( x_{1 \bar{3}}, x_{\bar{1} 3} ) \, , 
\end{align}
\esubeq
in addition to the result
\begin{align}
    \frac{(x_{\bar{1}3 }{}^{2})^{q_{3}} (x_{\bar{3} 1}{}^{2})^{\bar{q}_{3}} (x_{\bar{1} 2}{}^{2})^{q_{2}} (x_{\bar{2} 1}{}^{2})^{\bar{q}_{2}}}{(x_{\bar{3}1 }{}^{2})^{q_{1}} (x_{\bar{1} 3}{}^{2})^{\bar{q}_{1}} (x_{\bar{3} 2}{}^{2})^{q_{2}} (x_{\bar{2} 3}{}^{2})^{\bar{q}_{2}}} = \frac{x_{1\bar{3}}^{-2(\bar{a} - 2a)} x_{\bar{1}3}^{-2(a - 2 \bar{a})} }{  X_{1}^{2 \bar{q}_{2}} \bar{X}_{1}^{2q_{2}} } \, ,
\end{align}
we obtain
\begin{align} \label{e1}
\begin{split}
    \tilde{\cH}_{\bar{\cA}_{3}\bar{\cA}_{2}\cA_{1}} (X_1, \Q_1, \bar{\Q}_1)
    &= (-1)^{i_{2} + j_{2} + i_{3} + j_{3}} \frac{x_{1\bar{3}}^{-2(\bar{a} - 2a)} x_{\bar{1}3}^{-2(a - 2 \bar{a})} }{ X_{1}^{2 \bar{q}_{2}} \bar{X}_{1}^{2q_{2}}  } \, \bar{\mathfrak{I}}_{\bar{\cA}_{2}}{}^{\cA''_{2}}( \bar{X}_{1}, X_{1} )\\
    & \hspace{10mm} \times \mathfrak{I}_{\cA_{1}}{}^{\bar{\cA}'_{1}}(x_{1\bar{3}}, x_{\bar{1}3}) \, \mathfrak{I}_{\cA''_{2}}{}^{\bar{\cA}'_{2}}(x_{1\bar{3}}, x_{\bar{1}3}) \\
    & \hspace{10mm} \times \bar{\mathfrak{I}}_{\bar{\cA}_{3}}{}^{\cA'_{3}}(x_{\bar{1}3}, x_{1\bar{3}}) \, 
    \cH_{\bar{\cA}'_{1} \bar{\cA}'_{2} \cA'_{3}}(\fn3)~.
\end{split}
\end{align}
To simplify \eqref{e1} further it is necessary to understand how the inversion tensors $\cI_{\cA}{}^{\bar{\cA}}(x_{1\bar{3}}, x_{\bar{1}3})$, $\bar{\cI}_{\bar{\cA}}{}^{\cA}(x_{\bar{1}3}, x_{1\bar{3}})$ act on $\cH(X_{3}, \Theta_{3}, \bar{\Theta}_{3})$. Consider the second and third lines of \eqref{e1}:
\begin{align} \label{e1-normalisedH}
    \mathfrak{I}_{\cA_{1}}{}^{\bar{\cA}'_{1}}(x_{1\bar{3}}, x_{\bar{1}3}) \, \mathfrak{I}_{\cA_{2}}{}^{\bar{\cA}'_{2}}(x_{1\bar{3}}, x_{\bar{1}3}) \, \bar{\mathfrak{I}}_{\bar{\cA}_{3}}{}^{\cA'_{3}}(x_{\bar{1}3}, x_{1\bar{3}}) \, 
    \cH_{\bar{\cA}'_{1} \bar{\cA}'_{2} \cA'_{3}}(\fn3)~.
\end{align}
%
Since $\cH$ is constructed from the building blocks \eqref{building blocks}, only the following fundamental products may appear in \eqref{e1-normalisedH}:
\begin{subequations}
\begin{align}
    \cI_{\a}{}^{\bd} (x_{1 \bar{3}}) \, \bar{\cI}_{\ad}{}^{\b} (x_{\bar{1} 3}) \, X_{3 \b \bd} = c \bar{c} \, \bar{X}_{1 \a \ad}^I~,
\end{align}
\vspace{-10mm}
\begin{align}
    \cI_{\a}{}^{\ad}(x_{1 \bar{3}}) \, \cI_{\b}{}^{\bd}(x_{1 \bar{3}}) \, \ve_{\ad \bd} = -\ve_{\a \b}~, \hspace{10mm} \bar{\cI}_{\ad}{}^{\a}(x_{\bar{1} 3}) \, \bar{\cI}_{\bd}{}^{\b}(x_{\bar{1}3}) \,\ve_{\a \b} = -\ve_{\ad \bd}~,
\end{align}
\vspace{-10mm}
\begin{align}
    \cI_{\a}{}^{\ad}(x_{1 \bar{3}}) \, \bar{\Q}_{3 \ad} = \bar{c}\, \Q^I_{1 \a}~, \hspace{10mm} \bar{\cI}_{\ad}{}^{\a}(x_{\bar{1} 3}) \, \Q_{3 \a} = c\, \bar{\Q}^{I}_{1 \ad}~,
\end{align}
\vspace{-10mm}
\begin{align}
    \cI_{\a}{}^{\ad}(x_{1 \bar{3}}) \, (X_3 \cdot \Q_{3})_{\ad} = -c^{2} \bar{c} \,(\bar{X}_1^I \cdot \bar{\Q}_{1}^I)_{\a}~, \hspace{5mm} \bar{\cI}_{\ad}{}^{\a}(x_{\bar{1} 3}) \, (X_3 \cdot \bar{\Q}_{3})_{\a} = - c \, \bar{c}^{2} \,(\bar{X}_1^I \cdot \Q_{1}^I)_{\ad}~,
\end{align}
\vspace{-10mm}
\begin{align}
    J_{3} = \Q_{3}^{\a} \bar{\Q}_{3}^{\ad} X_{3 \a \ad} = c^{2} \bar{c}^{2} \, 
    \bar{\Q}_{1}^{I \ad} \Q^{I\a}_{1} 
    \bar{X}_{1 \a \ad}^{I} = c^{2} \bar{c}^{2} \bar{J}_{1}^{I} ~.
\end{align}
\end{subequations} 
where
\begin{align}
    c = \frac{ x_{\bar{1} 3}}{ x^{2}_{1 \bar{3}} } \frac{1}{X_{1}} \, , \hspace{10mm} \bar{c} = \frac{ x_{1 \bar{3}}}{ x^{2}_{\bar{1} 3} } \frac{1}{ \bar{X}_{1} } \, , \hspace{10mm} c \, \bar{c} = \frac{ 1 }{ x_{1 \bar{3}} x_{\bar{1} 3} X_{1} \bar{X}_{1} } \, .
\end{align}
Recall that the ``inverted" building blocks $X^{I}, \bar{X}^{I}, \Q^I$, $\bar{\Q}^I$ are defined as in \eqref{inverted-Z3}. Denoting the above transformations by $\cI_{(1 \bar{3}, \bar{1} 3)}$, its action on $\cH(\fn3)$ is therefore equivalent to the following replacements:  
\begin{subequations}
\begin{align} \label{even-13}  
    X_{3\,\a \ad} \stackrel{\cI_{(1 \bar{3}, \bar{1} 3)}}{\longrightarrow} c \bar{c} \, \bar{X}^I_{1\,\a \ad}~, && \Q_{3\a} \stackrel{\cI_{(\bar{1} 3)}}{\longrightarrow} c\, {\bar{\Q}}^I_{1\ad}~, &&
    \bar{\Q}_{3\ad} \stackrel{\cI_{(1 \bar{3})}}{\longrightarrow} \bar{c}\, \Q^I_{1\a}~, && J_{3} \stackrel{\cI_{(1 \bar{3}, \bar{1} 3)}}{\longrightarrow} c^{2} \bar{c}^{2} \, \bar{J}_{1}^I~,~~~~~~~ 
\end{align} 
\vspace{-12mm}
\begin{align} \label{odd-13}
\begin{split}
    & \hspace{30mm} \ve_{\a \b} \stackrel{\cI_{(\bar{1} 3)}}{\longrightarrow} -\ve_{\ad \bd}~, \hspace{10mm}
    \ve_{\ad \bd} \stackrel{\cI_{({1} \bar{3})}}{\longrightarrow} -\ve_{\a \b}~, \\[1mm]
    &(X_3 \cdot \Q_3)_{\ad} \stackrel{\cI_{(1 \bar{3})}}{\longrightarrow} - c^{2} \bar{c}\,(\bar{X}_1^I \cdot \bar{\Q}_{1}^I)_{\a}~, \hspace{10mm}
    (X_3 \cdot \bar{\Q}_3)_{\a} \stackrel{\cI_{(\bar{1} 3)}}{\longrightarrow} - c \, \bar{c}^{2} \,(\bar{X}_1^I \cdot \Q_{1}^I)_{\ad}~. 
\end{split}
\end{align}
\end{subequations}
Hence, the action of $\cI_{(1 \bar{3}, \bar{1} 3)}$ is equivalent to $\cI_{X,\bar{X}}$ combined with a scale transformation. As a consequence of \eqref{even-13}, \eqref{odd-13} and the homogeneity property \eqref{Homogeneity property}, we obtain the following transformation formula for \eqref{e1-normalisedH}:
\begin{align}
    \eqref{e1-normalisedH} &= \cH_{\cA_{1} \cA_{2} \bar{\cA}_{3}}( \ve \rightarrow - \bar{\ve}, \bar{\ve} \rightarrow - \ve, c \bar{c} \,\bar{X}_{1}^{I}, c \, \bar{\Q}_{1}^{I}, \bar{c} \, \Q_{1}^{I} ) \\
    &= c^{2a} \, \bar{c}^{2 \bar{a}} \cH_{\cA_{1} \cA_{2} \bar{\cA}_{3}} ( \ve \rightarrow - \bar{\ve}, \bar{\ve} \rightarrow - \ve, \bar{X}_{1}^{I}, \bar{\Q}_{1}^{I}, \Q_{1}^{I} ) \\
    &= \frac{x_{1\bar{3}}^{-2(a - 2 \bar{a})} x_{\bar{1}3}^{-2(\bar{a} - 2 a)} }{ X_{1}^{2 a} \bar{X}_{1}^{2 \bar{a}}} \cH^{I}_{\cA_{1} \cA_{2} \bar{\cA}_{3}}( X_{1}, \Q_{1}, \bar{\Q}_{1} ) \, . \non
\end{align}
Hence, we obtain the final result
\begin{align}
    \frac{x_{1\bar{3}}^{-2(a - 2 \bar{a})} x_{\bar{1}3}^{-2(\bar{a} - 2 a)} }{ X_{1}^{2 a} \bar{X}_{1}^{2 \bar{a}}} \cH^{I}_{\cA_{1} \cA_{2} \bar{\cA}_{3}}( X_{1}, \Q_{1}, \bar{\Q}_{1} ) &= \mathfrak{I}_{\cA_{1}}{}^{\bar{\cA}'_{1}}(x_{1\bar{3}}, x_{\bar{1}3}) \, \mathfrak{I}_{\cA_{2}}{}^{\bar{\cA}'_{2}}(x_{1\bar{3}}, x_{\bar{1}3}) \\
    & \hspace{10mm} \times \bar{\mathfrak{I}}_{\bar{\cA}_{3}}{}^{\cA'_{3}}(x_{\bar{1}3}, x_{1\bar{3}}) \, \cH_{\bar{\cA}'_{1} \bar{\cA}'_{2} \cA'_{3}}(\fn3)  \, . \non
\end{align}
%
We now substitute the relation above into \eqref{e1}, and we obtain
\begin{align} \label{Htilde-HI-relation}
    \tilde{\cH}_{\bar{\cA}_{3}\bar{\cA}_{2} \cA_{1}} (X, \Q, \bar{\Q}) = \frac{(-1)^{i_{2} + j_{2} + i_{3} + j_{3}}}{  X^{2 (a + \bar{q}_{2})} \bar{X}^{2( \bar{a} + q_{2})} } \, \bar{\mathfrak{I}}_{\bar{\cA}_{2}}{}^{\cA'_{2}}( \bar{X}, X ) \, \cH^I_{\cA_{1} \cA'_{2} \bar{\cA}_{3}}(X,\Q,\bar{\Q})~.
\end{align}
For vector-like superfields (i.e. $i = j$) the expression \eqref{Htilde-HI-relation} is equivalent to eq.\,(3.37) of \cite{Osborn:1998qu}. Note that by explicitly expanding out the indices, we obtain
\begin{align}
\tilde{\cH}_{\g(j_3) \gd(i_3) \b(j_2)\bd(i_2) \a(i_1) \ad(j_1)} (X, \Q, \bar{\Q})
&= \frac{(-1)^{i_{2} + j_{2} + i_{3} + j_{3}}}{X^{2(a + \bar{q}_{2})} \bar{X}^{2(\bar{a} + q_{2})}} \bar{\cI}_{\bd(i_2)}{}^{\b'(i_2)} (\bar{X}) \, \cI_{\b(j_2)}{}^{\bd'(j_2)} (X) \non\\
&\times \cH^{I}_{\a(i_1) \ad(j_1)\b'(i_2) \bd'(j_2) \g(j_3) \gd(i_3)} (X, \Q, \bar{\Q})~.
\label{Htilde-HI-relation1}
\end{align}
Once $\tilde{\cH}$ is obtained, conservation on $J''$ can be imposed as if it were located at the ``first point", using the superconformal covariant derivative identities \eqref{Cderivs}. 
Analogous to the general form of $\cH$ which can always be expanded in $\cH = F + \Theta \bar{\Theta} G$, here we can also employ a similar argument \cite{Buchbinder:2022kmj}. 
The conservation equations on $J''$ also imply that $\tilde{\cH} = \tilde{F} + \Theta \bar{\Theta} \tilde{G}$. This will further simplify future computations, as one simply 
needs to keep terms up to $O(\Q \bar{\Q})$ and higher-order terms in $\Q$ and $\bar{\Q}$ may be discarded.


\subsection{Generating function formalism}\label{GeneratingFunctionFormalism}

Analogous to the approach of \cite{Buchbinder:2022mys} we utilise auxiliary spinors to streamline the calculations. Consider a general spin-tensor $\cH_{\cA_{1} \cA_{2} \cA_{3}}(X, \Q, \bar{\Q})$, where $\cA_{1} = \{ \a(i_{1}), \ad(j_{1}) \}, \cA_{2} = \{ \b(i_{2}), \bd(j_{2}) \}, \cA_{3} = \{ \g(i_{3}), \gd(j_{3}) \}$ represent sets of totally symmetric spinor indices associated with the superfields at points $z_{1}$, $z_{2}$ and $z_{3}$ respectively. We introduce sets of commuting auxiliary spinors for each point; $ U = \{ u, \bar{u} \}$ at $z_{1}$, $ V = \{ v, \bar{v} \}$ at $z_{2}$, and $W = \{ w, \bar{w} \}$ at $z_{3}$, where the spinors satisfy 
\begin{align}
	u^2 &= \varepsilon_{\a \b} \, u^{\a} u^{\b}=0\,, & \bar{u}^2& = \varepsilon_{\ad \bd} \, \bar{u}^{\ad} \bar{u}^{\bd}=0\,,  &
	v^{2} &= \bar{v}^{2} = 0\,, & w^{2} &= \bar{w}^{2} = 0\,. 
	\label{extra1}
\end{align}
Now if we define the objects
\begin{subequations}
	\begin{align}
		\boldsymbol{U}^{\cA_{1}} &\equiv \boldsymbol{U}^{\a(i_{1}) \ad(j_{1})} = u^{\a_{1}} \dots u^{\a_{i_{1}}} \bar{u}^{\ad_{1}} \dots \bar{u}^{\ad_{j_{1}}} \, , \\
		\boldsymbol{V}^{\cA_{2}} &\equiv \boldsymbol{V}^{\b(i_{2}) \bd(j_{2})} = v^{\b_{1}} \dots v^{\b_{i_{2}}} \bar{v}^{\bd_{1}} \dots \bar{v}^{\bd_{j_{2}}} \, , \\
		\boldsymbol{W}^{\cA_{3}} &\equiv \boldsymbol{W}^{\g(i_{3}) \gd(j_{3})} = w^{\g_{1}} \dots w^{\g_{i_{3}}} \bar{w}^{\gd_{1}} \dots \bar{w}^{\gd_{j_{3}}} \, ,
	\end{align}
\end{subequations}
then the generating polynomial for $\cH$ is constructed as follows:
\begin{equation} \label{H - generating polynomial}
	\cH(X, \Q, \bar{\Q}; U(i_{1},j_{1}), V(i_{2},j_{2}), W(i_{3},j_{3})) = \cH_{ \cA_{1} \cA_{2} \cA_{3} }(X, \Q, \bar{\Q}) \, \boldsymbol{U}^{\cA_{1}} \boldsymbol{V}^{\cA_{2}} \boldsymbol{W}^{\cA_{3}} \, , \\
\end{equation}
where we have used the notation $U(i_{1},j_{1})$, $V(i_{2},j_{2})$, $W(i_{3},j_{3})$ to keep track of homogeneityies of the auxiliary spinors $(u, \bar{u})$, $(v,\bar{v})$ and $(w,\bar{w})$. For compactness of notation we will drop the homogeneities and simply write $\cH(X, \Q, \bar{\Q}; U, V, W)$.

The tensor $\cH$ can then be extracted from the polynomial by acting on it with the following partial derivative operators:
\begin{subequations}
	\begin{align}
		\frac{\pa}{\pa \boldsymbol{U}^{\cA_{1}} } &\equiv \frac{\pa}{\pa \boldsymbol{U}^{\a(i_{1}) \ad(j_{1})} } = \frac{1}{i_{1}!j_{1}!} \frac{\pa}{\pa u^{\a_{1}} } \dots \frac{\pa}{\pa u^{\a_{i_{1}}}} \frac{\pa}{\pa \bar{u}^{\ad_{1}}} \dots \frac{\pa }{\pa \bar{u}^{\ad_{j_{1}}}} \, , \\
		\frac{\pa}{\pa \boldsymbol{V}^{\cA_{2}} } &\equiv \frac{\pa}{\pa \boldsymbol{V}^{\b(i_{2}) \bd(j_{2})} } = \frac{1}{i_{2}!j_{2}!} \frac{\pa}{\pa v^{\b_{1}} } \dots \frac{\pa}{\pa v^{\b_{i_{2}}}} \frac{\pa}{\pa \bar{v}^{\bd_{1}}} \dots \frac{\pa }{\pa \bar{v}^{\bd_{j_{2}}}} \, , \\
		\frac{\pa}{\pa \boldsymbol{W}^{\cA_{3}} } &\equiv \frac{\pa}{\pa \boldsymbol{W}^{\g(i_{3}) \gd(j_{3})} } = \frac{1}{i_{3}!j_{3}!} \frac{\pa}{\pa w^{\g_{1}} } \dots \frac{\pa}{\pa w^{\g_{i_{3}}}} \frac{\pa}{\pa \bar{w}^{\gd_{1}}} \dots \frac{\pa }{\pa \bar{w}^{\gd_{j_{3}}}} \, . 
	\end{align}
\end{subequations}
The tensor $\cH$ is then extracted from the polynomial as follows:
\begin{equation}
	\cH_{\cA_{1} \cA_{2} \cA_{3}}(X, \Q, \bar{\Q}) = \frac{\pa}{ \pa \boldsymbol{U}^{\cA_{1}} } \frac{\pa}{ \pa \boldsymbol{V}^{\cA_{2}}} \frac{\pa}{ \pa \boldsymbol{W}^{\cA_{3}} } \, \cH(X, \Q, \bar{\Q}; U, V, W) \, .
\end{equation}
To construct $\cH$, it is often useful to introduce the ``normalised" tensor $\hat{\cH}$, which is defined to be homogeneous degree 0 and is obtained by rescaling $\cH$ by a suitable power of $X$. Suppressing the indices on $\cH$, we let
\bea
\cH (X, \Theta, \bar{\Theta}) = X^{a+\bar{a}} \hat{\cH} (X, \Theta, \bar{\Theta})~,
\label{defn-normalisedH}
\eea
where $\hat{\cH} (X, \Theta, \bar{\Theta})$ is homogeneous degree 0 in $(X, \Theta, \bar{\Theta})$, i.e.
\bea
\hat{\cH} (\l \bar{\l}X, \l \Theta, \bar{\l}\bar{\Theta}) = \hat{\cH} (X, \Theta, \bar{\Theta})~.\label{scaling-Hhat}
\eea
An ansatz for the homogeneous degree 0 tensor $\hat{\cH}_{\cA_{1} \cA_{2} \cA_{3}}(X,\Q,\bar{\Q})$ is constructed out of symmetric and traceless combinations of the following building blocks:
\begin{align}
    \ve_{\a \b} \, , \hspace{8mm} \ve_{\ad \bd} \, , \hspace{8mm} \hat{X}_{\a \ad} \, , \hspace{8mm} \hat{\Q}_{\a} \, , \hspace{8mm} \hat{\bar{\Q}}_{\ad} \, , \hspace{8mm} (\hat{X} \cdot \hat{\bar{\Q}})_{\a} \, , \hspace{8mm} (\hat{X} \cdot \hat{\Q})_{\ad} \, , \hspace{8mm} J \, .
\end{align}
Hence, to construct the corresponding polynomial $\hat{\cH}(X,\Q,\bar{\Q}; U,V,W)$ we encode the tensor structures above in the following monomials:\\[2mm]
\textbf{Bosonic:}
\begin{subequations} \label{Monomials-bosonic}
	\begin{align} 
		P_{1} &= \ve_{\a \b} v^{\a} w^{\b} \, , & P_{2} &= \ve_{\a \b} w^{\a} u^{\b} \, , & P_{3} &= \ve_{\a \b} u^{\a} v^{\b} \, , \\
		\tilde{P}_{1} &= \ve_{\ad \bd} \bar{v}^{\ad} \bar{w}^{\bd} \, , & \tilde{P}_{2} &= \ve_{\ad \bd} \bar{w}^{\ad} \bar{u}^{\bd} \, , & \tilde{P}_{3} &= \ve_{\ad \bd} \bar{u}^{\ad} \bar{v}^{\bd} \, , \\[1mm]
		Q_{1} &= \hat{X}_{\a \ad} \, v^{\a} \bar{w}^{\ad} \, ,  &  Q_{2} &= \hat{X}_{\a \ad} \, w^{\a} \bar{u}^{\ad} \, ,  &  Q_{3} &= \hat{X}_{\a \ad} \, u^{\a} \bar{v}^{\ad} \, , \\
		\tilde{Q}_{1} &= \hat{X}_{\a \ad} \, w^{\a} \bar{v}^{\ad} \, ,  &  \tilde{Q}_{2} &= \hat{X}_{\a \ad} \, u^{\a} \bar{w}^{\ad} \, ,  &  \tilde{Q}_{3} &= \hat{X}_{\a \ad} \, v^{\a} \bar{u}^{\ad} \, , \\[1mm]
		Z_{1} &= \hat{X}_{\a \ad} \, u^{\a} \bar{u}^{\ad}  \, , & Z_{2} &= \hat{X}_{\a \ad} \, v^{\a} \bar{v}^{\ad} \, , & Z_{3} &= \hat{X}_{\a \ad} \, w^{\a} \bar{w}^{\ad} \, .
	\end{align}
\end{subequations}
\textbf{Fermionic:}
\begin{subequations} \label{Monomials-fermionic}
	\begin{align} 
		R_{1} &= \ve_{\a \b} u^{\a} \hat{\Q}^{\b} \, , & R_{2} &= \ve_{\a \b} v^{\a} \hat{\Q}^{\b} \, , & R_{3} &= \ve_{\a \b} u^{\a} \hat{\Q}^{\b} \, , \\
		\tilde{R}_{1} &= \ve_{\ad \bd} \bar{u}^{\ad} \hat{\bar{\Q}}^{\bd} \, , & \tilde{R}_{2} &= \ve_{\ad \bd} \bar{v}^{\ad} \hat{\bar{\Q}}^{\bd} \, , & \tilde{R}_{3} &= \ve_{\ad \bd} \bar{w}^{\ad} \hat{\bar{\Q}}^{\bd} \, , \\[1mm]
		S_{1} &= \hat{X}_{\a \ad} \, u^{\a} \hat{\bar{\Q}}^{\ad} \, ,  &  S_{2} &= \hat{X}_{\a \ad} \, v^{\a} \hat{\bar{\Q}}^{\ad} \, ,  &  S_{3} &= \hat{X}_{\a \ad} \, u^{\a} \hat{\bar{\Q}}^{\ad} \, , \\
		\tilde{S}_{1} &= \hat{X}_{\a \ad} \, \hat{\Q}^{\a} \bar{u}^{\ad} \, ,  &  \tilde{S}_{2} &= \hat{X}_{\a \ad} \, \hat{\Q}^{\a} \bar{v}^{\ad} \, ,  &  \tilde{S}_{3} &= \hat{X}_{\a \ad} \, \hat{\Q}^{\a} \bar{w}^{\ad} \, .
	\end{align}
\end{subequations}
To construct a linearly independent ansatz for a given three-point function, one must also take into account numerous non-linear relations between the monomials. A list of some fundamental relations involving the monomials are presented in Appendix \ref{AppA}.

The task now is to construct a complete list of possible (linearly independent) solutions for the polynomial $\cH$ for a given set of spins. This process is simplified by introducing a generating function, $\cG(X,\Q,\bar{\Q}; U, V, W \, | \, \G)$, defined as follows:
\begin{align} \label{Generating function}
	\cG(X,\Q,\bar{\Q}; U,V,W \, | \, \G) &= P_{1}^{k_{1}} P_{2}^{k_{2}} P_{3}^{k_{3}} \, 	\tilde{P}_{1}^{\bar{k}_{1}} \tilde{P}_{2}^{\bar{k}_{2}} \tilde{P}_{3}^{\bar{k}_{3}} Q_{1}^{l_{1}} Q_{2}^{l_{2}} Q_{3}^{l_{3}} \, \tilde{Q}_{1}^{\bar{l}_{1}} \tilde{Q}_{2}^{\bar{l}_{2}} \tilde{Q}_{3}^{\bar{l}_{3}} Z_{1}^{r_{1}} Z_{2}^{r_{2}} Z_{3}^{r_{3}} \nonumber \\
    & \hspace{10mm} \times R_{1}^{p_{1}} R_{2}^{p_{2}} R_{3}^{p_{3}} \tilde{S}_{1}^{\bar{m}_{1}} \tilde{S}_{2}^{\bar{m}_{2}} \tilde{S}_{3}^{\bar{m}_{3}} \tilde{R}_{1}^{\bar{p}_{1}} \tilde{R}_{2}^{\bar{p}_{2}} \tilde{R}_{3}^{\bar{p}_{3}} S_{1}^{m_{1}} S_{2}^{m_{2}} S_{3}^{m_{3}} J^{\s} \, ,
\end{align}
where the non-negative integers, $ \G =  \bigcup_{i \in \{1,2,3\} }  \{ k_{i}, \bar{k}_{i}, l_{i}, \bar{l}_{i}, r_{i}, p_{i}, \bar{p}_{i}, m_{i}, \bar{m}_{i}, \s \}$, are solutions to the following linear system:
\begin{subequations} \label{Diophantine equations}
	\begin{align}
		k_{2} + k_{3} + r_{1} + l_{3} + \bar{l}_{2} + p_{1} + m_{1} &= i_{1} \, , &  \bar{k}_{2} + \bar{k}_{3} + r_{1} + \bar{l}_{3} + l_{2} + \bar{p}_{1} + \bar{m}_{1} &= j_{1} \, , \\
		k_{1} + k_{3} + r_{2} + l_{1} + \bar{l}_{3} + p_{2} + m_{2} &= i_{2} \, , &  \bar{k}_{1} + \bar{k}_{3} + r_{2} + \bar{l}_{1} + l_{3} + \bar{p}_{2} + \bar{m}_{2} &= j_{2} \, , \\
		k_{1} + k_{2} + r_{3} + l_{2} + \bar{l}_{1} + p_{3} + m_{3} &= i_{3} \, , &  \bar{k}_{1} + \bar{k}_{2} + r_{3} + \bar{l}_{2} + l_{1} + \bar{p}_{3} + \bar{m}_{3} &= j_{3} \, .
	\end{align}
\end{subequations}
Here $i_{1}, i_{2}, i_{3}$, $j_{1}, j_{2}, j_{3}$ are fixed integers corresponding to the spin representations of the fields in the three-point function. To obtain solutions which are at most $O(\Theta \bar{\Theta})$, we also require
\begin{align}
	p_{1} + p_{2} + p_{3} + \bar{m}_{1} + \bar{m}_{2} + \bar{m}_{3} + \s &\leq 1 \, , &  \bar{p}_{1} + \bar{p}_{2} + \bar{p}_{3} + m_{1} + m_{2} + m_{3} + \s &\leq 1  \, .
\end{align}

Given a finite number of solutions $\G_{I}$, $I = 1, ..., N$ to \eqref{Diophantine equations} for a particular choice of $i_{1}, i_{2}, i_{3}, j_{1}, j_{2}, j_{3}$, the most general ansatz for the polynomial $\cH$ in \eqref{H - generating polynomial} is as follows:
\begin{equation}
	\cH(X, \Q, \bar{\Q}; U, V, W) = X^{\D_{3} - \D_{2} - \D_{1}} \sum_{I=1}^{N} c_{I} \cG(X, \Q, \bar{\Q}; U, V, W \, | \, \G_{I}) \, ,
\end{equation}
where $a_{I}$ are a set of complex constants. Hence, constructing the most general ansatz for the generating polynomial $\cH$ is now equivalent to finding all non-negative integer solutions $\G_{I}$ of \eqref{Diophantine equations}. Once this ansatz has been obtained, the linearly independent structures can be found by systematically applying the linear dependence relations \eqref{Linear dependence 1}--\eqref{Linear dependence 16} using replacement rules in Mathematica.

It is also important to note that the auxiliary spinor formalism greatly simplifies the computation of $\bar{\cH}(X,\Q, \bar{\Q})$, $\cH^{I}(X,\Q, \bar{\Q})$ and $\tilde{\cH}(X,\Q, \bar{\Q})$. The map $\cH(X,\Q, \bar{\Q}) \longrightarrow \bar{\cH}(X,\Q, \bar{\Q})$ is equivalent to the following replacement rules for the monomials
\begin{subequations}
	\begin{align} 
		P_{1} &\rightarrow \tilde{P}_{1} \, , & P_{2} &\rightarrow \tilde{P}_{2} \, , & P_{3} &\rightarrow \tilde{P}_{3} \, , \\
		\tilde{P}_{1} &\rightarrow P_{1} \, , & \tilde{P}_{2} &\rightarrow P_{2} \, , & \tilde{P}_{3} &\rightarrow P_{3} \, , 
	\end{align}
    \vspace{-10mm}
    \begin{align}
        Q_1 &\to (1-2 \text{i} J) \, \tilde{Q}_1 - 4 \text{i} R_3 \tilde{R}_2 \, , & \tilde{Q}_1 &\to (1-2 \text{i} J) \, Q_1 - 4 \text{i} R_2 \tilde{R}_3 \, , \\
        Q_2 &\to (1-2 \text{i} J) \, \tilde{Q}_2 - 4 \text{i} R_1 \tilde{R}_3 \, , & \tilde{Q}_2 &\to (1-2 \text{i} J) \, Q_2 - 4 \text{i} R_3 \tilde{R}_1 \, , \\
        Q_3 &\to (1-2 \text{i} J) \,  \tilde{Q}_3 - 4 \text{i} R_2 \tilde{R}_1 \, , & \tilde{Q}_3 &\to (1-2 \text{i} J) \,  Q_3 - 4 \text{i} R_1 \tilde{R}_2 \, ,
    \end{align}
    \vspace{-10mm}
    \begin{align} 
		R_{1} &\rightarrow \tilde{R}_{1} \, , & R_{2} &\rightarrow \tilde{R}_{2} \, , & R_{3} &\rightarrow \tilde{R}_{3} \, , \\
		\tilde{R}_{1} &\rightarrow R_{1} \, , & \tilde{R}_{2} &\rightarrow R_{2} \, , & \tilde{R}_{3} &\rightarrow R_{3} \, , \\
        S_{1} &\rightarrow \tilde{S}_{1} \, , & S_{2} &\rightarrow \tilde{S}_{2} \, , & S_{3} &\rightarrow \tilde{S}_{3} \, , \\
		\tilde{S}_{1} &\rightarrow S_{1} \, , & \tilde{S}_{2} &\rightarrow S_{2} \, , & \tilde{S}_{3} &\rightarrow S_{3} \, , 
	\end{align}
    \vspace{-10mm}
    \begin{align}
        Z_1 &\to (1-2 \text{i} J) \,  Z_1 - 4 \text{i} R_1 \tilde{R}_1 \, , \\
        Z_2 &\to (1-2 \text{i} J) \,  Z_2 - 4 \text{i} R_2 \tilde{R}_2 \, , \\
        Z_3 &\to (1-2 \text{i} J) \,  Z_3 - 4 \text{i} R_3 \tilde{R}_3 \, ,
    \end{align}
\end{subequations}
Next, by using the properties transformation properties \eqref{inversion replacements} of the superconformal building blocks under superinversion, the map $\cH(X,\Q, \bar{\Q}; U,V,W) \stackrel{\cI}{\longrightarrow} \cH^{I}(X,\Q, \bar{\Q}; U,V,W)$ is equivalent to the following replacement rules for the monomials: 
\begin{subequations} \label{H_I replacement rules}
	\begin{align} 
		P_{1} &\rightarrow - \tilde{P}_{1} \, , & P_{2} &\rightarrow - \tilde{P}_{2} \, , & P_{3} &\rightarrow - \tilde{P}_{3} \, , \\
		\tilde{P}_{1} &\rightarrow - P_{1} \, , & \tilde{P}_{2} &\rightarrow - P_{2} \, , & \tilde{P}_{3} &\rightarrow - P_{3} \, , 
	\end{align}
    \vspace{-10mm}
    \begin{align}
        Q_1 &\to -(1-2 \text{i} J) \, \tilde{Q}_1 + 4 \text{i} R_3 \tilde{R}_2 \, , & \tilde{Q}_1 &\to -(1-2 \text{i} J) \, Q_1 + 4 \text{i} R_2 \tilde{R}_3 \, , \\
        Q_2 &\to -(1-2 \text{i} J) \, \tilde{Q}_2 + 4 \text{i} R_1 \tilde{R}_3 \, , & \tilde{Q}_2 &\to -(1-2 \text{i} J) \, Q_2 + 4 \text{i} R_3 \tilde{R}_1 \, , \\
        Q_3 &\to -(1-2 \text{i} J) \,  \tilde{Q}_3 + 4 \text{i} R_2 \tilde{R}_1 \, , & \tilde{Q}_3 &\to -(1-2 \text{i} J) \,  Q_3 + 4 \text{i} R_1 \tilde{R}_2 \, ,
    \end{align}
    \vspace{-10mm}
    \begin{align} 
		R_{1} &\rightarrow - \tilde{S}_{1} \, , & R_{2} &\rightarrow - \tilde{S}_{2} \, , & R_{3} &\rightarrow - \tilde{S}_{3} \, , \\
		\tilde{R}_{1} &\rightarrow - S_{1} \, , & \tilde{R}_{2} &\rightarrow - S_{2} \, , & \tilde{R}_{3} &\rightarrow - S_{3} \, , \\
        S_{1} &\rightarrow - \tilde{R}_{1} \, , & S_{2} &\rightarrow - \tilde{R}_{2} \, , & S_{3} &\rightarrow - \tilde{R}_{3} \, , \\
		\tilde{S}_{1} &\rightarrow - R_{1} \, , & \tilde{S}_{2} &\rightarrow - R_{2} \, , & \tilde{S}_{3} &\rightarrow - R_{3} \, , 
	\end{align}
    \vspace{-10mm}
    \begin{align}
        Z_1 &\to -(1-2 \text{i} J) \,  Z_1 + 4 \text{i} R_1 \tilde{R}_1 \, , \\
        Z_2 &\to -(1-2 \text{i} J) \,  Z_2 + 4 \text{i} R_2 \tilde{R}_2 \, , \\
        Z_3 &\to -(1-2 \text{i} J) \,  Z_3 + 4 \text{i} R_3 \tilde{R}_3 \, ,
    \end{align}
\end{subequations}
Similarly, the map $\cH^{I}(X,\Q, \bar{\Q}; U,V,W) \stackrel{\cI_{(2)}}{\longrightarrow} \tilde{\cH}(X,\Q, \bar{\Q}; U,V,W)$ is equivalent to the replacement rules below: 
\begin{subequations} \label{Htilde replacement rules}
	\begin{align} 
		P_{1} &\rightarrow -(1-2 \text{i} J) \, \tilde{Q}_1 + 4 \text{i} R_3 \tilde{R}_2 \, , & 
		\tilde{P}_{1} &\rightarrow - Q_{1} \, , \\
        P_{3} &\rightarrow (1-2 \text{i} J) \, Q_3 - 4 \text{i} R_1 \tilde{R}_2 \, , & \tilde{P}_{3} &\rightarrow \tilde{Q}_{3} \, , \\[2mm]
        Q_1 &\to -(1-2 \text{i} J) \, \tilde{P}_1 - 4 \text{i} \tilde{S}_3 R_2 \, , & \tilde{Q}_1 &\to -P_{1} \, , \\
        Q_3 &\to P_{3} \, , & \tilde{Q}_3 &\to (1-2 \text{i} J) \, \tilde{P}_3 - 4 \text{i} \tilde{S}_1 R_2  \, , \\[2mm]
		R_{2} &\rightarrow - \tilde{S}_{2} \, , & \tilde{R}_{2} &\rightarrow - S_{2} \, , \\
        S_{2} &\rightarrow - \tilde{R}_{2} \, , & \tilde{S}_{2} &\rightarrow - R_{2} \, , \\[2mm]
        Z_2 &\to -(1-2 \text{i} J) \,  Z_2 + 4 \text{i} R_2 \tilde{R}_2 \, ,
    \end{align}
\end{subequations}
We may also derive replacement rules for the transformation of the monomials under $z_{1} \leftrightarrow z_{2}$ and $z_{1} \leftrightarrow z_{3}$. For $z_{1} \leftrightarrow z_{2}$ we obtain
\begin{subequations} \label{1-2 point switch rules}
	\begin{align} 
		P_{1} &\rightarrow - P_{2} \, , & P_{2} &\rightarrow - P_{1} \, , & P_{3} &\rightarrow - P_{3} \, , \\
		\tilde{P}_{1} &\rightarrow - \tilde{P}_{2} \, , & \tilde{P}_{2} &\rightarrow - \tilde{P}_{1} \, , & \tilde{P}_{3} &\rightarrow - \tilde{P}_{3} \, , 
	\end{align}
    \vspace{-10mm}
    \begin{align}
        Q_1 &\to -(1-2 \text{i} J) \, \tilde{Q}_2 + 4 \text{i} R_1 \tilde{R}_3 \, , & \tilde{Q}_1 &\to -(1-2 \text{i} J) \, Q_2 + 4 \text{i} R_3 \tilde{R}_1 \, , \\
        Q_2 &\to -(1-2 \text{i} J) \, \tilde{Q}_1 + 4 \text{i} R_3 \tilde{R}_2 \, , & \tilde{Q}_2 &\to -(1-2 \text{i} J) \, Q_1 + 4 \text{i} R_2 \tilde{R}_3 \, , \\
        Q_3 &\to -(1-2 \text{i} J) \,  \tilde{Q}_3 + 4 \text{i} R_2 \tilde{R}_1 \, , & \tilde{Q}_3 &\to -(1-2 \text{i} J) \,  Q_3 + 4 \text{i} R_1 \tilde{R}_2 \, ,
    \end{align}
    \vspace{-10mm}
    \begin{align} 
		R_{1} &\rightarrow - R_{2} \, , & R_{2} &\rightarrow - R_{1} \, , & R_{3} &\rightarrow - R_{3} \, , \\
		\tilde{R}_{1} &\rightarrow - \tilde{R}_{2} \, , & \tilde{R}_{2} &\rightarrow - \tilde{R}_{1} \, , & \tilde{R}_{3} &\rightarrow - \tilde{R}_{3} \, ,
    \end{align}
    \vspace{-10mm}
    \begin{align}
        S_{1} &\rightarrow S_{2} \, , & S_{2} &\rightarrow S_{1} \, , \\
		\tilde{S}_{1} &\rightarrow \tilde{S}_{2} \, , & \tilde{S}_{2} &\rightarrow \tilde{S}_{1} \, , 
	\end{align}
    \vspace{-10mm}
    \begin{align}
        Z_1 &\to -(1-2 \text{i} J) \,  Z_2 + 4 \text{i} R_2 \tilde{R}_2 \, , \\
        Z_2 &\to -(1-2 \text{i} J) \,  Z_1 + 4 \text{i} R_2 \tilde{R}_1 \, , \\
        Z_3 &\to -(1-2 \text{i} J) \,  Z_3 + 4 \text{i} R_3 \tilde{R}_3 \, ,
    \end{align}
    \vspace{-10mm}
    \begin{align}
        J \rightarrow - J
    \end{align}
\end{subequations}
A similar list of rules is obtained for $z_{1} \leftrightarrow z_{3}$. Any monomials not included in the replacement rules above are inert with respect to the transformation. 

It is important to note that in the transformations above, only terms which are at most $O(\Q \bar{\Q})$ are kept, while all higher-order terms in $\Q$ and $\bar{\Q}$ are dropped. The reason for this is that any formulation of a given three-point function of conserved supercurrent multiplets may be described by a tensor $\cH$, which can be expanded in the form $\cH = F + \Q \bar{\Q} G$ after imposing conservation on two of the superspace points.

\subsection{Summary of constraints}


To conclude this section, we list out all the constraints that we must impose on a given three-point correlation function in the generating function formalism. Recalling that $s_{1} = \tfrac{1}{2}(i_{1} + j_{1})$, $s_{2} = \tfrac{1}{2}(i_{2} + j_{2})$, $s_{3} = \tfrac{1}{2}( i_{3} + j_{3})$, first we define:
\begin{subequations}
	\begin{align}
		J^{}_{s_{1}}(z_{1}; U) & = J^{}_{\a(i_{1}) \ad(j_{1})}(z_{1}) \, \boldsymbol{U}^{\a(i_{1}) \ad(j_{1})} \, , \\
		J'_{s_{2}}(z_{2}; V) &= J'_{\b(i_{2}) \bd(j_{2})}(z_{2}) \, \boldsymbol{V}^{\b(i_{2}) \bd(j_{2})} \, , \\
		J''_{s_{3}}(z_{3}; W) &= J''_{\g(i_{3}) \gd(j_{3})}(z_{3}) \, \boldsymbol{W}^{\g(i_{3}) \gd(j_{3})} \, ,
	\end{align}
\end{subequations}
The general ansatz can be converted into the auxiliary spinor formalism as follows
\begin{align}
	\begin{split}
		\langle J^{}_{s_{1}}(z_{1}; U) \, J'_{s_{2}}(z_{2}; V) \, J''_{s_{3}}(z_{3}; W) \rangle &= \frac{ \cI^{(i_{1}, j_{1})}(x_{1 \bar{3}}, x_{\bar{1} 3}; U, \bar{U}') \,  \cI^{(i_{2}, j_{2})}(x_{2 \bar{3}}, x_{\bar{2} 3}; V, \bar{V}') }{ (x_{1\bar{3}}{}^2)^{q_1} (x_{\bar{1}3}{}^2)^{\bar{q}_1}
(x_{2\bar{3}}{}^2)^{q_2} (x_{\bar{2}3}{}^2)^{\bar{q}_2} } \\
		& \hspace{5mm} \times \cH(X_{3}, \Q_{3}, \bar{\Q}_{3}; \bar{U}',\bar{V}', W) \, ,
	\end{split}
\end{align} 
where $\D_{i} = q_{i} + \bar{q}_{i} = s_{i} + 2$. The generating polynomial, $\cH(X, \Q, \bar{\Q}; U, V, W)$, is defined as follows:
\begin{align}
    \begin{split}
    	\cH(X,\Q,\bar{\Q}; U,V,W) &= \cH_{\a(i_{1}) \ad(j_{1}) \b(i_{2}) \bd(j_{2}) \g(i_{3}) \gd(j_{3})}(X, \Q, \bar{\Q}) \\
        & \hspace{20mm} \times \boldsymbol{U}^{\a(i_{1}) \ad(j_{1})} \boldsymbol{V}^{\b(i_{2}) \bd(j_{2})} \boldsymbol{W}^{\g(i_{3}) \gd(j_{3})} \, ,
    \end{split}
\end{align}
and
\begin{align}
	\cI^{(i,j)}(x, \bar{x}; U, \bar{U}) \equiv \cI^{(i,j)}_{x, \bar{x}}( U, \bar{U}) =\boldsymbol{U}^{\a(i) \ad(j)} \cI_{\a(i)}{}^{\ad'(i)}(x) \, \bar{\cI}_{\ad(j)}{}^{\a'(j)}(\bar{x}) \, \frac{\pa}{\pa \bar{\boldsymbol{U}}'^{\a'(j) \ad'(i)}} \, ,
\end{align}
is the inversion operator acting on polynomials degree $(i,j)$ in $(\tilde{\bar{u}}, \tilde{u})$. Note that $\bar{\boldsymbol{U}}'$ has index structure conjugate to $\boldsymbol{U}$, and sometimes the indices $(i,j)$ will be omitted to streamline the notation. 

After converting the constraints summarised in the previous subsection into the auxiliary spinor formalism, we obtain:
\begin{enumerate}
	\item[\textbf{(i)}] {\bf Homogeneity:} \\
	Recall that $\cH$ is a homogeneous polynomial satisfying the following scaling property:
	\begin{align}
		\begin{split}
			&\cH(\l \bar{\l} X, \l \Q, \bar{\l} \bar{\Q}; U, V, W) = \l^{2a} \bar{\l}^{2\bar{a}} \, \cH(X, \Q, \bar{\Q}; U, V, W) \, ,
		\end{split}
	\end{align}
    where superconformal invariance requires
    \begin{align}
    a = \bar{a} = \frac{1}{2}(\D_{3} - \D_{1} - \D_{2}) \, .
    \end{align}
    This condition ensures that the three-point function possesses vanishing total R-charge and the appropriate transformation properties under scale transformations.
	
	\item[\textbf{(ii)}] {\bf Differential constraints:} \\
	First, define the following differential operators:
    \begin{subequations}
	\begin{align} \label{Derivative operators-1}
		D_{1} &= \cD^{\a} \frac{\pa}{\pa u^{\a}} \, , & D_{2} &= \cQ^{\a} \frac{\pa}{\pa v^{\a}} \, , & D_{3} &= \cQ^{\a} \frac{\pa}{\pa w^{\a}} \, , \\
        \bar{D}_{1} &= \bar{\cD}^{\ad} \frac{\pa}{\pa \bar{u}^{\ad}} \, , & \bar{D}_{2} &= \bar{\cQ}^{\ad} \frac{\pa}{\pa \bar{v}^{\ad}} \, , & \bar{D}_{3} &= \bar{\cQ}^{\ad} \frac{\pa}{\pa \bar{w}^{\ad}} \, ,
	\end{align}
    \end{subequations}
    For imposing conservation on superfields which obey linearity conditions, we must also introduce the operators
    \begin{subequations}
    \begin{align} \label{Derivative operators-2}
		D^{2}_{1} &= \cD^{\a} \cD_{\a}  \, , & D^{2}_{2} &= \cQ^{\a} \cQ_{\a}  \, , & D^{2}_{3} &= \cQ^{\a} \cQ_{\a} \, , \\
        \bar{D}^{2}_{1} &= \bar{\cD}_{\ad} \bar{\cD}^{\ad}  \, , & \bar{D}^{2}_{2} &= \bar{\cQ}_{\ad} \bar{\cQ}^{\ad}  \, , & \bar{D}^{2}_{3} &= \bar{\cQ}_{\ad} \bar{\cQ}^{\ad} \, ,
	\end{align}
    \end{subequations}
	Conservation on all three points may be imposed using the following constraints (note that below $\cH \equiv \cH(X, \Q, \bar{\Q}; U,V,W)$, $\tilde{\cH} \equiv \tilde{\cH}(X, \Q, \bar{\Q};U,V,W)$ to streamline presentation):
    \begin{center}
    \begingroup
    \renewcommand*{\arraystretch}{1.25}
    \newcolumntype{P}[1]{>{\centering\arraybackslash}p{#1}}
    \begin{tabular}{ P{4cm} | P{2cm} P{2cm} | P{2cm} P{2cm} }
        \hline
        & \multicolumn{2}{ P{4cm} | }{$i, j > 0$} & $i = 0$ & $j = 0$ \\ 
        \hline
        \text{Conservation at $z_{1}$:} & $D_{1} \cH = 0$ & $\bar{D}_{1} \cH = 0$ & $D^{2}_{1} \cH = 0$ & $\bar{D}^{2}_{1} \cH = 0 $ \\[1mm]
        \text{Conservation at $z_{2}$:} & $D_{2} \cH = 0$ & $\bar{D}_{2} \cH = 0$ & $D^{2}_{2} \cH = 0$ & $\bar{D}^{2}_{2} \cH = 0 $ \\[1mm]
        \text{Conservation at $z_{3}$:} & $D_{3} \tilde{\cH} = 0$ & $\bar{D}_{3} \tilde{\cH} = 0$ & $D^{2}_{3} \tilde{\cH} = 0$ & $\bar{D}^{2}_{3} \tilde{\cH} = 0 $ \\[1mm]
        \hline
    \end{tabular}
    \endgroup
    \end{center}
    Note that in the auxiliary spinor formalism, $\tilde{\cH}$ is computed as follows:
	\begin{equation}
		\tilde{\cH}(X,\Q,\bar{\Q}; U,V,W) = \frac{1}{X^{2(a + \bar{q}_{2})} \bar{X}^{2(\bar{a} + q_{2})}} \, \cI^{(i_{2},j_{2})}_{X,\bar{X}}( V,\bar{V}') \, \cH^{I}(X,\Q,\bar{\Q}; U, \bar{V}', W) \, .
	\end{equation}
    For imposing the conservation equations, the identities given in Appendix \ref{AppA} are useful.
\end{enumerate}

\begin{itemize}
	\item[\textbf{(iii)}] {\bf Point switch symmetries:} \\
	If the superfields $J$ and $J'$ coincide (hence $i_{1} = i_{2}$, $j_{1} = j_{2}$), then we obtain the following point-switch constraint
	\begin{equation} \label{Point switch A}
		\cH(X, \Q, \bar{\Q}; U,V,W) = (-1)^{\e(J)} \cH(-\bar{X}, -\Q, -\bar{\Q}; V,U,W) \, ,
	\end{equation}
	where $\e(J)$ is the Grassmann parity of $J$. Similarly, if the superfields $J$ and $J''$ coincide (hence $i_{1} = i_{3}$, $j_{1} = j_{3}$) then we obtain the constraint
	\begin{equation} \label{Point switch B}
		\tilde{\cH}(X, \Q, \bar{\Q}; U, V, W) = (-1)^{\e(J)} \cH(-\bar{X}, -\Q, - \bar{\Q}; W, V, U) \, .
	\end{equation}

	\item[\textbf{(iv)}] {\bf Reality conditions:} \\
	If the superfields in the correlation function belong to the $(s,s)$ representation, then the three-point function must satisfy the reality condition
	\begin{equation} \label{Reality condition}
		\bar{\cH}(X, \Q, \bar{\Q}; U, V, W) = \cH(X,\Q,\bar{\Q}; U, V, W) \, .
	\end{equation}
	Similarly, if the superfields at $J$, $J'$ at $z_{1}$ and $z_{2}$ respectively are conjugate to each other, i.e. $J' = \bar{J}$ (and, hence, $i_{1} = j_{2}$, $j_{1} = i_{2}$), we must impose a combined reality/point-switch condition using the following constraint
	\begin{equation} \label{Reality/switch condition}
		\bar{\cH}(X, \Q, \bar{\Q}; U, V, W) = \cH(-\bar{X}, -\Q, -\bar{\Q}; V, U, W) \, ,
	\end{equation}
\end{itemize}
In the remaining sections we apply our formalism to specific examples. First, we analyse the three-point correlation functions involving combinations of supercurrent and flavour current multiplets. We then propose a general classification of the results for arbitrary superspins based on a ``case-by-case" analysis for some particular configurations of $s_{1}$, $s_{2}$ and $s_{3}$. In general, since the number of independent structures grows dramatically with the minimum superspin, this determines the computational limitations of our analysis. The most computationally demanding cases are those for which $s_{i} = s$, and in these cases we were limited to $s=3$. However, for cases where e.g. $\min(s_{i}) = s_{3} = 1$ we could explore the general structure for $s_{1}, s_{2} \leq 5$. With further optimisation of the Mathematica code and/or more powerful hardware it is possible to overcome these computational limits. The available evidence is sufficient to propose the general structure of the three-point functions for arbitrary superspins, which we present in Subsection \ref{section4.2}. The proposal is further supported by direct comparison with the results for three-point functions of conserved currents in 4D CFT \cite{Buchbinder:2023coi}, see Subsection \ref{section4.3} for a detailed discussion. 


\section{Three-point functions of conserved supercurrents}\label{section4}

In the next subsections we analyse the structure of three-point functions involving conserved currents in 4D $\cN=1$ SCFT. 
We classify, using a computational approach, the general structure of three-point functions involving the conserved currents $J_{\a(s) \ad(s)}$, with our main goal being to determine the number of independent structures after imposing the reality and conservation conditions on all superspace points. 
As pointed out in the introduction, the number of independent conserved structures generically grows linearly with the minimum superspin and the solution for $\cH(X, \Q, \bar{\Q}; U, V, W)$ quickly becomes too complicated to present even for relatively low superspins. 
Thus, although our method allows us to find $\cH(X, \Q, \bar{\Q}; U, V, W)$ in an explicit form for arbitrary spins (limited only by computer hardware), we find it practical to present the solutions when 
there are a small number of structures. Examples involving low superspins are discussed in Subsection~\ref{section4.1}, while in Subsection~\ref{section4.2} we propose the classification for arbitrary spins. Some additional higher-superspin examples are presented in the appendices \ref{AppB}, \ref{AppC}.

\subsection{Supercurrent and flavour current correlators}\label{section4.1}

We begin our analysis by considering the three-point correlation functions involving low superspin supercurrents, such as the flavour current and supercurrent multiplets. Some of these results are known throughout the literature (see e.g. \cite{Osborn:1998qu,Buchbinder:2022kmj}), but we derive them again here to demonstrate our formalism.

The flavour current multiplet is a real scalar superfield $L(z)$ with weights $(1,1)$ and satisfies the superfield linearity conditions $D^{2} L = \bar{D}^{2} L = 0$. The supercurrent multiplet is a real vector superfield $J_{\a \ad}(z)$ with weights $(\tfrac{3}{2}, \tfrac{3}{2})$ and satisfies the conservation equations $D^{\a} J_{\a \ad} = \bar{D}^{\ad} J_{\a \ad} = 0$. These objects possess fundamental information associated with internal and spacetime (super)symmetries and, hence, it is essentialy that we understand their general structure. The possible three-point functions involving the flavour current and supercurrent multiplets are:
\begin{subequations}
	\begin{align} \label{Low-spin component correlators}
		\langle L(z_{1}) \, L(z_{2}) \, L(z_{3}) \rangle \, , &&  \langle L(z_{1}) \, L(z_{2}) \, J_{\g \gd}(z_{3})  \rangle \, , \\
		\langle J_{\a \ad}(z_{1}) \, J_{\b \bd}(z_{2}) \, L(z_{3}) \rangle \, , &&  \langle J_{\a \ad}(z_{1}) \, J_{\b \bd}(z_{2}) \, J_{\g \gd}(z_{3}) \rangle \, .
	\end{align}
\end{subequations}
In the following subsections we systematically analyse these correlation functions and obtain their general form consistent with superconformal symmetry, conservation equations, and symmetries under permutations of superspace points. We then explicitly check the superinversion condition \eqref{Parity condition} and classify the conserved structures as parity-even or parity-odd under superinversion after imposing all the constraints.

\subsubsection{ \texorpdfstring{ $\langle L(z_{1}) L(z_{2}) L(z_{3}) \rangle$ }{ < L L L >} }

Let us first consider $\langle L L L \rangle$, hence, we will study the general three-point function $\langle J J' J'' \rangle$. The general ansatz for this correlation function, according to \eqref{3ptgen} is
\begin{align}
	\langle J(z_{1}) \, J'(z_{2}) \, J''(z_{3}) \rangle = \frac{ 1 }{ x_{1\bar{3}}{}^2  x_{\bar{1}3}{}^2 x_{2\bar{3}}{}^2  x_{\bar{2}3}{}^2 }
	\; \cH(X_{3}, \Q_{3}, \bar{\Q}_{3}) \, .
\end{align} 
where $\cH( \l \lb X, \l \Q, \lb \bar{\Q}) = (\l \lb)^{-2} \cH(X,\Q,\bar{\Q})$. Recall that it is usually more convenient to work with $\hat{\cH}(X, \Q, \bar{\Q})$, which is defined to be homogeneous degree 0 and is obtained by rescaling $\cH$ by a suitable power of $X$. 

Using the formalism outlined in Subsection \ref{GeneratingFunctionFormalism} the correlation function is encoded in the polynomial $\hat{\cH}$ constructed from the monomials given in \eqref{Monomials-bosonic}, \eqref{Monomials-fermionic}. Since $\hat{\cH}$ is a scalar superfield, there are only two structures to consider, which are linearly independent. The general ansatz for $\cH$ is
\begin{align}
    \hat{\cH}(X, \Q, \bar{\Q}) = A_{1} + A_{2} J \, ,
\end{align}
where $A_{i} = a_{i} + \text{i} b_{i}$ are complex coefficients.

The task now is to impose the conservation equations on each of the currents. Since the flavour current is a linear superfield, we must impose linearity conditions on all three superspace points. The constraints are:
\begin{subequations}
\begin{align}
    D_{1}^{2} \cH(X, \Q, \bar{\Q}) &= 0 \, , & \bar{D}_{1}^{2} \cH(X, \Q, \bar{\Q}) &= 0 \, , \\
    D_{2}^{2} \cH(X, \Q, \bar{\Q}) &= 0 \, , & \bar{D}_{2}^{2} \cH(X, \Q, \bar{\Q}) &= 0 \, , \\
    D_{3}^{2} \tilde{\cH}(X, \Q, \bar{\Q}) &= 0 \, , & \bar{D}_{3}^{2} \tilde{\cH}(X, \Q, \bar{\Q}) &= 0 \, .
\end{align}
\end{subequations}
After imposing each constraint we obtain a linear system in the coefficients $a_{i}$, $b_{i}$ which can be solved computationally. In this case, all conservation conditions are automatically satisfied for the given ansatz. Since the flavour currents are real superfields, the next constrain to impose is the reality condition:
\begin{align}
    \cH(X,\Q,\bar{\Q}) = \bar{\cH}(X,\Q,\bar{\Q}) \, , 
\end{align}
which results in the following relations between the coefficients $a_{i}$, $b_{i}$:
\begin{align}
    \{ a_{1} \rightarrow a_{1}, \,  a_{2} \rightarrow a_{2}, \, b_{1} \rightarrow 0,  \, b_{2} \rightarrow - 2 a_{1} \}
\end{align}
Hence, after imposing the conservation and reality conditions the polynomial $\cH(X,\Q,\bar{\Q})$ (and therefore the entire three-point function) is fixed up to two independent real parameters:
\begin{align}
    \cH(X,\Q,\bar{\Q}) = \frac{a_{1}}{X^{2}} (1 - 2 \text{i} J ) + \frac{a_{2}}{X^{2}} J \, . 
\end{align}
By explicitly checking the superinversion condition \eqref{Parity condition}, the structures above are classified as parity-even in superspace. 

The only remaining constraints to impose are those arising from symmetries under permutations of spacetime points, which apply when the currents in the three-point function are identical, i.e. when $J=J'=J''$. After imposing invariance under $z_{1} \leftrightarrow z_{2}$, only the structure corresponding to the coefficient $a_{1}$ survives. The remaining structure is manifestly compatible with the constraint resulting from invariance under $z_{1} \leftrightarrow z_{3}$. Hence, the three-point function of three flavour currents is fixed by superconformal symmetry up to a single real parameter, $a_{1}$.

It is important to note that if the flavour currents are non-abelian (and, hence, possess flavour indices), then the correlation function is described by a tensor $\cH_{ijk}$ with general solution\footnote{Note that these structures are identical to those appearing in eqn. (6.10) of \cite{Osborn:1998qu} after expanding $\bar{X}$.}
\begin{align}
    \cH_{ijk}(X,\Q,\bar{\Q}) = \frac{a_{1}}{X^{2}} d_{ijk} ( 1 - 2 \text{i} J ) + \frac{a_{2}}{X^{2}} f_{ijk} J \, .
\end{align}
Here, the $a_{1}$ structure is proportionate to a totally symmetric structure constant, $d_{ijk}$, while the $a_{2}$ structure is proportionate to a totally anti-symmetric structure constant, $f_{ijk}$. This implies that $\cN=1$ superconformal invariance leads to unique totally symmetric or totally antisymmetric expressions for the three-point function of conserved vector currents.

\subsubsection{ \texorpdfstring{ $\langle L(z_{1}) L(z_{2}) J_{\g \gd}(z_{3}) \rangle$ }{< L L J >} }

The next example to consider is $\langle L L J \rangle$, hence, we will study the correlation function $\langle J J' J''_{\g \gd} \rangle$. The general ansatz for this correlation function, according to \eqref{3ptgen} is
\begin{align}
	\langle J(z_{1}) \, J'(z_{2}) \, J''_{\g \gd}(z_{3}) \rangle = \frac{ 1 }{x_{1\bar{3}}{}^2  x_{\bar{1}3}{}^2 x_{2\bar{3}}{}^2  x_{\bar{2}3}{}^2}
	\; \cH_{\g \gd}(X_{3}, \Q_{3}, \bar{\Q}_{3}) \, .
\end{align} 
where $\cH$ is homogeneous degree $-1$, i.e. $\cH_{\g \gd}( \l \lb X, \l \Q, \lb \bar{\Q}) = (\l \lb)^{-1} \cH_{\g \gd}(X,\Q,\bar{\Q})$. Using the formalism outlined in Subsection \ref{GeneratingFunctionFormalism} the correlation function is encoded in the polynomial
\begin{align}
     \cH(X, \Q, \bar{\Q}; W) = \cH_{\g \gd}(X, \Q, \bar{\Q}) \, \boldsymbol{W}^{\g \gd} \, , 
\end{align}
which is constructed from the monomials given in \eqref{Monomials-bosonic},\eqref{Monomials-fermionic}. In this case there are four structures which may be used to construct $\hat{\cH}$ as a result of \eqref{Diophantine equations}:
\begin{align}
    \{ Z_{3}, J Z_{3}, R_{3} \tilde{R}_{3}, S_{3} \tilde{S}_{3} \}
\end{align}
However, these structures are linearly dependent due to the relation 
\begin{align}
    J Z_{3} + R_{3} \tilde{R}_{3} + S_{3} \tilde{S}_{3} = 0 \, ,
\end{align}
and, hence, the $S_{3} \tilde{S}_{3}$ structure may be eliminated. The general ansatz for $\hat{\cH}$ is therefore
\begin{align}
    \hat{\cH}(X, \Q, \bar{\Q}; W) = A_{1} Z_{3} + A_{2} J Z_{3}  + A_{3} R_{3} \tilde{R}_{3} \, ,
\end{align}
where $A_{i} = a_{i} + \text{i} b_{i}$ are complex coefficients. By successively applying the replacement rules \eqref{H_I replacement rules}, \eqref{Htilde replacement rules} to compute $\hat{\tilde{\cH}}$, the corresponding expression for $\hat{\tilde{\cH}}(X, \Q, \bar{\Q}; W)$ is:
\begin{align}
    \hat{\tilde{\cH}}(X, \Q, \bar{\Q}; W) = A_{1} ( - Z_{3} + 4 \text{i} J Z_{3}  + 4 \text{i} R_{3} \tilde{R}_{3} ) - A_{2} J Z_{3}  - A_{3} S_{3} \tilde{S}_{3} \, .
\end{align}
The task now is to impose the conservation equations on each of the currents. Since the flavour current multiplet, $L$, is a linear superfield, we must impose linearity conditions on the first two superspace points. Overall the constraints are:
\begin{subequations}
\begin{align}
    D_{1}^{2} \cH(X, \Q, \bar{\Q}; W) &= 0 \, , & \bar{D}_{1}^{2} \cH(X, \Q, \bar{\Q}; W) &= 0 \, , \\
    D_{2}^{2} \cH(X, \Q, \bar{\Q}; W) &= 0 \, , & \bar{D}_{2}^{2} \cH(X, \Q, \bar{\Q}; W) &= 0 \, , \\
    D_{3} \tilde{\cH}(X, \Q, \bar{\Q}; W) &= 0 \, , & \bar{D}_{3} \tilde{\cH}(X, \Q, \bar{\Q}; W) &= 0 \, .
\end{align}
\end{subequations}
After imposing each constraint we obtain a linear system in the coefficients $a_{i}$, $b_{i}$ which can be solved computationally. In this case, after conservation on $z_{1}$ and $z_{2}$ we obtain the following constraints on the coefficients:
\begin{align}
    \big\{ a_{1} \rightarrow a_{1}, \, a_{2} \rightarrow a_{2}, \, a_{3} \rightarrow a_{3}, \, b_{1} \rightarrow \frac{1}{4} ( a_{2} + a_{3} ), \, b_{2} \rightarrow b_{2} , \, b_{3} \rightarrow - 4a_{1} - b_{2} \big\} 
\end{align}
Conservation on $z_{3}$ is then automatically satisfied.

The next constrain to impose is the superfield reality condition \eqref{Reality condition}, from which we obtain the following relations between the coefficients:
\begin{align}
    \big\{ a_{1} \rightarrow a_{1}, \, a_{2} \rightarrow a_{2}, \, a_{3} \rightarrow - a_{2}, \, b_{1} \rightarrow 0, \, b_{2} \rightarrow - 2a_{1} , \, b_{3} \rightarrow - 2 a_{1} \big\} 
\end{align}
Therefore, after imposing conservation on all three superspace points and the superfield reality condition, the three-point function is fixed up to two independent real parameters, $a_{1}$ and $a_{2}$. At this stage the solution for $\cH(X,\Q,\bar{\Q};W)$ is
\begin{align}
    \cH(X,\Q,\bar{\Q};W) = \frac{a_{1}}{X} ( Z_{3} - 2 \text{i} J Z_{3} - 2 \text{i} R_{3} \tilde{R}_{3} ) + \frac{a_{2}}{X} ( J Z_{3} - R_{3} \tilde{R}_{3} ) \, .
\end{align}
Using the superinversion condition \eqref{Parity condition}, these structures are classified as parity-even under superinversion in superspace. Note that the structures are also hermitian, however, to show this explicitly one must make use of the linear dependence relations \eqref{Linear dependence 1}-\eqref{Linear dependence 16}. 

The last constraint to impose is the symmetry under permutation of superspace points $z_{1}$ and $z_{2}$, from which we obtain $a_{1} \to 0$. Hence, the three-point function $\langle L(z_{1}) L(z_{2}) J_{\g \gd}(z_{3}) \rangle$ is fixed by $\cN=1$ superconformal invariance up to one independent real parameter.

\noindent\textbf{Generalisation: $ \langle L(z_{1}) L(z_{2}) J_{\g(s) \gd(s)}(z_{3}) \rangle$}

One can also consider the case where the supercurrent $J$ is promoted to a ``vector-like" supercurrent $J_{s}$ of arbitrary superspin-$s$. In this case the solution for $\cH(X,\Q,\bar{\Q};W)$ is fixed up to two independent parameters: 
\begin{align}
    &\cH(X,\Q,\bar{\Q};W) = \frac{a_{1}}{X^{2-s}} ( Z_{3} - 2 \text{i} J Z_{3} - 2 s \text{i}R_{3} \tilde{R}_{3} ) Z_{3}^{s-1}  + \frac{a_{2}}{X^{2-s}} ( J Z_{3} - s R_{3} \tilde{R}_{3} ) Z_{3}^{s-1} \, .
\end{align}
In this case, after imposing the symmetry under permutation of superspace points $z_{1}$ and $z_{2}$ we find for even $s$, $a_{2} \to 0$, while for odd $s$, $a_{1} \to 0$. Hence, the three-point function $\langle L(z_{1}) L(z_{2}) J_{\g(s) \gd(s)}(z_{3}) \rangle$ is fixed up to one independent real parameter in general.

\subsubsection{ \texorpdfstring{ $\langle J_{\a \ad}(z_{1}) J_{\b \bd}(z_{2}) L(z_{3}) \rangle$ }{< J J L >} }
The next three-point function to consider is $\langle J J L \rangle$. To determine the structure of this correlation function we may analyse $\langle J_{\a \ad}(z_{1}) J'_{\b \bd}(z_{2}) J''(z_{3}) \rangle$, which has the following general ansatz in accordance with \eqref{3ptgen}:
\begin{align}
	\langle J_{\a \ad}(z_{1}) \, J'_{\b \bd}(z_{2}) \, J''(z_{3}) \rangle = \frac{ \cI_{\a}{}^{\ad'}(x_{1\bar{3} }) \, \bar{\cI}_{\ad}{}^{\a'}(x_{\bar{1}3}) \, \cI_{\b}{}^{\bd'}(x_{2 \bar{3}}) \, \bar{\cI}_{\bd}{}^{\b'}(x_{\bar{2} 3}) }{ (x_{1\bar{3}}{}^2)^{3/2} (x_{\bar{1}3}{}^2)^{3/2} (x_{2\bar{3}}{}^2)^{3/2}  (x_{\bar{2}3}{}^2)^{3/2} }
	\; \cH_{\a' \ad' \b' \bd'}(X_{3}, \Q_{3}, \bar{\Q}_{3}) \, .
\end{align} 
Using the general formalism, the structure of this three-point function is encoded in the polynomial
\begin{align}
    \cH(X,\Q,\bar{\Q}; U, V) = \cH_{\a \ad \b \bd}(X,\Q,\bar{\Q}) \, \boldsymbol{U}^{\a \ad} \boldsymbol{V}^{\b \bd} \, .
\end{align}
After solving \eqref{Diophantine equations} for this configuration of superspins, we obtain 18 possible structures from which the polynomial $\cH$ may be constructed:
\begin{align}
    \begin{split}
		& \big\{Z_1 Z_2,J Z_1 Z_2,P_3 \tilde{P}_3,J P_3
        \tilde{P}_3,R_2 S_1 \tilde{P}_3,R_1 S_2
        \tilde{P}_3,Q_3 \tilde{Q}_3,J Q_3 \tilde{Q}_3,Q_3 R_2
        \tilde{R}_1, \\
        & \hspace{5mm} Z_2 R_1 \tilde{R}_1,Z_1 R_2
        \tilde{R}_2, \tilde{Q}_3 R_1 \tilde{R}_2,Q_3 S_2
        \tilde{S}_1, Z_2 S_1 \tilde{S}_1, P_3 \tilde{R}_2
        \tilde{S}_1,Z_1 S_2 \tilde{S}_2,\tilde{Q}_3 S_1 
        \tilde{S}_2,P_3 \tilde{R}_1 \tilde{S}_2\big\}
    \end{split}
\end{align}
There is, however, a high degree of linear dependence present. Indeed, after systematic application of the linear dependence relations \eqref{Linear dependence 1}--\eqref{Linear dependence 16} we obtain 7 linearly independent structures:
\begin{align}
    \big\{P_3 \tilde{P}_3,J P_3 \tilde{P}_3,R_2 S_1
    \tilde{P}_3,Q_3 \tilde{Q}_3,J Q_3 \tilde{Q}_3,Q_3 R_2
    \tilde{R}_1,R_1 \tilde{Q}_3 \tilde{R}_2\big\}
\end{align}
A general ansatz for $\hat{\cH}$ is therefore
\begin{align}
   \hat{\cH}(X,\Q,\bar{\Q}; U, V) &= A_{1} P_3 \tilde{P}_3 + A_{2} J P_3 \tilde{P}_3 + A_{3} R_2 S_1 \tilde{P}_3 + A_{4} Q_3 \tilde{Q}_3 \non \\
   & \hspace{10mm} + A_{5} J Q_3 \tilde{Q}_3 + A_{6} Q_3 R_2 \tilde{R}_1 + A_{7} R_1 \tilde{Q}_3 \tilde{R}_2 \, .
\end{align}
After applying the replacement rules \eqref{H_I replacement rules}, \eqref{Htilde replacement rules} the corresponding expression for $\hat{\tilde{\cH}}$ is
\begin{align}
   \hat{\tilde{\cH}}(X,\Q,\bar{\Q}; U, V) &= A_{1} ( Q_3 \tilde{Q}_3-4 \text{i} R_1 \tilde{Q}_3 \tilde{R}_2 ) + A_{2} J
   Q_3 \tilde{Q}_3 + A_{3} Q_3 R_2 \tilde{R}_1 \\
   & \hspace{5mm} + A_{4}( P_3 \tilde{P}_3 -4 \text{i} J P_3
   \tilde{P}_3+4 \text{i} R_1 S_2 \tilde{P}_3+4 \text{i} P_3
   \tilde{R}_2 \tilde{S}_1-4 \text{i} P_3 \tilde{R}_1
   \tilde{S}_2 ) \non \\
   & \hspace{10mm} + A_{5} J P_3 \tilde{P}_3 + A_{6} R_2 S_1
   \tilde{P}_3 - A_{7} P_3 \tilde{R}_2 \tilde{S}_1 \, . \non
\end{align}
Given the linearly independent ansatz, the task now is to impose the superfield conservation and reality conditions. The conservation conditions which we must impose in this case are
\begin{subequations}
\begin{align}
    D_{1} \cH(X, \Q, \bar{\Q}; U,V) &= 0 \, , & \bar{D}_{1} \cH(X, \Q, \bar{\Q}; U,V) &= 0 \, , \\
    D_{2} \cH(X, \Q, \bar{\Q}; U, V) &= 0 \, , & \bar{D}_{2} \cH(X, \Q, \bar{\Q}; U,V) &= 0 \, , \\
    D_{3}^{2} \tilde{\cH}(X, \Q, \bar{\Q}; U,V) &= 0 \, , & \bar{D}_{3}^{2} \tilde{\cH}(X, \Q, \bar{\Q}; U,V) &= 0 \, .
\end{align}
\end{subequations}
After imposing conservation on $z_{1}$, we obtain the following relations between the coefficients:
\begin{align}
    &\big\{a_1\to a_1, \, a_2\to a_2, \, a_3\to 0, \, a_4\to -2
   a_1, \, a_5\to -3 a_2, \, a_6\to -a_2, \, a_7\to -a_2, \\
   & \hspace{10mm} b_1\to b_1,\, b_2\to b_2,\, b_3\to 0, \, b_4\to -2 b_1,\, b_5\to -3 b_2,\, b_6\to -b_2,\, b_7\to -b_2\big\} \non
\end{align}
Hence, at this stage the solution is fixed up to four independent real parameters $a_{1}, a_{2}, b_{1}, b_{2}$. It turns out that conservation on $z_{2}$ and $z_{3}$ automatically follows from the above relations. Next, we impose the superfield reality condition \eqref{Reality condition}, which results in $b_{1} \to 0$, $b_{2} \to - 4 a_{1}$. Therefore, after imposing conservation on all three superspace points and the reality condition, $\cH$ is fixed up to two independent real parameters $a_{1}$ and $a_{2}$:
\begin{align}
\cH(X,\Q,\bar{\Q}; U,V) &= \frac{a_1}{X^{4}}
   \big(P_3 \tilde{P}_3 -2 Q_3 \tilde{Q}_3 -4 \text{i} J P_3 \tilde{P}_3 +4  \text{i} Q_3 R_2 \tilde{R}_1 \non \\
   & \hspace{20mm} +4
    \text{i} R_1 \tilde{Q}_3 \tilde{R}_2 +12  \text{i} J Q_3
   \tilde{Q}_3 \big) \non \\
   & \hspace{5mm}+ \frac{a_2}{X^{4}} \big(J P_3 \tilde{P}_3-3 J Q_3 \tilde{Q}_3-Q_3 R_2 \tilde{R}_1-R_1 \tilde{Q}_3 \tilde{R}_2\big) \, .
\end{align}
By explicitly checking the superinversion condition \eqref{Parity condition}, these structures are classified as parity-even in superspace (and are hermitian). After setting $J = J'$ and imposing the required symmetries under the exchange of $z_{1}$ and $z_{2}$ it may be shown that $a_{2} \to 0$. Hence, the three-point function $\langle J J L \rangle$ is fixed by $\cN = 1$ superconformal symmetry up to a single independent real parameter.

We can also consider the case where the supercurrent $J$ at $z_{1}$ is promoted to a ``vector-like" supercurrent $J_{s}$ of arbitrary superspin $s>1$. In this case, after solving \eqref{Diophantine equations} there are 24 possible structures, only 8 of which are linearly independent:
\begin{align}
    \begin{split}
        &\big\{P_3 \tilde{P}_3 Z_1^{s-1},J P_3 \tilde{P}_3 Z_1^{s-1},Q_3 \tilde{Q}_3 Z_1^{s-1},J Q_3 \tilde{Q}_3
        Z_1^{s-1},R_1 S_1 \tilde{P}_3 \tilde{Q}_3 Z_1^{s-2}, \non \\
        & \hspace{15mm} P_3 R_1 \tilde{P}_3 \tilde{R}_1 Z_1^{s-2},Q_3 R_1
        \tilde{Q}_3 \tilde{R}_1 Z_1^{s-2},P_3 Q_3 \tilde{R}_1 \tilde{S}_1 Z_1^{s-2}\big\}
    \end{split}
\end{align}
After forming an ansatz from the above structures and imposing the conservation equations and the superfield reality condition, the solution for $\cH(X,\Q,\bar{\Q};U,V)$ is 
\begin{align}
   \cH(X,\Q,\bar{\Q}; U,V) &= \frac{a_1}{X^{s+3}} \Big( P_3 \tilde{P}_3 Z_1 - (s+1) Q_3
   \tilde{Q}_3 Z_1 - 2 (s+1)\text{i} J P_3  \tilde{P}_3
   Z_1 \non \\
   & +2 (s^{2} + 3s + 2) \text{i} J Q_3  \tilde{Q}_3 Z_1 +2 (s^{2} + 2s + 1) \text{i} Q_3 R_1 \tilde{Q}_3 \tilde{R}_1  \non \\
   & -2 (s+1) \text{i} R_1  S_1 \tilde{P}_3 \tilde{Q}_3 +2 (s+1) \text{i} P_3 Q_3 \tilde{R}_1 \tilde{S}_1 -2 (s-1) \text{i} P_3 R_1  \tilde{P}_3 \tilde{R}_1  \Big) Z_1^{s-2} \non \\
   &+ \frac{a_2}{X^{s+3}} \Big(J P_3 \tilde{P}_3 Z_1- (s+2) J Q_3 \tilde{Q}_3 Z_1+R_1 S_1 \tilde{P}_3 \tilde{Q}_3
   \non \\
   & \hspace{5mm} +\frac{s-1}{s+1} P_3 R_1 \tilde{P}_3 \tilde{R}_1
    - (s+1) Q_3 R_1 \tilde{Q}_3 \tilde{R}_1 -P_3 Q_3 \tilde{R}_1 \tilde{S}_1 \Big) \, Z_1^{s-2} \, .
\end{align}
Hence, the three-point function $\langle J_{s} J L \rangle$ is fixed up to two independent real parameters for arbitrary $s$. 

\noindent\textbf{Generalisation: $\langle J^{}_{\a(s_1) \ad(s_1)}(z_{1}) J'_{\b(s_2) \bd(s_2)}(z_{2}) L(z_{3}) \rangle$}

In \cite{Buchbinder:2022kmj}, a closed form expression for the correlator $\langle J_{s_1} J'_{s_2} L \rangle$ for arbitrary superspins $s_1$ and $s_2$ was presented using a similar index-free formalism.\footnote{Here, without loss of generality, one can assume $s_1 \leq s_2$.}
The general ansatz for this three-point function in accordance with \eqref{3ptgen} is
\bea 
&& \langle J_{\a(s_1) \ad(s_1)}(s_1)  J'_{\b(s_2) \bd(s_2)}(z_2) L(z_3) \rangle \non\\
&=&
\frac{1}{(x_{1 \bar{3}} x_{\bar{1} 3} )^{s_1+2} (x_{2 \bar{3} } x_{\bar{2} 3} )^{s_2+2}}  \cI_{\a(s_1)}{}^{\ad'(s_1)} (x_{1 \bar{3}})\, \bar{\cI}_{\ad(s_1)}{}^{\a'(s_1)} (x_{\bar{1} 3}) \non\\
&&\times~ \cI_{\b(s_2)}{}^{ \bd'(s_2)} (x_{2 \bar{3}})\, \bar{\cI}_{\bd(s_2)}{}^{ \b'(s_2)} (x_{\bar{2} 3})  \,\, \cH_{\a'(s_1)\ad'(s_1), \, \b'(s_2) \bd'(s_2)} (X_3, {\Q}_3 , \bar{\Q}_3)~.
\eea 
In our notation, the solution for the polynomial
\begin{align}
    \cH(X,\Q,\bar{\Q}; U, V) = \cH_{\a(s_1) \ad(s_1),\, \b(s_2) \bd(s_2)}(X,\Q,\bar{\Q}) \, \boldsymbol{U}^{\a(s_1) \ad(s_1)} \boldsymbol{V}^{\b(s_2) \bd(s_2)} \, .
\end{align}
reads
\bea
&&\cH(X,\Q,\bar{\Q}; U, V) = \frac{1}{X^{s_1+s_2+2}} \bigg\{
\sum_{k=0}^{s_1} \Big(-\hf\Big) a_k \,Q_1^{s_1-k} Q_2^{s_2-k} Z_1^{s_1-k} Z_2^{s_2-k} P_3^k \tilde{P}_3^k
\non\\
&&-  
\sum_{k=0}^{s_1} \Big(b_k- \ri(s_1+s_2+1-k) a_k \Big) J Z_1^{s_1-k} Z_2^{s_2-k} P_3^k \tilde{P}_3^k
\non\\
&&+ \sum_{k=0}^{s_1-1} \Big(\frac{s_1-k}{s_1+s_2+1-k}b_k+ \ri (s_1-k) a_k \Big)
R_1 \tilde{R}_1 Z_1^{s_1-(k+1)} Z_2^{s_2-k} P_3^k \tilde{P}_3^k
\non\\
&&+ 
\sum_{k=0}^{s_1} \Big(\frac{s_1-k}{s_1+s_2+1-k}b_k+ \ri (s_2-k) a_k \Big)
R_2 \tilde{R}_2 Z_1^{s_1-k} Z_2^{s_2-(k+1)} P_3^k \tilde{P}_3^k \bigg\}~,
\eea
where, for $k=1,2, \dots s_1$, the real parameters $a_k$ and $b_k$ satisfy the recursive relations
\bea
a_k = -\frac{(s_1+1-k) (s_2+1-k)}{2k(s_1+s_2+1-k)} a_{k-1}~, \quad
b_k = -\frac{(s_1+1-k) (s_2+1-k)}{2k(s_1+s_2+2-k)} b_{k-1}~.
\eea
This implies that the three-point function $\langle J_{s_1}J'_{s_2} L \rangle$ is fixed up to two independent real parameters $a_0$ and $b_0$ for arbitrary $s_1$ and $s_2$.

In the special case where $s_1= s_2= s$, the correlator is subject to an extra constraint due to the $1 \leftrightarrow 2$ symmetry. As discussed in \cite{Buchbinder:2022kmj}, this leads to $b_k = 0$, for $k=0,1,\dots, s$. The three-point function $\langle J_{s}J_{s} L \rangle$ is thus fixed up to one real parameter $a_0$.

\subsubsection{ \texorpdfstring{ $\langle J_{\a \ad}(z_{1}) J_{\b \bd}(z_{2}) J_{\g \gd}(z_{3}) \rangle$ }{< J J J >} }

The final fundamental correlation function to study is the three-point function of three supercurrents $\langle J J J \rangle$, and so we analyse the structure of $\langle J_{\a \ad}(z_{1}) J'_{\b \bd}(z_{2}) J''_{\g \gd}(z_{3}) \rangle$. The general ansatz for this three-point function in accordance with \eqref{3ptgen} is
\begin{align}
	\langle J_{\a \ad}(z_{1}) \, J'_{\b \bd}(z_{2}) \, J''_{\g \gd}(z_{3}) \rangle = \frac{ \cI_{\a}{}^{\ad'}(x_{1 \bar{3}}) \, \bar{\cI}_{\ad}{}^{\a'}(x_{\bar{1} 3}) \, \cI_{\b}{}^{\bd'}(x_{2 \bar{3}}) \, \bar{\cI}_{\bd}{}^{\b'}(x_{\bar{2} 3}) }{ (x_{1\bar{3}}{}^2)^{3/2} (x_{\bar{1}3}{}^2)^{3/2} (x_{2\bar{3}}{}^2)^{3/2}   (x_{\bar{2}3}{}^2)^{3/2}}
	\; \cH_{\a' \ad' \b' \bd' \g \gd}(X_{3}, \Q_{3}, \bar{\Q}_{3}) \, .
\end{align} 
Using the general formalism presented in \ref{GeneratingFunctionFormalism}, the structure of this three-point function is encoded in the polynomial
\begin{align}
    \cH(X,\Q,\bar{\Q}; U, V, W) = \cH_{\a \ad \b \bd \g \gd}(X,\Q,\bar{\Q}) \, \boldsymbol{U}^{\a \ad} \boldsymbol{V}^{\b \bd} \boldsymbol{W}^{\g \gd} \, .
\end{align}
After solving \eqref{Diophantine equations} for this configuration of superspins we find there are 120 linearly dependent structures - too many to present here. However, only 19 of the structures are linearly independent:
\begin{align}
    &\big\{Q_1 Q_2 Q_3,J Q_1 Q_2 Q_3,P_3 Q_2 \tilde{P}_1,J
   P_3 Q_2 \tilde{P}_1,P_1 Q_3 \tilde{P}_2,J P_1 Q_3
   \tilde{P}_2,P_2 Q_1 \tilde{P}_3, J P_2 Q_1 \tilde{P}_3, \non \\
   & \hspace{10mm} \tilde{Q}_1 \tilde{Q}_2 \tilde{Q}_3,J
   \tilde{Q}_1 \tilde{Q}_2 \tilde{Q}_3,Q_1 Q_3 R_3
   \tilde{R}_1,P_1 R_1 \tilde{P}_1 \tilde{R}_1,P_2 R_2
   \tilde{P}_1 \tilde{R}_1, Q_1 R_1 \tilde{Q}_1
   \tilde{R}_1, \\
   & \hspace{15mm} Q_1 Q_2 R_1 \tilde{R}_2,P_2 R_2
   \tilde{P}_2 \tilde{R}_2,Q_2 R_2 \tilde{Q}_2
   \tilde{R}_2,Q_2 Q_3 R_2 \tilde{R}_3,Q_3 R_3
   \tilde{Q}_3 \tilde{R}_3\big\} \non
\end{align}
We now form a general ansatz for $\cH(X,\Q,\bar{\Q}; U,V,W)$ from the above list such that the `$i$'th element is associated with a complex coefficient $A_{i} = a_{i} + \text{i} b_{i}$ as follows:
\begin{align}
    \hat{\cH}(X,\Q,\bar{\Q};U,V,W) &= A_{1} Q_1 Q_2 Q_3 + A_{2} J Q_1 Q_2 Q_3 + A_{3} P_3 Q_2 \tilde{P}_1 \non \\
    & \hspace{5mm} + A_{4} J
    P_3 Q_2 \tilde{P}_1 + A_{5} P_1 Q_3 \tilde{P}_2 + A_{6} J P_1 Q_3
    \tilde{P}_2 + \dots \, .
\end{align}
The task now is to impose the superfield conservation equations and the reality condition. The conservation equations are summarised below:
\begin{subequations}
\begin{align}
    D_{1} \cH(X, \Q, \bar{\Q}; U,V, W) &= 0 \, , & \bar{D}_{1} \cH(X, \Q, \bar{\Q}; U,V, W) &= 0 \, , \\
    D_{2} \cH(X, \Q, \bar{\Q}; U, V, W) &= 0 \, , & \bar{D}_{2} \cH(X, \Q, \bar{\Q}; U,V,W) &= 0 \, , \\
    D_{3} \tilde{\cH}(X, \Q, \bar{\Q}; U,V, W) &= 0 \, , & \bar{D}_{3} \tilde{\cH}(X, \Q, \bar{\Q}; U,V, W) &= 0 \, .
\end{align}
\end{subequations}
After imposing conservation on $z_{1}$ there are 16 independent real parameters remaining. Conservation on $z_{2}$ then reduces this down to 8 real parameters, with conservation on $z_{3}$ following automatically. Hence, after imposing conservation on all three-points, the three-point function is fixed up to 8 independent real parameters.

Next, we impose the reality condition \eqref{Reality condition}, which fixes the three-point function up to 4 real parameters. The relations between the coefficients are
\begin{align}
    &\big\{a_1\to a_1,a_2\to a_2,a_3\to a_3,a_4\to a_4,a_5\to
    2 a_1-3 a_3,a_6\to a_2-\tfrac{a_4}{6}, \non \\
    & \hspace{5mm} a_7\to a_3,a_8\to
    a_4,a_9\to a_3-a_1,a_{10}\to
    \tfrac{a_4}{2}-a_2,a_{11}\to -\tfrac{a_4}{3},a_{12}\to
    -\tfrac{a_4}{2}, \non \\
    & \hspace{5mm} a_{13}\to \tfrac{a_2}{2}-\tfrac{2
    a_4}{3},a_{14}\to \tfrac{2 a_4}{3}, a_{15}\to
    -\tfrac{a_4}{3},a_{16}\to -\tfrac{a_4}{2},a_{17}\to
    \tfrac{2 a_4}{3}, \non \\
    & \hspace{5mm} a_{18}\to -\tfrac{a_4}{3},a_{19}\to
    -\tfrac{a_4}{6},b_1\to
    \tfrac{a_2}{8}-\tfrac{a_4}{4},b_2\to 4 a_3-8 a_1,b_3\to
    0, \\
    & \hspace{5mm} b_4\to -4 a_3,b_5\to 0,b_6\to 10 a_3-8 a_1,b_7\to
    0,b_8\to -4 a_3,\non \\
    & \hspace{5mm}  b_9\to
    \tfrac{a_4}{4}-\tfrac{a_2}{8},b_{10}\to 8 a_1-10
    a_3, b_{11}\to 0,b_{12}\to 2 a_3,b_{13}\to 8 a_3-4
    a_1, \non \\
    & \hspace{5mm} b_{14}\to 0,b_{15}\to -4 a_3,b_{16}\to 2
    a_3,b_{17}\to 0,b_{18}\to 0,b_{19}\to -2 a_3\big\} \non
\end{align}
At this stage the remaining free parameters are $a_{1}$, $a_{2}$, $a_{3}$ and $a_{4}$. After making the relabeling $a_{2} \to b_{1}$, $a_{3} \to a_{2}$ and $a_{4} \to b_{2}$, the general solution for $\cH(X,\Q,\bar{\Q}; U,V,W)$ after imposing conservation and reality is
\begin{align}
    & \frac{a_1}{X^{3}} \big(Q_1 Q_2 Q_3+2 P_1 Q_3 \tilde{P}_2 -\tilde{Q}_1 \tilde{Q}_2 \tilde{Q}_3-8 \text{i} J P_1 Q_3 \tilde{P}_2 \non \\
    & \hspace{15mm} +8 \text{i} J \tilde{Q}_1 \tilde{Q}_2 \tilde{Q}_3-4 \text{i} P_2
    R_2 \tilde{P}_1 \tilde{R}_1 -8 \text{i} J Q_1 Q_2 Q_3 \big) \non \\[2mm]
    & + \frac{a_2}{X^{3}}
    \big(P_3 Q_2 \tilde{P}_1  -3 P_1 Q_3
    \tilde{P}_2 +P_2 Q_1 \tilde{P}_3 + \tilde{Q}_1 \tilde{Q}_2
    \tilde{Q}_3 + 10 \text{i} J P_1 Q_3 \tilde{P}_2-4 \text{i} J P_3 Q_2
    \tilde{P}_1  \non \\
    & \hspace{10mm} -4 \text{i} J P_2 Q_1 \tilde{P}_3-10 \text{i} J
    \tilde{Q}_1 \tilde{Q}_2 \tilde{Q}_3 +2
    \text{i} P_1 R_1 \tilde{P}_1 \tilde{R}_1 +8 \text{i} P_2 R_2
    \tilde{P}_1 \tilde{R}_1 \\[1mm]
    & \hspace{15mm} +2 \text{i} P_2 R_2 \tilde{P}_2
    \tilde{R}_2 -2 \text{i} Q_3 R_3 \tilde{Q}_3 \tilde{R}_3-4 \text{i} Q_1 Q_2 R_1 \tilde{R}_2+4 \text{i} J Q_1 Q_2 Q_3\big) \non \\[2mm]
    & + \frac{\text{i} b_1}{X^{3}}
    \big(\tfrac{1}{8} Q_1 Q_2
    Q_3 -\tfrac{1}{8} \tilde{Q}_1 \tilde{Q}_2
    \tilde{Q}_3 - \text{i} J P_1 Q_3 \tilde{P}_2 + \text{i} J \tilde{Q}_1 \tilde{Q}_2
    \tilde{Q}_3 - \tfrac{\text{i} }{2} P_2 R_2 \tilde{P}_1
    \tilde{R}_1 - \text{i} J Q_1 Q_2 Q_3\big) \non \\[2mm]
    & + \frac{\text{i} b_2}{X^{3}} \big( \tfrac{1}{4} \tilde{Q}_1 \tilde{Q}_2
    \tilde{Q}_3-\tfrac{1}{4} Q_1 Q_2 Q_3 +\tfrac{\text{i} }{6} J P_1 Q_3
    \tilde{P}_2 - \text{i} J P_3 Q_2 \tilde{P}_1 - \text{i} J P_2 Q_1
    \tilde{P}_3\non \\
    & \hspace{10mm} - \tfrac{\text{i} }{2} J \tilde{Q}_1 \tilde{Q}_2
    \tilde{Q}_3 + \tfrac{\text{i} }{2} P_1 R_1 \tilde{P}_1
    \tilde{R}_1 + \tfrac{2\text{i} }{3} P_2 R_2 \tilde{P}_1
    \tilde{R}_1 + \tfrac{\text{i} }{2} P_2 R_2 \tilde{P}_2
    \tilde{R}_2 + \tfrac{\text{i} }{3} Q_1 Q_3 R_3
    \tilde{R}_1 \non \\
    & \hspace{15mm}  + \tfrac{\text{i} }{3} Q_2 Q_3 R_2
    \tilde{R}_3 + \tfrac{\text{i} }{6} Q_3 R_3 \tilde{Q}_3
    \tilde{R}_3 - \tfrac{2\text{i} }{3} Q_1 R_1 \tilde{Q}_1
    \tilde{R}_1 + \tfrac{\text{i} }{3} Q_1 Q_2 R_1
    \tilde{R}_2 - \tfrac{2\text{i} }{3} Q_2 R_2 \tilde{Q}_2
    \tilde{R}_2 \big) \non \, .
\end{align}
Using the superinversion condition \eqref{Parity condition}, the structures with coefficients $a_{1}$, $a_{2}$ are classified as parity-even in superspace, while the structures with coefficients $b_{1}$, $b_{2}$ are classified as parity-odd in superspace. Note that the structures with coefficients $a_{i}$ and $b_{i}$ are hermitian and anti-hermitian respectively, however, to show this explicitly one must make use of the linear dependence relations \eqref{Linear dependence 1}-\eqref{Linear dependence 16}. 

The last constraints to impose are those resulting from symmetries under permutations of superspace points. In particular, we require (separately) invariance under $z_{1} \leftrightarrow z_{2}$ and $z_{1} \leftrightarrow z_{3}$. After imposing these constraints we obtain $a_{1} \to 0$, $a_{2} \to 0$ and, hence, the three-point function of three supercurrent multiplets is fixed by $\cN=1$ superconformal symmetry up to two independent real parameters, which is in accord with the results presented in \cite{Osborn:1998qu,Buchbinder:2022kmj}. However, according to the superinversion property \eqref{Parity condition} they are classified as parity-odd in superspace.


\subsection{General structure of three-point functions of higher-spin (vector-like) supercurrents}\label{section4.2}

Based on our computational analysis we propose that the general structure of three-point correlation functions involving real ``vector-like" supercurrent multiplets $J_{s}(z) := J_{\a(s) \ad(s)}(z)$ of superspin-$s$ are of the following form
\begin{align}
	\langle  J^{}_{s_{1}}(z_{1}) \, J'_{s_{2}}(z_{2}) \, J''_{s_{3}}(z_{3}) \rangle &= \sum_{i=1}^{\min(s_{1}, s_{2}, s_{3}) + 1} a_{i} \, \langle  J^{}_{s_{1}}(z_{1}) \, J'_{s_{2}}(z_{2}) \,  J''_{s_{3}}(z_{3}) \rangle_{i}^{E} \\
    & \hspace{10mm} + \sum_{i=1}^{\min(s_{1}, s_{2}, s_{3}) + 1} \text{i} b_{i} \, \langle  J^{}_{s_{1}}(z_{1}) \, J'_{s_{2}}(z_{2}) \, J''_{s_{3}}(z_{3}) \rangle_{i}^{O} \, , \non
\end{align}
where $a_{i}$, $b_{i}$ are real coefficients and $\langle J^{}_{s_{1}}(z_{1}) \, J'_{s_{2}}(z_{2}) \, J''_{s_{3}}(z_{3}) \rangle_{i}^{E/O}$ are linearly independent conserved structures which are classified as parity-even/odd in superspace according to the definition \eqref{Parity condition}. Furthermore, for three-point functions which possess symmetries under permutation of superspace points (i.e. in the case where any of the superfields are identical), the following classification holds:
\begin{itemize}
	\item For three-point functions $\langle J^{}_{s} \, J'_{s} \, J''_{s} \rangle$ there are $2 s + 2$ conserved structures, where $s+1$ are parity-even and $s+1$ are parity-odd in superspace according to the superinversion condition \eqref{Parity condition}. When the superfields coincide, i.e. $J = J' $, for even $s$ the $s+1$ parity-even structures survive, while for odd $s$ the $s+1$ parity-odd structures survive.
	
	\item For three-point functions $\langle J^{}_{s_{1}} \, J'_{s_{1}} \, J''_{s_{2}} \rangle$, there are $2 \min(s_{1}, s_{2}) + 2$ conserved structures, where $\min(s_{1}, s_{2}) + 1$ are parity-even and $\min(s_{1}, s_{2}) + 1$ are parity-odd in superspace. When the superfields coincide, i.e. $J = J'$, we obtain $\min(s_{1}, s_{2})+1$ parity-even structures in the case when the superspin $s_2$ is even, and $\min(s_{1}, s_{2}) + 1$ parity-odd structures when $s_2$ is odd.
\end{itemize}
Since the independent structures in the supersymmetric three-point functions quickly become large for increasing superspins, we found it feasible to present only some simple cases in Appendix \ref{AppC}. The simplest example which demonstrates the pattern for the number of independent structures is the superspin-2 multiplet three-point function, see \ref{superspin2}.


\subsection{Comments on superspace reduction of three-point functions and the classification of parity-even/odd structures}\label{section4.3}

In this subsection we will discuss how $\cN=1$ supersymmetry constrains the component three-point functions contained 
in $\langle  J^{}_{s_{1}}(z_{1}) \, J'_{s_{2}}(z_{2}) \, J''_{s_{3}}(z_{3}) \rangle$. First, let us recall some results regarding the structure of 
three-point functions of conserved currents. In four dimensions, three-point correlation functions of real ``vector-like" higher-spin conserved 
currents, $J_{s}(x) := J_{\a(s) \ad(s)}(x)$, have been analysed in the following 
publications~\cite{Stanev:2012nq, Zhiboedov:2012bm,Buchbinder:2023coi, Karapetyan:2023zdu}. The general structure of the three-point function $\langle J^{}_{s_{1}}(x_{1}) \, J'_{s_{2}}(x_{2}) \, J''_{s_{3}}(x_{3}) \rangle$ was found to be fixed up to the 
following form \cite{Zhiboedov:2012bm, Stanev:2012nq, Buchbinder:2023coi, Karapetyan:2023zdu}:
\begin{align} \label{4D correlators results}
	\langle  J^{}_{s_{1}}(x_{1}) \, J'_{s_{2}}(x_{2}) \, J''_{s_{3}}(x_{3}) \rangle &= \sum_{i=1}^{\min(s_{1}, s_{2}, s_{3}) + 1} a_{i} \, \langle  J^{}_{s_{1}}(x_{1}) \, J'_{s_{2}}(x_{2}) \, J''_{s_{3}}(x_{3}) \rangle_{i}^{E} \\
    & \hspace{10mm} + \sum_{i=1}^{\min(s_{1}, s_{2}, s_{3}) } \text{i} b_{i} \, \langle  J^{}_{s_{1}}(x_{1}) \, J'_{s_{2}}(x_{2}) \, J''_{s_{3}}(x_{3}) \rangle_{i}^{O} \, , \non
\end{align}
where $a_{i}$, $b_{i}$ are real coefficients and $\langle J^{}_{s_{1}}(x_{1}) \, J'_{s_{2}}(x_{2}) \, J''_{s_{3}}(x_{3}) \rangle_{i}^{E/O}$ are linearly 
independent conserved structures which are classified as parity-even/odd under inversion.\footnote{Note that if the reality condition is not imposed, 
the three-point function is fixed up to $2 \min(s_{i}) + 1$ structures with complex coefficients.} 
For correlation functions which possess symmetries under permutation of spacetime points (i.e. in the case where any of the fields are identical), 
the following classification holds:
\begin{itemize}
	\item For three-point functions $\langle J^{}_{s} \, J'_{s} \, J''_{s} \rangle$ there are $2 s + 1$ conserved structures, $s+1$ being parity-even and $s$ being parity-odd.  
	When the fields coincide, i.e. $J = J' $ the number of structures is reduced to the $s+1$ parity-even structures in the case when the spin $s$ is even,
	or to the $s$ parity-odd structures in the case when $s$ is odd. 
	
	\item For three-point functions $\langle J^{}_{s_{1}} \, J'_{s_{1}} \, J''_{s_{2}} \rangle$, there are $2 \min(s_{1}, s_{2}) + 1$ conserved structures, 
	$\min(s_{1}, s_{2}) +1$ are parity-even and $\min(s_{1}, s_{2}) $ are parity-odd. For $J = J'$, the number of structures is reduced to the $\min(s_{1}, s_{2})+1$ parity-even structures in the case when the spin $s_2$ is even, or to the $\min(s_{1}, s_{2})$ parity-odd structures in the case when $s_2$ is odd.
\end{itemize}
In the remainder of this section we will make a direct comparison of the results for 4D $\cN=1$ SCFT three-point functions against the non-supersymmetric results. 

Recall that a supercurrent multiplet $J_{\a(s) \ad(s)}(z)$ in 4D SCFT contains independent conserved component currents $R_{s}(x) := R_{\a(s) \ad(s)}(x)$, $Q_{s+\frac{1}{2}}(x) := Q_{\a(s+1) \ad(s)}(x)$ (along with its conjugate $\bar{Q}_{s+\frac{1}{2}}(x)$), and $T_{s+1}(x) := T_{\a(s+1) \ad(s+1)}(x)$. First let us consider the leading order component three-point function $\langle R_{s_{1}}(x_{1}) \, R'_{s_{2}}(x_{2}) \,R''_{s_{3}}(x_{3}) \rangle$, which is obtained by the following bar-projection:
\begin{align} \label{bar-projection-correlator}
    \langle R_{s_{1}}(x_{1}) \, R'_{s_{2}}(x_{2}) \,R''_{s_{3}}(x_{3}) \rangle = \langle J_{s_{1}}(z_{1}) \, J'_{s_{2}}(z_{2}) \, J''_{s_{3}}(z_{3}) \rangle |_{ \q_{i} = \bar{\q}_{i} = 0} \, .
\end{align}
The three-point function $\langle R_{s_{1}}(x_{1}) \, R'_{s_{2}}(x_{2}) \,R''_{s_{3}}(x_{3}) \rangle$ is completely described by a tensor $\cH$ which is obtained as follows:
\begin{align} \label{bar-projection}
    \cH(X; U,V,W) = \cH(X, \Q, \bar{\Q}; U,V,W)|_{\Q = \bar{\Q} = 0} \, .
\end{align}
Let us compare $\cH(X; U, V, W)$ in \eqref{bar-projection} for the results we have obtained in section \ref{section4} against those obtained 
in the 4D CFT case presented in \cite{Buchbinder:2023coi}. First, we note that the RHS of \eqref{bar-projection} is fixed up to $2 \min(s_{i}) + 2$ parameters, 
while the LHS is fixed up to $2 \min(s_{i}) + 1$ parameters based on the general structure of 4D CFT correlators \eqref{4D correlators results}. 
The resolution to this apparent contradiction is that the additional parity-odd contribution in the supersymmetric three-point functions (which is linearly 
independent in superspace) becomes linearly dependent upon bar-projection. Some examples of this are presented in Appendix \ref{AppB}, and one can also check 
it explicitly for the low-superspin correlators in section \ref{section4}. Hence, there is complete agreement between the superfield analysis and the 
component analysis of \cite{Buchbinder:2023coi} after bar-projection.

Let us now consider the general structure of the component correlator
\begin{align}
    \langle T^{}_{s_{1}+1}(x_{1}) \, T'_{s_{2}+1}(x_{2}) \, T''_{s_{3}+1}(x_{3}) \rangle \, .
\end{align}
We recall that the case $s_{i} = 1$ corresponds to the energy-momentum tensor three-point function $\langle T^{}_{2}(x_{1}) \, T'_{2}(x_{2}) \, T''_{2}(x_{3}) \rangle$ which resides in the three-point function of the supercurrent multiplet. Based on our classification of the conserved structures in superspace, this component correlator is fixed up to $2 \min(s_{i}) + 2 = 2 \min(s_{i} + 1)$ conserved structures, where $\min(s_{i} + 1)$ are parity-even, and $\min(s_{i} + 1)$ 
are parity-odd under superinversion. This is \textit{one less} structure than expected based on the classification of 4D CFT three-point functions \eqref{4D correlators results}. Hence, the parity-even sector of the component three-point function 
$\langle T^{}_{s_{1}+1}(x_{1}) \, T'_{s_{2}+1}(x_{2}) \, T''_{s_{3}+1}(x_{3}) \rangle$ is further constrained by $\cN = 1$ supersymmetry compared to its non-supersymmetric counterpart. 

Finally, let us comment on some differences between the classification of parity-even/odd structures based on their transformation properties 
under \textit{superinversion} and the classification of parity-even/odd structures in component correlators based on their transformation properties 
under \textit{inversion}. For this it is instructive to consider three-point functions of the 
form $\langle J^{}_{s_{1}}(z_{1}) \, J^{}_{s_{1}}(z_{2}) \, J'_{s_{2}}(z_{3}) \rangle$. According to our general classification presented in the previous subsection,
for $s_{2}$ even, after imposing the $z_{1} \leftrightarrow z_{2}$ point-switch symmetry, these correlation functions are fixed up 
to $\min(s_{i}) + 1 = \min(s_{i}+1)$ conserved structures which are classified as parity-even under {\it superinversion}. 
Let us consider the component three-point function $\langle R^{}_{s_{1}}(x_{1}) \, R_{s_{1}}(x_{2}) \, R'_{s_{2}}(x_{3}) \rangle$. Since $s_{2}$ is even, 
from the general structure of 4D CFT three-point functions these three-point functions are fixed up to $\min(s_{i} + 1)$ structures, 
which are classified as even under {\it inversion}. Hence, structures which are classified as parity-even under superinversion in superspace are also 
classified as parity-even in $\langle R^{}_{s_{1}}(x_{1}) \, R_{s_{1}}(x_{2}) \, R'_{s_{2}}(x_{3}) \rangle$. 
On the other hand, for $\langle T^{}_{s_{1}+1}(x_{1}) \, T_{s_{1}+1}(x_{2}) \, T'_{s_{2}+1}(x_{3}) \rangle$, since $s_{2} + 1$ is odd these 
three-point functions are fixed up to $\min(s_{i} + 1)$ structures which are classified as parity-odd under {\it inversion}. This implies that structures 
in $\langle J^{}_{s_{1}}(z_{1}) J^{}_{s_{1}}(z_{2}) J'_{s_{2}}(z_{3}) \rangle$ which are classified as even under superinversion are instead classified as 
odd under inversion in the component correlator $\langle T^{}_{s_{1}+1}(x_{1}) \, T_{s_{1}+1}(x_{2}) \, T'_{s_{2}+1}(x_{3}) \rangle$. 
Similar results are obtained if we were to consider the case where $s_{2}$ is odd; in this case the structures 
in $\langle J^{}_{s_{1}}(z_{1}) J^{}_{s_{1}}(z_{2}) J'_{s_{2}}(z_{3}) \rangle$ are classified as parity-odd under superinversion. In the corresponding component 
three-point functions, the structures in $\langle R^{}_{s_{1}}(x_{1}) \, R_{s_{1}}(x_{2}) \, R'_{s_{2}}(x_{3}) \rangle$ are classified as parity-odd, 
while the structures in $\langle T^{}_{s_{1}+1}(x_{1}) \, T_{s_{1}+1}(x_{2}) \, T'_{s_{2}+1}(x_{3}) \rangle$ are classified as parity-even. 
This demonstrates that an eigenstate of {\it the  superinversion operator} (that is a contribution with definite parity $\pm 1$) is not, in general, 
an eigenstate of {\it the inversion operator}. The eigenstates of the inversion operator are contained in 
$\langle J^{}_{s_{1}}(z_{1}) \, J^{}_{s_{1}}(z_{2}) \, J'_{s_{2}}(z_{3}) \rangle$ at various orders in the Grassmann variables but with 
different eigenvalues. Hence, the classification of structures as parity-even/odd under \textit{superinversion} in superspace 
is not necessarily the same as the classification 
of structures as parity-even/odd under \textit{inversion} in the corresponding component three-point functions.


\section*{Acknowledgements}
The authors are grateful to Sergei Kuzenko for valuable discussions. The work of E.I.B. and B.J.S. is supported in part by the Australian Research Council, project DP230101629. The work of J.H. was also supported in part by the Australian Research Council project DP230101629. Since April 2024, the work of J.H. is supported by the European Union under the Marie Sklodowska-Curie grant agreement number 101107602.\footnote{Views and opinions expressed are however those of the author(s) only and do not necessarily reflect those of the European Union or European Research Executive Agency. Neither the European Union nor the granting authority can be held responsible for them.}

\appendix
\section{Linear dependence relations and useful identities}\label{AppA}
First we present the linear dependence relations amongst the Bosonic monomials, which are identical to those appearing in \cite{Buchbinder:2023coi}:
\begin{subequations} \label{Linear dependence 1}
	\begin{align} 
		Z_{2} Z_{3} + P_{1} \tilde{P}_{1} - Q_{1} \tilde{Q}_{1} &= 0 \, , \\
		Z_{1} Z_{3} + P_{2} \tilde{P}_{2} - Q_{2} \tilde{Q}_{2} &= 0 \, , \\
		Z_{1} Z_{2} + P_{3} \tilde{P}_{3} - Q_{3} \tilde{Q}_{3} &= 0 \, ,
	\end{align}
\end{subequations}
\vspace{-10mm}
\begin{subequations} \label{Linear dependence 2}
	\begin{align} 
		Z_{1} P_{1} + P_{2} \tilde{Q}_{3} + P_{3} Q_{2} &= 0 \, , & Z_{1} \tilde{P}_{1} + \tilde{P}_{2} Q_{3} + \tilde{P}_{3} \tilde{Q}_{2} &= 0  \, , \\
		Z_{2} P_{2} + P_{3} \tilde{Q}_{1} + P_{1} Q_{3} &= 0 \, , & Z_{2} \tilde{P}_{2} + \tilde{P}_{3} Q_{1} + \tilde{P}_{1} \tilde{Q}_{3} &= 0 \, , \\
		Z_{3} P_{3} + P_{1} \tilde{Q}_{2} + P_{2} Q_{1} &= 0 \, , & Z_{3} \tilde{P}_{3} + \tilde{P}_{1} Q_{2} + \tilde{P}_{2} \tilde{Q}_{1} &= 0 \, .
	\end{align}
\end{subequations}
\vspace{-10mm}
\begin{subequations} \label{Linear dependence 3}
	\begin{align} 
		Z_{1} Q_{1} + \tilde{P}_{2} P_{3} - \tilde{Q}_{2} \tilde{Q}_{3} &= 0 \, , & Z_{1} \tilde{Q}_{1} + P_{2} \tilde{P}_{3} - Q_{2} Q_{3} &= 0  \, , \\
		Z_{2} Q_{2} + \tilde{P}_{3} P_{1} - \tilde{Q}_{3} \tilde{Q}_{1} &= 0 \, , & Z_{2} \tilde{Q}_{2} + P_{3} \tilde{P}_{1} - Q_{3} Q_{1} &= 0 \, , \\
		Z_{3} Q_{3} + \tilde{P}_{1} P_{2} - \tilde{Q}_{1} \tilde{Q}_{2} &= 0 \, , & Z_{3} \tilde{Q}_{3} + P_{1} \tilde{P}_{2} - Q_{1} Q_{2} &= 0 \, .
	\end{align}
\end{subequations}
These allow elimination of the combinations $Z_{i} Z_{j}$, $Z_{i} P_{i}$, $Z_{i} \tilde{P}_{i}$, $Z_{i} Q_{i}$, $Z_{i} \tilde{Q}_{i}$. There are also the following relations involving triple products:
\begin{subequations} \label{Linear dependence 4}
	\begin{align} 
		P_{1} \tilde{P}_{2} \tilde{P}_{3} + \tilde{P}_{1} Q_{2} \tilde{Q}_{3} + \tilde{P}_{2} \tilde{Q}_{3} \tilde{Q}_{1} + \tilde{P}_{3} Q_{1} Q_{2} &= 0 \, , \\
		P_{2} \tilde{P}_{3} \tilde{P}_{1} + \tilde{P}_{2} Q_{3} \tilde{Q}_{1} + \tilde{P}_{3} \tilde{Q}_{1} \tilde{Q}_{2} + \tilde{P}_{1} Q_{2} Q_{3} &= 0 \, , \\
		P_{3} \tilde{P}_{1} \tilde{P}_{2} + \tilde{P}_{3} Q_{1} \tilde{Q}_{2} + \tilde{P}_{1} \tilde{Q}_{2} \tilde{Q}_{3} + \tilde{P}_{2} Q_{3} Q_{1} &= 0 \, ,
	\end{align}
\end{subequations}
\vspace{-10mm}
\begin{subequations} \label{Linear dependence 5}
	\begin{align} 
		\tilde{P}_{1} P_{2} P_{3} + P_{1} \tilde{Q}_{2} Q_{3} + P_{2} Q_{3} Q_{1} + P_{3} \tilde{Q}_{1} \tilde{Q}_{2} &= 0 \, , \\
		\tilde{P}_{2} P_{3} P_{1} + P_{2} \tilde{Q}_{3} Q_{1} + P_{3} Q_{1} Q_{2} + P_{1} \tilde{Q}_{2} \tilde{Q}_{3} &= 0 \, , \\
		\tilde{P}_{3} P_{1} P_{2} + P_{3} \tilde{Q}_{1} Q_{2} + P_{1} Q_{2} Q_{3} + P_{2} \tilde{Q}_{3} \tilde{Q}_{1} &= 0 \, ,
	\end{align}
\end{subequations}
\vspace{-10mm}
\begin{subequations} \label{Linear dependence 6}
	\begin{align} 
		\tilde{P}_{1} P_{2} \tilde{Q}_{3} - P_{1} \tilde{P}_{2} Q_{3} + \tilde{Q}_{1} \tilde{Q}_{2} \tilde{Q}_{3} - Q_{1} Q_{2} Q_{3} &= 0 \, , \\
		\tilde{P}_{2} P_{3} \tilde{Q}_{1} - P_{2} \tilde{P}_{3} Q_{1} + \tilde{Q}_{1} \tilde{Q}_{2} \tilde{Q}_{3} - Q_{1} Q_{2} Q_{3} &= 0 \, , \\
		\tilde{P}_{3} P_{1} \tilde{Q}_{2} - P_{3} \tilde{P}_{1} Q_{2} + \tilde{Q}_{1} \tilde{Q}_{2} \tilde{Q}_{3} - Q_{1} Q_{2} Q_{3} &= 0 \, ,
	\end{align}
\end{subequations}
which allow for elimination of the products $P_{i} \tilde{P}_{j} \tilde{P}_{k}$, $\tilde{P}_{i} P_{j} P_{k}$, $\tilde{P}_{i} P_{j} \tilde{Q}_{k}$. There are also relations involving the fermionic monomials. For instance, at quadratic order we have the cyclic relations
\begin{subequations} \label{Linear dependence 7}
	\begin{align} 
		P_{1} R_{1} + P_{2} R_{2} + P_{3} R_{3} &= 0 \, , \\
		\tilde{P}_{1} \tilde{R}_{1} + \tilde{P}_{2} \tilde{R}_{2} + \tilde{P}_{3} \tilde{R}_{3} &= 0 \, , \\[2mm]
		P_{1} S_{1} + P_{2} S_{2} + P_{3} S_{3} &= 0 \, , \\
		\tilde{P}_{1} \tilde{S}_{1} + \tilde{P}_{2} \tilde{S}_{2} + \tilde{P}_{3} \tilde{S}_{3} &= 0 \, .
	\end{align}
\end{subequations}
There are, however, many more relations which are quadratic in the monomials which are linear in $\Q \; (\bar{\Q})$:
\begin{subequations} \label{Linear dependence 8}
	\begin{align} 
		S_{1} \tilde{P}_{1} - \tilde{Q}_{2} \tilde{R}_{2} + Q_{3} \tilde{R}_{3} &= 0 \, , & \tilde{S}_{1} P_{1} - Q_{2} R_{2} + \tilde{Q}_{3} R_{3} &= 0  \, , \\
		S_{2} \tilde{P}_{2} - \tilde{Q}_{3} \tilde{R}_{3} + Q_{1} \tilde{R}_{1} &= 0 \, , & \tilde{S}_{2} P_{2} - Q_{3} R_{3} + \tilde{Q}_{1} R_{1} &= 0 \, , \\
		S_{3} \tilde{P}_{3} - \tilde{Q}_{1} \tilde{R}_{1} + Q_{2} \tilde{R}_{2} &= 0 \, , & \tilde{S}_{3} P_{3} - Q_{1} R_{1} + \tilde{Q}_{2} R_{2} &= 0 \, ,
	\end{align}
\end{subequations}
\vspace{-10mm}
\begin{subequations} \label{Linear dependence 9}
	\begin{align} 
		S_{1} \tilde{Q}_{1} + P_{2} \tilde{R}_{2} - S_{3} Q_{3} &= 0 \, , & \tilde{S}_{1} Q_{1} + \tilde{P}_{2} R_{2} - \tilde{S}_{3} \tilde{Q}_{3}  &= 0  \, , \\
		S_{2} \tilde{Q}_{2} + P_{3} \tilde{R}_{3} - S_{1} Q_{1} &= 0 \, , & \tilde{S}_{2} Q_{2} + \tilde{P}_{3} R_{3} - \tilde{S}_{1} \tilde{Q}_{1}  &= 0  \, , \\
		S_{3} \tilde{Q}_{3} + P_{1} \tilde{R}_{1} - S_{2} Q_{2} &= 0 \, , & \tilde{S}_{3} Q_{3} + \tilde{P}_{1} R_{1} - \tilde{S}_{2} \tilde{Q}_{2}  &= 0  \, , 
	\end{align}
\end{subequations}
\vspace{-8mm}
\begin{subequations} \label{Linear dependence 10}
	\begin{align} 
		Z_{2} R_{3} + P_{1} \tilde{S}_{2} - \tilde{Q}_{1} R_{2} &= 0 \, , & Z_{3} R_{2} - P_{1} \tilde{S}_{3} - Q_{1} R_{3} &= 0  \, , \\
		Z_{3} R_{1} + P_{2} \tilde{S}_{3} - \tilde{Q}_{2} R_{3} &= 0 \, , & Z_{1} R_{3} - P_{2} \tilde{S}_{1} - Q_{2} R_{1} &= 0  \, , \\
        Z_{1} R_{2} + P_{3} \tilde{S}_{1} - \tilde{Q}_{3} R_{1} &= 0 \, , & Z_{2} R_{1} - P_{3} \tilde{S}_{2} - Q_{3} R_{2} &= 0  \, ,
	\end{align}
\end{subequations}
\vspace{-10mm}
\begin{subequations} \label{Linear dependence 11}
	\begin{align} 
		Z_{2} \tilde{R}_{3} + \tilde{P}_{1} S_{2} - Q_{1} \tilde{R}_{2} &= 0 \, , & Z_{3} \tilde{R}_{2} - \tilde{P}_{1} S_{3} - \tilde{Q}_{1} \tilde{R}_{3} &= 0  \, , \\
		Z_{3} \tilde{R}_{1} + \tilde{P}_{2} S_{3} - Q_{2} \tilde{R}_{3} &= 0 \, , & Z_{1} \tilde{R}_{3} - \tilde{P}_{2} S_{1} - \tilde{Q}_{2} \tilde{R}_{1} &= 0  \, , \\
        Z_{1} \tilde{R}_{2} + \tilde{P}_{3} S_{1} - Q_{3} \tilde{R}_{1} &= 0 \, , & Z_{2} \tilde{R}_{1} - \tilde{P}_{3} S_{2} - \tilde{Q}_{3} \tilde{R}_{2} &= 0  \, ,
	\end{align}
\end{subequations}
\vspace{-8mm}
\begin{subequations} \label{Linear dependence 12}
	\begin{align} 
		Z_{2} S_{3} - \tilde{Q}_{1} S_{2} + P_{1} \tilde{R}_{2} &= 0 \, , & Z_{3} S_{2} - Q_{1} S_{3} - P_{1} \tilde{R}_{3} &= 0  \, , \\
		Z_{3} S_{1} - \tilde{Q}_{2} S_{3} + P_{2} \tilde{R}_{3} &= 0 \, , & Z_{1} S_{3} - Q_{2} S_{1} - P_{2} \tilde{R}_{1} &= 0  \, , \\
        Z_{1} S_{2} - \tilde{Q}_{3} S_{1} + P_{3} \tilde{R}_{1} &= 0 \, , & Z_{2} S_{1} - Q_{3} S_{2} - P_{3} \tilde{R}_{2} &= 0 \, ,
	\end{align}
\end{subequations}
\vspace{-10mm}
\begin{subequations} \label{Linear dependence 13}
	\begin{align} 
		Z_{2} \tilde{S}_{3} - Q_{1} \tilde{S}_{2} + \tilde{P}_{1} R_{2} &= 0 \, , & Z_{3} \tilde{S}_{2} - \tilde{Q}_{1} \tilde{S}_{3} - \tilde{P}_{1} R_{3} &= 0  \, , \\
		Z_{3} \tilde{S}_{1} - Q_{2} \tilde{S}_{3} + \tilde{P}_{2} R_{3} &= 0 \, , & Z_{1} \tilde{S}_{3} - \tilde{Q}_{2} \tilde{S}_{1} - \tilde{P}_{2} R_{1} &= 0  \, , \\
        Z_{1} \tilde{S}_{2} - Q_{3} \tilde{S}_{1} + \tilde{P}_{3} R_{1} &= 0 \, , & Z_{2} \tilde{S}_{1} - \tilde{Q}_{3} \tilde{S}_{2} - \tilde{P}_{3} R_{2} &= 0 \, .
	\end{align}
\end{subequations}
There are also relations which are quadratic in the monomials, yet are $O(\Q \bar{\Q})$:
\begin{subequations} \label{Linear dependence 14}
	\begin{align} 
		J P_{1} + R_{3} S_{2} - R_{2} S_{3}  &= 0 \, , & J \tilde{P}_{1} + \tilde{R}_{2} \tilde{S}_{3} - \tilde{R}_{3} \tilde{S}_{2}  &= 0  \, , \\
		J P_{2} + R_{1} S_{3} - R_{3} S_{1}  &= 0 \, , & J \tilde{P}_{2} + \tilde{R}_{3} \tilde{S}_{1} - \tilde{R}_{1} \tilde{S}_{3}  &= 0  \, , \\
        J P_{3} + R_{2} S_{1} - R_{1} S_{2}  &= 0 \, , & J \tilde{P}_{3} + \tilde{R}_{1} \tilde{S}_{2} - \tilde{R}_{2} \tilde{S}_{1}  &= 0  \, ,
	\end{align}
\end{subequations}
\vspace{-10mm}
\begin{subequations} \label{Linear dependence 15}
	\begin{align} 
		J Q_{1} + R_{2} \tilde{R}_{3} + S_{2} \tilde{S}_{3}  &= 0 \, , & J \tilde{Q}_{1} + R_{3} \tilde{R}_{2} + S_{3} \tilde{S}_{2}  &= 0  \, , \\
		J Q_{2} + R_{3} \tilde{R}_{1} + S_{3} \tilde{S}_{1}  &= 0 \, , & J \tilde{Q}_{2} + R_{1} \tilde{R}_{3} + S_{1} \tilde{S}_{2}  &= 0  \, , \\
        J Q_{3} + R_{1} \tilde{R}_{2} + S_{1} \tilde{S}_{2}  &= 0 \, , & J \tilde{Q}_{3} + R_{2} \tilde{R}_{1} + S_{2} \tilde{S}_{3}  &= 0   \, ,
	\end{align}
\end{subequations}
\vspace{-10mm}
\begin{subequations} \label{Linear dependence 16}
	\begin{align}  
		J Z_{1} + R_{1} \tilde{R}_{1} + S_{1} \tilde{S}_{1} &= 0 \, , \\
		J Z_{2} + R_{2} \tilde{R}_{2} + S_{2} \tilde{S}_{2} &= 0 \, , \\
        J Z_{3} + R_{3} \tilde{R}_{3} + S_{3} \tilde{S}_{3} &= 0 \, .
	\end{align}
\end{subequations}
We also found relations which are cubic, quartic and quintic in the monomials.

For imposing conservation equations on three-point functions, one must act with the operators \eqref{Derivative operators-1}, \eqref{Derivative operators-2} on the generating function \eqref{Generating function}. For this, the following identities are useful:
\begin{subequations}
	\begin{align}
		\cD_{\a} Q_{1} &= - \frac{2 \text{i}}{ X^{1/2}} \big( 2 v_{\a} \tilde{R}_{3} - Q_{1} (\hat{X}\cdot\hat{\bar{\Q}})_{\a} \big) \, ,\\
		\cD_{\a} Q_{2} &= - \frac{2 \text{i}}{ X^{1/2}} \big( 2 w_{\a} \tilde{R}_{1} - Q_{2} (\hat{X}\cdot\hat{\bar{\Q}})_{\a} \big) \, ,\\
        \cD_{\a} Q_{3} &= - \frac{2 \text{i}}{ X^{1/2}} \big( 2 u_{\a} \tilde{R}_{2} - Q_{3} (\hat{X}\cdot\hat{\bar{\Q}})_{\a} \big) \, ,
	\end{align}
\end{subequations}
\vspace{-5mm}
\begin{subequations}
	\begin{align}
		\cD_{\a} \tilde{Q}_{1} &= - \frac{2 \text{i}}{ X^{1/2}} \big( 2 w_{\a} \tilde{R}_{2} - \tilde{Q}_{1} (\hat{X}\cdot\hat{\bar{\Q}})_{\a} \big) \, ,\\
		\cD_{\a} \tilde{Q}_{2} &= - \frac{2 \text{i}}{ X^{1/2}} \big( 2 u_{\a} \tilde{R}_{3} - \tilde{Q}_{2} (\hat{X}\cdot\hat{\bar{\Q}})_{\a} \big) \, ,\\
        \cD_{\a} \tilde{Q}_{3} &= - \frac{2 \text{i}}{ X^{1/2}} \big( 2 v_{\a} \tilde{R}_{1} - \tilde{Q}_{3} (\hat{X}\cdot\hat{\bar{\Q}})_{\a} \big) \, ,
	\end{align}
\end{subequations}
\vspace{-5mm}
\begin{subequations}
	\begin{align}
		\cD_{\a} Z_{1} &= - \frac{2 \text{i}}{ X^{1/2}} \big( 2 u_{\a} \tilde{R}_{1} - Z_{1} (\hat{X}\cdot\hat{\bar{\Q}})_{\a} \big) \, ,\\
		\cD_{\a} Z_{2} &= - \frac{2 \text{i}}{ X^{1/2}} \big( 2 v_{\a} \tilde{R}_{2} - Z_{2} (\hat{X}\cdot\hat{\bar{\Q}})_{\a} \big) \, ,\\
        \cD_{\a} Z_{3} &= - \frac{2 \text{i}}{ X^{1/2}} \big( 2 w_{\a} \tilde{R}_{3} - Z_{3} (\hat{X}\cdot\hat{\bar{\Q}})_{\a} \big) \, ,
	\end{align}
\end{subequations}
\vspace{-5mm}
\begin{subequations}
	\begin{align}
		\cD_{\a} R_{1} &= - \frac{1}{ X^{1/2}} \big( u_{\a} - \text{i}(\hat{X}\cdot\hat{\bar{\Q}})_{\a} R_{1}  \big) \, ,\\
		\cD_{\a} R_{2} &= - \frac{1}{ X^{1/2}} \big( v_{\a} - \text{i}(\hat{X}\cdot\hat{\bar{\Q}})_{\a} R_{2}  \big) \, ,\\
        \cD_{\a} R_{3} &= - \frac{1}{ X^{1/2}} \big( w_{\a} - \text{i}(\hat{X}\cdot\hat{\bar{\Q}})_{\a} R_{3}  \big)  \, ,
	\end{align}
\end{subequations}
\vspace{-5mm}
\begin{subequations}
	\begin{align}
		\cD_{\a} \tilde{R}_{1} &= - \frac{\text{i}}{ 2 X^{1/2}} \hat{\bar{\Q}}^{2} (\hat{X} \cdot \bar{u})_{\a} \, ,\\
		\cD_{\a} \tilde{R}_{2} &= - \frac{\text{i}}{ 2 X^{1/2}} \hat{\bar{\Q}}^{2} (\hat{X} \cdot \bar{v})_{\a} \, ,\\
        \cD_{\a} \tilde{R}_{3} &= - \frac{\text{i}}{ 2 X^{1/2}} \hat{\bar{\Q}}^{2} (\hat{X} \cdot \bar{w})_{\a} \, ,
	\end{align}
\end{subequations}
\vspace{-5mm}
\begin{subequations}
	\begin{align}
		\cD_{\a} S_{1} &= - \frac{5 \text{i}}{ 2 X^{1/2}} \hat{\bar{\Q}}^{2} u_{\a} \, ,\\
		\cD_{\a} S_{2} &= - \frac{5 \text{i}}{ 2 X^{1/2}} \hat{\bar{\Q}}^{2} v_{\a} \, ,\\
        \cD_{\a} S_{3} &= - \frac{5 \text{i}}{ 2 X^{1/2}} \hat{\bar{\Q}}^{2} w_{\a} \, ,
	\end{align}
\end{subequations}
\vspace{-5mm}
\begin{subequations}
	\begin{align}
		\cD_{\a} \tilde{S}_{1} &= \frac{1}{ X^{1/2}} \big( (\hat{X} \cdot \bar{u})_{\a} + 4 \text{i} \hat{\Q}_{\a} \tilde{R}_{1} + 3 \text{i} (\hat{X} \cdot \hat{\bar{\Q}})_{\a} \tilde{S}_{1} \big) \, ,\\
		\cD_{\a} \tilde{S}_{2} &=  \frac{1}{ X^{1/2}} \big( (\hat{X} \cdot \bar{v})_{\a} + 4 \text{i} \hat{\Q}_{\a} \tilde{R}_{2} + 3 \text{i} (\hat{X} \cdot \hat{\bar{\Q}})_{\a} \tilde{S}_{2} \big) \, ,\\
        \cD_{\a} \tilde{S}_{3} &= \frac{1}{ X^{1/2}} \big( (\hat{X} \cdot \bar{w})_{\a} + 4 \text{i} \hat{\Q}_{\a} \tilde{R}_{3} + 3 \text{i} (\hat{X} \cdot \hat{\bar{\Q}})_{\a} \tilde{S}_{3} \big) \, .
	\end{align}
\end{subequations}
Similarly, for the action of $\bar{\cD}_{\ad}$ on the monomials we obtain
\begin{subequations}
	\begin{align}
		\bar{\cD}_{\ad} \tilde{R}_{1} &= \frac{\bar{u}_{\ad}}{ X^{1/2}} \, , & \bar{\cD}_{\ad} S_{1} &= - \frac{1}{ X^{1/2}} (\hat{X} \cdot u)_{\ad} \, , \\
		\bar{\cD}_{\ad} \tilde{R}_{2} &= \frac{\bar{v}_{\ad}}{ X^{1/2}} \, , & \bar{\cD}_{\ad} S_{1} &= - \frac{1}{ X^{1/2}} (\hat{X} \cdot v)_{\ad} \, , \\
        \bar{\cD}_{\ad} \tilde{R}_{3} &= \frac{\bar{w}_{\ad}}{ X^{1/2}} \, , & \bar{\cD}_{\ad} S_{1} &= - \frac{1}{ X^{1/2}} (\hat{X} \cdot w)_{\ad} \, .
	\end{align}
\end{subequations}
Analogous identities may be derived for the action of $\cQ_{\a}$, $\bar{\cQ}_{\ad}$ on the monomials \eqref{Monomials-bosonic}, \eqref{Monomials-fermionic}. They take a similar form to the identities above.


\section{Implications of superinversion pseudo-covariance}\label{AppD}

Consider a generic tensor superfield $\cH(X,\Q,\bar{\Q})$ which satisfies the pseudo-superinversion covariance condition \eqref{Parity condition}. Such a tensor can be written in the form:
\begin{align} \label{even odd expansion}
    \cH_{\cA_{1} \cA_{2} \cA_{3}}(X,\Q,\bar{\Q}) = \cH^{E}_{\cA_{1} \cA_{2} \cA_{3}}(X,\Q,\bar{\Q}) + \cH^{O}_{\cA_{1} \cA_{2} \cA_{3}}(X,\Q,\bar{\Q}) \, .
\end{align}
Here $\cH^{E}$ is a linear combination of tensor structures with complex coefficients satisfying \eqref{Parity condition} with a ``$+$", while $\cH^{O}$ is a linear combination of tensor structures with complex coefficients satisfying \eqref{Parity condition} with a ``$-$". Due to the transformation properties \eqref{inversion replacements} of the tensorial building blocks under superinversion, we can decompose our solution $\cH^{E}$ further into $\{ \cH^{E(+)}, \cH^{E(-)} \}$, with complex coefficients $\{ c^{E(+)}, c^{E(-)} \}$ respectively. The solution $\cH^{E(+)}$ is an overall even function of \eqref{odd-X}, while $\cH^{E(-)}$ is an overall odd function of \eqref{odd-X}. Similarly we decompose $\cH^{O}$ into $\{ \cH^{O(+)}, \cH^{O(-)} \}$, with complex coefficients $\{ c^{O(+)}, c^{O(-)} \}$. The expansion \eqref{even odd expansion} then becomes
\begin{align} \label{even odd expansion 2}
\begin{split}
    \cH_{\cA_{1} \cA_{2} \cA_{3}}(X,\Q,\bar{\Q}) &= c^{E(+)} \cH^{E(+)}_{\cA_{1} \cA_{2} \cA_{3}}(X,\Q,\bar{\Q}) + c^{E(-)} \cH^{E(-)}_{\cA_{1} \cA_{2} \cA_{3}}(X,\Q,\bar{\Q})  \\
    & \hspace{5mm} + c^{O(+)} \cH^{O(+)}_{\cA_{1} \cA_{2} \cA_{3}}(X,\Q,\bar{\Q}) + c^{O(-)} \cH^{O(-)}_{\cA_{1} \cA_{2} \cA_{3}}(X,\Q,\bar{\Q}) \, .
\end{split}
\end{align}
For vector-like supercurrents, it turns out the contributions $\cH^{E(-)}$, $\cH^{O(-)}$ must vanish due to the linear system \eqref{Diophantine equations}. The reasoning for this is as follows. Recall that the solutions $\cH^{(-)}$ correspond to containing an odd number of the objects $P_{i}$, $\tilde{P}_{i}$, $S_{i}$, $\tilde{S}_{i}$. By examining \eqref{Diophantine equations} we may define
\begin{align}
    \Omega &= k_{1} + k_{2} + k_{3} + \bar{k}_{1} + \bar{k}_{2} + \bar{k}_{3} + m_{1} + m_{2} + m_{3} + \bar{m}_{1} + \bar{m}_{2} + \bar{m}_{3} \, .
\end{align}
We obtain a solution of type $\cH^{(+)}$ when $\Omega$ is even, and a solution of type $\cH^{(-)}$ when $\Omega$ is odd. However, one can show that for three-point functions of vector-like supercurrents $\Omega$ is always even. Hence, $\cH^{(-)} = 0$ in such three-point functions and we may remove them from the expansion \eqref{even odd expansion 2} so that we obtain 
\begin{align}
    \cH_{\cA_{1} \cA_{2} \cA_{3}}(X,\Q,\bar{\Q}) &= c^{E(+)} \cH^{E(+)}_{\cA_{1} \cA_{2} \cA_{3}}(X,\Q,\bar{\Q}) + c^{O(+)} \cH^{O(+)}_{\cA_{1} \cA_{2} \cA_{3}}(X,\Q,\bar{\Q}) \, .
\end{align}
Now consider the superinversion formula \eqref{inversion formula}. We recall that the solutions of type $\cH^{(+)}$ are even functions of the tensorial building blocks \eqref{odd-X}. As a result, we obtain
\begin{subequations} \label{superinversion simplification}
    \begin{align}
    \cH^{I(+)}_{\cA_{1} \cA_{2} \bar{\cA}_{3}}(X, \Q, \bar{\Q} ) &= \cH^{(+)}_{\cA_{1} \cA_{2} \bar{\cA}_{3}}(\ve \rightarrow - \bar{\ve}, \bar{\ve} \rightarrow - \ve, X \rightarrow \bar{X}^{I}, \Q \rightarrow \bar{\Q}^{I}, \bar{\Q} \rightarrow \Q^{I} ) \non \\
    &= \cH^{(+)}_{\cA_{1} \cA_{2} \bar{\cA}_{3}}(\ve \rightarrow \bar{\ve}, \bar{\ve} \rightarrow \ve, \bar{X}^{I}, \bar{\Q}^{I}, \Q^{I} ) \, .
\end{align}
\end{subequations}
Now recall that the operation of complex conjugation, $\star$, acts on the building blocks $\ve$, $\bar{\ve}$, $X$, $\Q$, $\bar{\Q}$ according to the replacement rules
\begin{subequations} \label{conjugation operation}
\begin{align}
    & \hspace{25mm} \ve \stackrel{\star}{\longrightarrow} \bar{\ve} \, , \hspace{10mm} \bar{\ve} \stackrel{\star}{\longrightarrow} \ve \, , \\[2mm]
    &X_{\a \ad} \stackrel{\star}{\longrightarrow} \bar{X}_{\a \ad} \, , \hspace{10mm} \Q_{\a} \stackrel{\star}{\longrightarrow} \bar{\Q}_{\ad} \, , \hspace{10mm} \bar{\Q}_{\ad} \stackrel{\star}{\longrightarrow} \Q_{\a} \, .
\end{align} 
\end{subequations}
Hence, $\star$ acts on a tensor $\cH_{\cA_{1} \cA_{2} \bar{\cA}_{3}}(X,\Q,\bar{\Q})$, as follows:
\begin{subequations} \label{conjugation operation b}
\begin{align}
    \cH_{\bar{\cA}_{1} \bar{\cA}_{2} \cA_{3}}(\fxq) \stackrel{\star}{\longrightarrow} \bar{\cH}_{\cA_{1} \cA_{2} \bar{\cA}_{3}}(X, \Q, \bar{\Q})
    &= \cH_{\cA_{1} \cA_{2} \bar{\cA}_{3}}(\ve \to \bar{\ve}, \bar{\ve} \to \ve, \bar{X}, \bar{\Q},\Q) \, .
\end{align}
\end{subequations}
We see from \eqref{superinversion simplification} that the action of superinversion on $\cH^{(+)}$ is now equivalent to that of \eqref{conjugation operation b}, except expressed in terms of the inverted variables \eqref{inverted-Z3}. We then obtain
\begin{align}
    \cH^{I(+)}_{\cA_{1} \cA_{2} \bar{\cA}_{3}}(X, \Q, \bar{\Q} ) &= \bar{\cH}^{(+)}_{\cA_{1} \cA_{2} \bar{\cA}_{3}}(X^{I}, \Q^{I}, \bar{\Q}^{I} ) \, .
\end{align}
By using this result, the superinversion covariance condition \eqref{Parity condition} for our solution $\cH^{E} = c^{E(+)} \cH^{E(+)}$ becomes 
\begin{align}
    c^{E(+)} \cH^{E(+)}_{\cA_{1} \cA_{2} \cA_{3}}(X^{I},\Q^{I},\bar{\Q}^{I}) &= c^{E(+)} \bar{\cH}^{E(+)}_{\cA_{1} \cA_{2} \cA_{3}}(X^{I},\Q^{I},\bar{\Q}^{I}) \non \\[2mm]
    \implies\hspace{5mm} \cH^{E(+)}_{\cA_{1} \cA_{2} \cA_{3}}(X,\Q,\bar{\Q}) &= \bar{\cH}^{E(+)}_{\cA_{1} \cA_{2} \cA_{3}}(X,\Q,\bar{\Q}) \, ,
\end{align}
which implies that our solution $\cH^{E(+)}_{\cA_{1} \cA_{2} \cA_{3}}(X,\Q,\bar{\Q})$ must be hermitian. Using a similar argument, one can show for the odd sector that the superinversion pseudo-covariance condition implies 
\begin{align}
    c^{O(+)} \cH^{E(+)}_{\cA_{1} \cA_{2} \cA_{3}}(X^{I},\Q^{I},\bar{\Q}^{I}) &= - c^{O(+)} \bar{\cH}^{E(+)}_{\cA_{1} \cA_{2} \cA_{3}}(X^{I},\Q^{I},\bar{\Q}^{I}) \non \\[2mm]
    \implies\hspace{5mm} \cH^{O(+)}_{\cA_{1} \cA_{2} \cA_{3}}(X,\Q,\bar{\Q}) &= - \bar{\cH}^{O(+)}_{\cA_{1} \cA_{2} \cA_{3}}(X,\Q,\bar{\Q}) \, .
\end{align}
Hence, $\cH^{O(+)}_{\cA_{1} \cA_{2} \cA_{3}}(X,\Q,\bar{\Q})$ must be anti-hermitian. As a result, after imposing the superinversion pseudo-covariance condition our solution for $\cH_{\cA_{1} \cA_{2} \cA_{3}}(X,\Q,\bar{\Q})$ may be written in the form
\begin{align}
    \cH_{\cA_{1} \cA_{2} \cA_{3}}(X,\Q,\bar{\Q}) &= c^{E(+)} \underbrace{\cH^{E(+)}_{\cA_{1} \cA_{2} \cA_{3}}(X,\Q,\bar{\Q})}_{\text{hermitian}} + c^{O(+)} \underbrace{\cH^{O(+)}_{\cA_{1} \cA_{2} \cA_{3}}(X,\Q,\bar{\Q})}_{\text{anti-hermitian}} \, .
\end{align}
Therefore parity-even structures are covariant under superinversion and must be hermitian, while parity-odd structures are pseudo-covariant under superinversion and must be anti-hermitian. Furthermore, if we impose the reality condition \eqref{Reality condition}, it may be shown that $\text{Im}[c^{E(+)}] = \text{Re}[c^{O(+)}] = 0$, so $c^{E(+)} = a^{E(+)}$, $c^{O(+)} = \text{i} b^{O(+)}$, where $a^{E(+)}$, $b^{O(+)}$ are real coefficients. Hence, when superinversion pseudo-covariance and reality are imposed on our solution $\cH_{\cA_{1} \cA_{2} \cA_{3}}(X,\Q,\bar{\Q})$, it can always be written in the form
\begin{align}
    \cH_{\cA_{1} \cA_{2} \cA_{3}}(X,\Q,\bar{\Q}) &= a^{E} \underbrace{\cH^{E}_{\cA_{1} \cA_{2} \cA_{3}}(X,\Q,\bar{\Q})}_{\text{hermitian}} + \text{i} b^{O} \underbrace{\cH^{O}_{\cA_{1} \cA_{2} \cA_{3}}(X,\Q,\bar{\Q})}_{\text{anti-hermitian}} \, ,
\end{align}
where $\cH^{E}$ is parity-even, and $\cH^{O}$ is parity-odd (note we have dropped the ``$+$" labels). In general there are many linearly independent solutions $\cH^{E}_{i}$, $\cH^{O}_{i}$ of even and odd type, so we can promote the above formula to a sum over structures $a^{E}_{i} \cH^{E}_{i}$ and $\text{i} b^{O}_{i} \cH^{O}_{i}$. Imposing conservation and point-switch symmetries then results in relations among the $a^{E}_{i}$, and relations among the $b^{O}_{i}$ (however, the $a^{E}_{i}$ and $b^{O}_{i}$ do not mix with each other).

\newpage

\section{Correlation functions involving higher-spin spinor current multiplets}\label{AppB}

The main goal of this work was to study the general structure of three-point functions involving ``vector-like" supercurrent multiplets, however, the formalism is perfectly capable of analysing three-point functions of conserved supercurrents belonging to an arbitrary $(i,j)$ Lorentz representation. In this appendix we will use our new formalism to revisit some of the three-point functions studied in \cite{Buchbinder:2021kjk}, which examined the three-point functions involving higher-spin spinor current multiplets. The main three-point functions of interest are\footnote{ One could also consider the three-point functions $\langle \bar{S}_{\ad(k) }(z_{1}) S_{\a(l) }(z_{2}) L(z_{3}) \rangle$ and $\langle \bar{S}_{\ad(k) }(z_{1}) S_{\a(l) }(z_{2}) J_{\g \gd}(z_{3}) \rangle$ for $k \neq l$. However, as was pointed out in \cite{Buchbinder:2021kjk}, these three-point functions must vanish as they possess a non-trivial $R$-symmetry charge.}
\begin{align}
    \langle \bar{S}_{\ad(k) }(z_{1}) S_{\a(k) }(z_{2}) L(z_{3}) \rangle \, ,  && \langle \bar{S}_{\ad(k) }(z_{1}) S_{\a(k) }(z_{2}) J_{\g \gd}(z_{3}) \rangle \, .
\end{align}
The three-point function $\langle \bar{S}_{\ad(k) }(z_{1}) S_{\a(k) }(z_{2}) L(z_{3}) \rangle$ was shown to be fixed up to two independent real parameters, while $\langle \bar{S}_{\ad(k) }(z_{1}) S_{\a(k) }(z_{2}) J_{\g \gd}(z_{3}) \rangle$ was shown to be fixed up to three independent real parameters.

\subsection{ \texorpdfstring{  $\langle \bar{S}_{\ad(k) }(z_{1}) S_{\a(k) }(z_{2}) L(z_{3}) \rangle$ }{< Sbar S L >} }

First we will consider the three-point function $\langle \bar{S} S L \rangle$, with general ansatz
\begin{align}
	\langle \bar{S}_{\ad(k)}(z_{1}) \, S_{\a(k)}(z_{2}) \, L(z_{3}) \rangle = \frac{ \bar{\cI}_{\ad(k)}{}^{\a'(k)}(x_{\bar{1} 3}) \, \cI_{\a(k)}{}^{\ad'(k)}(x_{2 \bar{3}}) }{ x_{1\bar{3}}{}^2 (x_{\bar{1}3}{}^2)^{k/2+1} (x_{2\bar{3}}{}^2)^{k/2+1} x_{\bar{2}3}{}^2}
	\; \cH_{\a'(k) \ad'(k)}(X_{3}, \Q_{3}, \bar{\Q}_{3}) \, .
\end{align} 
where $\cH$ is homogeneous degree $-(k+2)$. Using the formalism outlined in Subsection \ref{GeneratingFunctionFormalism} the correlation function is encoded in the polynomial
\begin{align}
     \cH(X, \Q, \bar{\Q}; U,V) = \cH_{\a(k) \ad(k)}(X, \Q, \bar{\Q}) \, \boldsymbol{U}^{\a(k)} \boldsymbol{V}^{\ad(k)} \, .
\end{align}
In this case there are four possible solutions to \eqref{Diophantine equations} from which $\hat{\cH}$ may be constructed:
\begin{align}
    \left\{ Q_3^k,J Q_3^k,Q_3^{k-1} R_1 \tilde{R}_2,Q_3^{k-1} S_1 \tilde{S}_2\right\}
\end{align}
However, these structures are linearly dependent, which allows us to remove the last structure. Hence, the general linearly independent ansatz for $\hat{\cH}$ is
\begin{align}
    \hat{\cH}(X, \Q, \bar{\Q}; U,V) = A_{1} Q_{3}^{k} + A_{2} J Q_{3}^{k}  + A_{3} Q_{3}^{k-1} R_{1} \tilde{R}_{2} \, ,
\end{align}
where $A_{i} = a_{i} + \text{i} b_{i}$ are complex coefficients. By successively applying the replacement rules \eqref{H_I replacement rules}, \eqref{Htilde replacement rules} to compute $\hat{\tilde{\cH}}$, the corresponding expression for $\hat{\tilde{\cH}}(X, \Q, \bar{\Q}; U,V)$ is:
\begin{align}
    \hat{\tilde{\cH}}(X, \Q, \bar{\Q}; U,V) &= A_{1} ( \tilde{P}_3^k - 4 \text{i} k  J \tilde{P}_3^k + 4 \text{i} k \tilde{P}_3^{k-1} \tilde{R}_2 \tilde{S}_1- 4 \text{i} k \tilde{P}_3^{k-1}
   \tilde{R}_1 \tilde{S}_2 ) \non \\
   & \hspace{15mm} + A_{2} J \tilde{P}_3^k - A_{3} \tilde{P}_3^{k-1} \tilde{R}_2
   \tilde{S}_1 \, ,
\end{align}
up to an overall factor of $(-1)^{k}$, which may be ignored in this case. The task now is to impose the conservation equations on each of the currents. The constraints on $\cH$ and $\tilde{\cH}$ due to conservation are as follows:
\begin{subequations}
\begin{align}
    D_{1} \cH(X, \Q, \bar{\Q}; U,V) &= 0 \, , & \bar{D}_{1}^{2} \cH(X, \Q, \bar{\Q}; U,V) &= 0 \, , \\
    D_{2}^{2} \cH(X, \Q, \bar{\Q}; U,V) &= 0 \, , & \bar{D}_{2} \cH(X, \Q, \bar{\Q}; U,V) &= 0 \, , \\
    D_{3}^{2} \tilde{\cH}(X, \Q, \bar{\Q}; U,V) &= 0 \, , & \bar{D}_{3}^{2} \tilde{\cH}(X, \Q, \bar{\Q}; U,V) &= 0 \, .
\end{align}
\end{subequations}
After imposing conservation on all three-points we obtain the following relations between the coefficients
\begin{align}
    \big\{a_1\to a_1,a_2\to a_2,a_3\to \frac{k}{k+1} a_2 ,b_1\to b_1,b_2\to b_2,b_3\to \frac{k
   }{k+1} b_2\big\}
\end{align}
The general solution for $\cH$ at this step is:
\begin{align}
    \cH(X, \Q, \bar{\Q}; U,V) &= \frac{a_1}{X^{k+2}} Q_3^k+\frac{ \text{i} b_1}{X^{k+2}} Q_3^k \\
    & + \frac{a_2}{X^{k+2}} \big( J Q_3^k + \tfrac{k}{k+1} Q_3^{k-1} R_1 \tilde{R}_2 )+ \frac{\text{i} b_2}{X^{k+2}} \big( J Q_3^k + \tfrac{k}{k+1} Q_3^{k-1} R_1
   \tilde{R}_2 \big) \, . \non
\end{align}
Hence, after imposing the superfield conservation equations the correlation function $\langle \bar{S} S L \rangle $ is fixed up to four independent real parameters.

The next constraint to impose is the combined $z_{1} \leftrightarrow z_{2}$ point-switch/reality condition. After imposing this constraint, for even $k$ we obtain $a_{2} \to 0$, $b_{1} \to 0$, while for odd $k$ we obtain $a_{1} \to 0$, $b_{2} \to 0$. In either case, there are two independent real parameters remaining after imposing the condition. Hence, the three-point function $\langle \bar{S}_{\ad(k)}(z_{1}) S_{\a(k)}(z_{2}) L(z_{3}) \rangle$ is fixed up to two independent real parameters, in agreement with \cite{Buchbinder:2021kjk}.

\subsection{ \texorpdfstring{  $\langle \bar{S}_{\ad(k) }(z_{1}) S_{\a(k) }(z_{2}) J_{\g \gd}(z_{3}) \rangle$ }{< Sbar S J >} }
Let's now analyse the three-point function $\langle \bar{S} S J \rangle$, with general ansatz
\begin{align}
	\langle \bar{S}_{\ad(k)}(z_{1}) \, S_{\a(k)}(z_{2}) \, J_{\g \gd}(z_{3}) \rangle = \frac{ \bar{\cI}_{\ad(k)}{}^{\a'(k)}(x_{\bar{1} 3}) \, \cI_{\a(k)}{}^{\ad'(k)}(x_{2 \bar{3}}) }{  x_{1\bar{3}}{}^2 (x_{\bar{1}3}{}^2)^{k/2+1}  (x_{2\bar{3}}{}^2)^{k/2+1} x_{\bar{2}3}{}^2}
	\; \cH_{\a'(k) \ad'(k) \g \gd}(X_{3}, \Q_{3}, \bar{\Q}_{3}) \, .
\end{align} 
where $\cH$ is homogeneous degree $-(k+1)$. Using the formalism outlined in Subsection \ref{GeneratingFunctionFormalism} the correlation function is encoded in the polynomial
\begin{align}
     \cH(X, \Q, \bar{\Q}; U,V, W) = \cH_{\a(k) \ad(k) \g \gd}(X, \Q, \bar{\Q}) \, \boldsymbol{U}^{\a(k)} \boldsymbol{V}^{\ad(k)} \boldsymbol{W}^{\g \gd} \, , 
\end{align}
It must be noted that the $k=1$ and $k>1$ cases must be treated separately as the number of possible solutions to \eqref{Diophantine equations} is different in each case. The results for $k=1$ are contained in \cite{Buchbinder:2021izb}, so instead we will focus our attention on $k>1$. For arbitrary $k > 1$ there are 24 possible solutions to \eqref{Diophantine equations} from which $\hat{\cH}$ may be constructed:
\begin{align}
    &\big\{Z_3 Q_3^k,J Z_3 Q_3^k,P_2 \tilde{P}_1 Q_3^{k-1},J P_2 \tilde{P}_1 Q_3^{k-1},R_3 S_1
   \tilde{P}_1 Q_3^{k-1},R_1 S_3 \tilde{P}_1 Q_3^{k-1}, R_1 S_1 \tilde{P}_1 \tilde{Q}_1
   Q_3^{k-2}, \non \\
   & \hspace{5mm} \tilde{Q}_1 \tilde{Q}_2 Q_3^{k-1},J \tilde{Q}_1 \tilde{Q}_2 Q_3^{k-1},R_1 Z_3
   \tilde{R}_2 Q_3^{k-1},P_2 R_1 \tilde{P}_1 \tilde{R}_2 Q_3^{k-2}, R_3 \tilde{Q}_2 \tilde{R}_2
   Q_3^{k-1}, R_1 \tilde{Q}_1 \tilde{Q}_2 \tilde{R}_2 Q_3^{k-2}, \non \\
   & \hspace{10mm} R_3 \tilde{R}_3 Q_3^k,R_1 \tilde{Q}_1 \tilde{R}_3 Q_3^{k-1},S_1 Z_3 \tilde{S}_2 Q_3^{k-1},P_2 S_1 \tilde{P}_1 \tilde{S}_2 Q_3^{k-2}, S_3
   \tilde{Q}_2 \tilde{S}_2 Q_3^{k-1},S_1 \tilde{Q}_1 \tilde{Q}_2 \tilde{S}_2 Q_3^{k-2}, \non \\
   & \hspace{15mm} P_2 \tilde{Q}_2 \tilde{R}_2 \tilde{S}_2 Q_3^{k-2},P_2 \tilde{R}_3 \tilde{S}_2 Q_3^{k-1},S_3
   \tilde{S}_3 Q_3^k,S_1 \tilde{Q}_1 \tilde{S}_3 Q_3^{k-1},P_2 \tilde{R}_2 \tilde{S}_3
   Q_3^{k-1}\big\} 
\end{align}
After systematically applying the linear dependence relations \eqref{Linear dependence 1}--\eqref{Linear dependence 16}, it may be shown that only 8  structures are linearly independent: 
\begin{align}
    &\big\{P_2 \tilde{P}_1 Q_3^{k-1},J P_2 \tilde{P}_1 Q_3^{k-1},\tilde{Q}_1 \tilde{Q}_2 Q_3^{k-1},J
   \tilde{Q}_1 \tilde{Q}_2 Q_3^{k-1}, \\
   & \hspace{10mm} P_2 R_1 \tilde{P}_1 \tilde{R}_2 Q_3^{k-2},R_3 \tilde{Q}_2
   \tilde{R}_2 Q_3^{k-1},R_3 \tilde{R}_3 Q_3^k,R_1 \tilde{Q}_1 \tilde{R}_3 Q_3^{k-1}\big\} \non
\end{align}
Hence, a general linearly independent ansatz for $\hat{\cH}$ is
\begin{align}
    \hat{\cH}(X, \Q, \bar{\Q}; U,V, W) &= A_{1} P_2 \tilde{P}_1 Q_3^{k-1} + A_{2} J P_2 \tilde{P}_1 Q_3^{k-1} \\
    & \hspace{10mm} + A_{3} \tilde{Q}_1 \tilde{Q}_2 Q_3^{k-1} + A_{4} J
   \tilde{Q}_1 \tilde{Q}_2 Q_3^{k-1} + \dots \, , \non
\end{align}
where $A_{i} = a_{i} + \text{i} b_{i}$ are complex coefficients. Similar to the previous examples, one may use the replacement rules \eqref{H_I replacement rules},\eqref{Htilde replacement rules} to compute $\hat{\tilde{\cH}}$.

The task now is to impose the conservation equations on the current multiplets. The constraints on $\cH$ and $\tilde{\cH}$ following from conservation are:
\begin{subequations}
\begin{align}
    D_{1} \cH(X, \Q, \bar{\Q}; U,V,W) &= 0 \, , & \bar{D}_{1}^{2} \cH(X, \Q, \bar{\Q}; U,V,W) &= 0 \, , \\
    D_{2}^{2} \cH(X, \Q, \bar{\Q}; U,V,W) &= 0 \, , & \bar{D}_{2} \cH(X, \Q, \bar{\Q}; U,V,W) &= 0 \, , \\
    D_{3} \tilde{\cH}(X, \Q, \bar{\Q}; U,V,W) &= 0 \, , & \bar{D}_{3} \tilde{\cH}(X, \Q, \bar{\Q}; U,V,W) &= 0 \, .
\end{align}
\end{subequations}
After imposing conservation on $z_{1}$ and $z_{2}$ we obtain the following relations between the coefficients:
\begin{align}
    \begin{split}
    &\big\{a_1\to a_1,a_2\to a_2,a_3\to a_3,a_4\to a_4,a_5\to \frac{k-1}{k} a_2 -\frac{ k-1}{k(k+1)}a_4, \\
    & \hspace{5mm} a_6\to \frac{2 k}{k+1}a_4 ,a_7\to a_7,a_8\to \frac{2 k}{k+1}a_4 ,b_1\to \frac{1}{4
    k}a_2+\frac{2 k-1}{4 k}a_4 +\frac{1}{4}a_7, \\
    & \hspace{5mm} b_2\to b_2,b_3\to \frac{3 k+1}{4
    (k+1)}a_4 +\frac{1}{4}a_7,b_4\to \frac{4 k (k+1)}{k^2+1} \left(a_1-a_3\right)+\frac{
    k+1}{k^2+1}b_2, \\
    & \hspace{5mm} b_5\to -\frac{4(k-1)}{k^2+1} \left(a_1-a_3\right) + \frac{k (k-1)}{k^2+1}b_2  ,b_6\to
    \frac{8 k^2}{k^2+1}\left(a_1-a_3\right) +\frac{2 k}{k^2+1}b_2 , \\
    & \hspace{10mm} b_7\to -\frac{4 k (3
    k+1)}{k^2+1}a_1 +\frac{4 \left(2 k^2+k-1\right)}{k^2+1}a_3 -\frac{3 k+1}{k^2+1} b_2 , \\
    & \hspace{10mm} b_8\to \frac{8 k^2}{k^2+1}\left(a_1-a_3\right)+\frac{2k}{k^2+1} b_2 \big\}
   \end{split}
\end{align}
It may be explicitly check that conservation on $z_{3}$ is automatically satisfied for the above relations. Hence, after imposing the superfield conservation equations, the correlation function $\langle \bar{S} S J \rangle $ is fixed up to six independent real parameters, $a_{1}$, $a_{2}$, $a_{3}$, $a_{4}$, $a_{7}$, $b_{2}$. 

The next constrain to impose is the combined $z_{1} \leftrightarrow z_{2}$ point-switch/reality condition. After imposing this constraint, for even $k$ we find $a_{1} \to 0$, $a_{3} \to 0$, $b_{2} \to 0$. On the other hand, for odd $k$ $a_{2} \to 0$, $a_{4} \to 0$, $a_{7} \to 0$. In either case, there are three independent real parameters remaining after imposing the condition. Hence, the three-point function $\langle \bar{S}_{\ad(k)}(z_{1}) S_{\a(k)}(z_{2}) L(z_{3}) \rangle$ is fixed up to three independent real parameters for arbitrary $k$, in agreement with the result reported in \cite{Buchbinder:2021kjk}.

\subsection{ \texorpdfstring{  $\langle \bar{S}_{\ad(s_1) }(z_{1}) S_{\b(s_1+s_2) }(z_{2}) \bar{S}'_{\gd(s_2)}(z_{3}) \rangle$ }{< Sbar S Sbar >} }
To end this section, it is useful to note that the general solution for the three-point function $\langle \bar{S}_{\ad(s_1) }(z_{1}) S_{\b(s_1+s_2) }(z_{2}) \bar{S}'_{\gd(s_2)}(z_{3}) \rangle$ for arbitrary superspins $s_1$ and $s_2$ has been constructed using an analytical method of \cite{Buchbinder:2021kjk}. The correlator 
$\langle \bar S_{\ad(s_1)}(z_1) S_{\b(k)}(z_2) \bar{S}_{\gd(s_2)} (z_3) \rangle$ with $k \neq s_1+s_2$ vanishes due to the non-compensating $R$-symmetry charge.

In our notation, the general form is given by
\bea
&&\langle \bar S_{\ad(s_1)}(z_1) S_{\b(s_1+s_2)}(z_2) \bar{S}'_{\gd(s_2)} (z_3) \rangle \non\\
&=& \frac{\cI_{\ad(s_1)}^{\,\,\,\,\,\,\a(s_1)} (x_{3 \bar{1}}) \,\cI_{\b(s_1+s_2)}^{\,\,\,\,\,\,\bd(s_1+s_2)} (x_{2 \bar{3}})} 
{(x_{\bar{1} 3}){}^{s_1+2} x_{\bar{3} 1}{}^2 x_{\bar{2} 3}{}^2  (x_{\bar{3} 2}){}^{s_1+s_2+2}}
\cH_{ \a(s_1),\,\bd(s_1+s_2),\,\gd(s_2)}
(X_3, \Theta_3, \bar{\Theta}_3) ~.
\eea
The explicit form for 
\bea
\cH(X, \Q, \bar{\Q}; U, V, W) = \cH_{\a(s_1),\, \bd(s_1+s_2),\, \gd(s_2)} (X, \Q, \bar{\Q}) u^{\a(s_1)} \bar{v}^{\bd(s_1+ s_2)} \bar{w}^{\gd(s_2)}
\eea
can be written as
\bea
\cH(X, \Theta, \bar{\Theta}; U, V, W) &=& \frac{1}{X^{s_1+2}}
\bigg\{\tilde{P}_1^{s_2-1} \tilde{Q}_2^{s_1} 
\Big[ a \tilde{P}_1 -b \tilde{S}_3 \tilde{R}_2\Big]
\non\\
&&+ \frac{4\ri s_2 a-b}{s_1+s_2+1} \tilde{P}_1^{s_2-1} \tilde{Q}_2^{s_1-1} \tilde{R}_3
\Big[s_1 \tilde{P}_1 R_1+(s_1+1) \tilde{S}_2 \tilde{Q}_2
\Big]
\bigg\}~,~~~~~~~~~
\eea
for \textit{two complex} parameters $a$ and $b$. If we further restrict $s_1= s_2= s$, there is another constraint from the $1 \leftrightarrow 3$ symmetry on $\langle \bar S_{\ad(s)}(z_1) S_{\b(2s)}(z_2) \bar{S}_{\gd(s)} (z_3) \rangle$. For odd values of $s$, one finds that $a=0$ and $b$ is arbitrary. For even values of $s$, one finds that $a= \frac{\ri b}{2}$. Hence, in any case there is a \textit{single} independent complex parameter.

\section{Examples of three-point functions of higher-spin supercurrents}\label{AppC}

\subsection{\texorpdfstring{$\langle J^{}_{\a(2) \ad(2)}(z_{1}) J'_{\b(2) \bd(2)}(z_{2}) J''_{\g \gd}(z_{3}) \rangle$}{ < J2 J'2 J''1 >} }
\noindent 
The solution for $\cH$ for this three-point function is:
\begin{flalign*}
	\hspace{5mm} \includegraphics[width=0.95\textwidth]{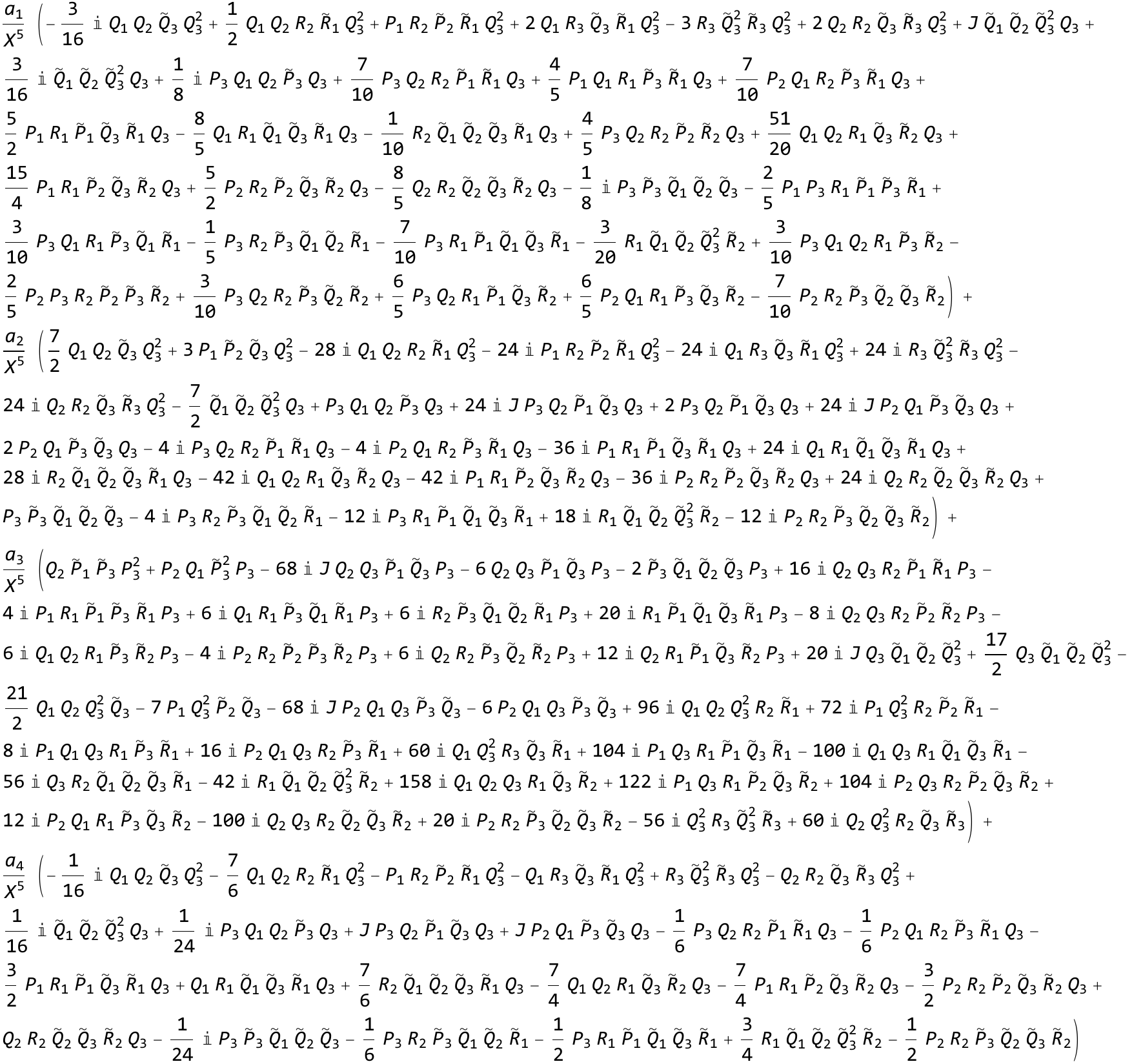} &&
\end{flalign*}
In this case there are 4 conserved structures which are linearly independent in superspace. However, after bar-projecting the solution, i.e. by computing $\cH(X,\Q,\bar{\Q}; U,V,W)|_{\Q = \bar{\Q} = 0}$, the parity-odd structures corresponding to the coefficients $a_{1}$ and $a_{4}$ become linearly dependent and combine to be proportional to an overall coefficient $b_{1} = \frac{1}{24} a_{1} + \frac{1}{8} a_{4}$. The result then coincides precisely with the parity-even and parity-odd structures in the three-point function $\langle J^{}_{2} J'_{2} J''_{1} \rangle $ presented in \cite{Buchbinder:2023coi}.

\subsection{\texorpdfstring{$\langle J^{}_{\a(3) \ad(3)}(z_{1}) J'_{\b(2) \bd(2)}(z_{2}) J''_{\g \gd}(z_{3}) \rangle$}{ < J3 J'2 J''1 >} }
\noindent 
The solution for $\cH$ for this three-point function is:
\begin{flalign*}
	\hspace{5mm} \includegraphics[width=0.95\textwidth]{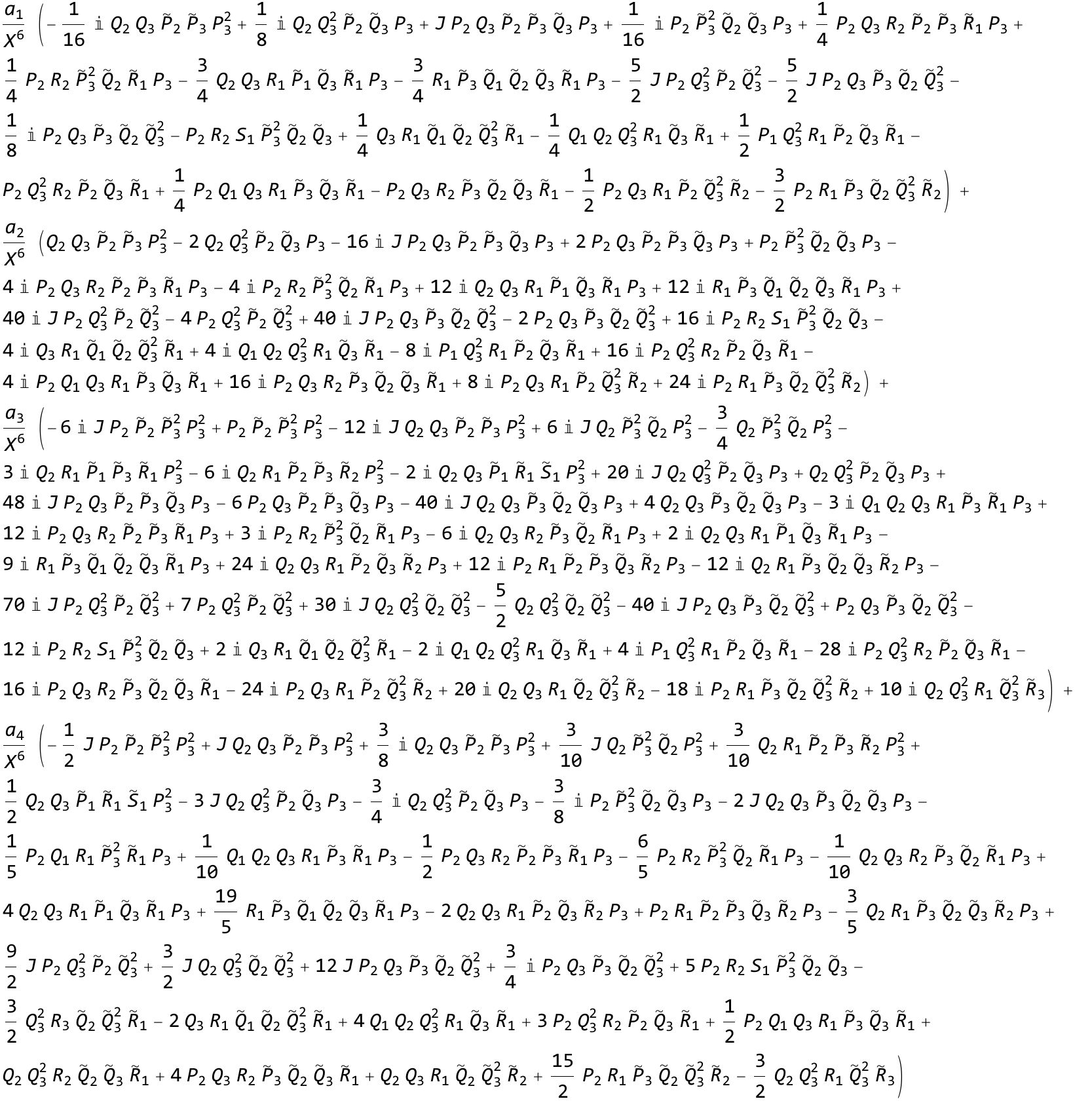} &&
\end{flalign*}
The 4 conserved structures are linearly independent in superspace, however, after bar-projection the parity-odd structures corresponding to the coefficients $a_{1}$ and $a_{4}$ become linearly dependent and combine to be proportional to an overall coefficient $b_{1} = -\frac{1}{16} a_{1} + \frac{3}{8} a_{4}$. The result then coincides with the parity-even and parity-odd structures in the three-point function $\langle J^{}_{3} J'_{2} J''_{1} \rangle $ presented in \cite{Buchbinder:2023coi}.

\subsection{\texorpdfstring{$\langle J^{}_{\a(2) \ad(2)}(z_{1}) J'_{\b(2) \bd(2)}(z_{2}) J''_{\g(2) \gd(2)}(z_{3}) \rangle$}{ < J2 J'2 J''2 >} }\label{superspin2}
\noindent 
The solution for $\cH$ for this three-point function is:
\begin{flalign*}
	\hspace{5mm} \includegraphics[width=0.95\textwidth]{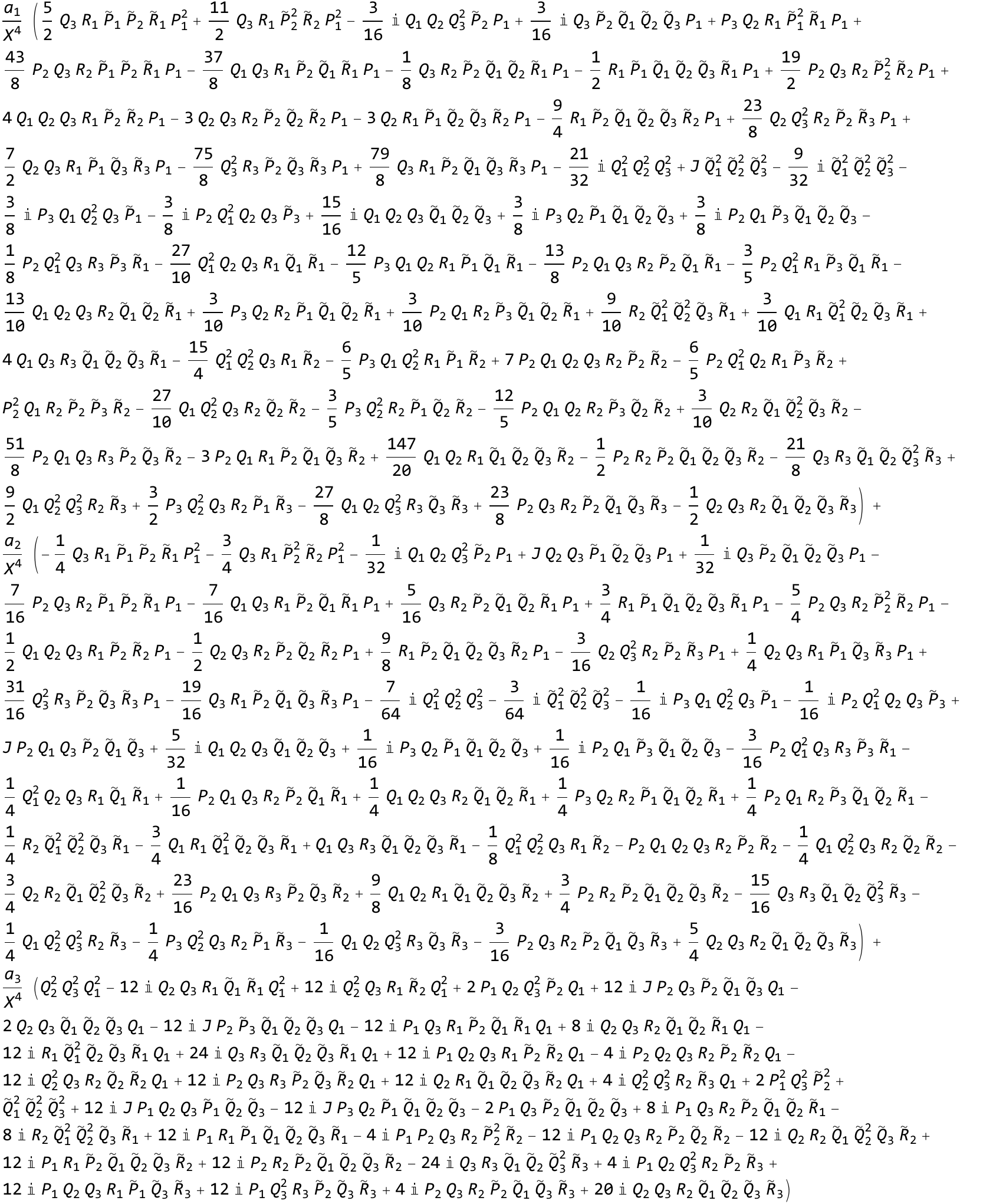} &&
\end{flalign*}
\begin{flalign*}
	\hspace{5mm} \includegraphics[width=0.9\textwidth]{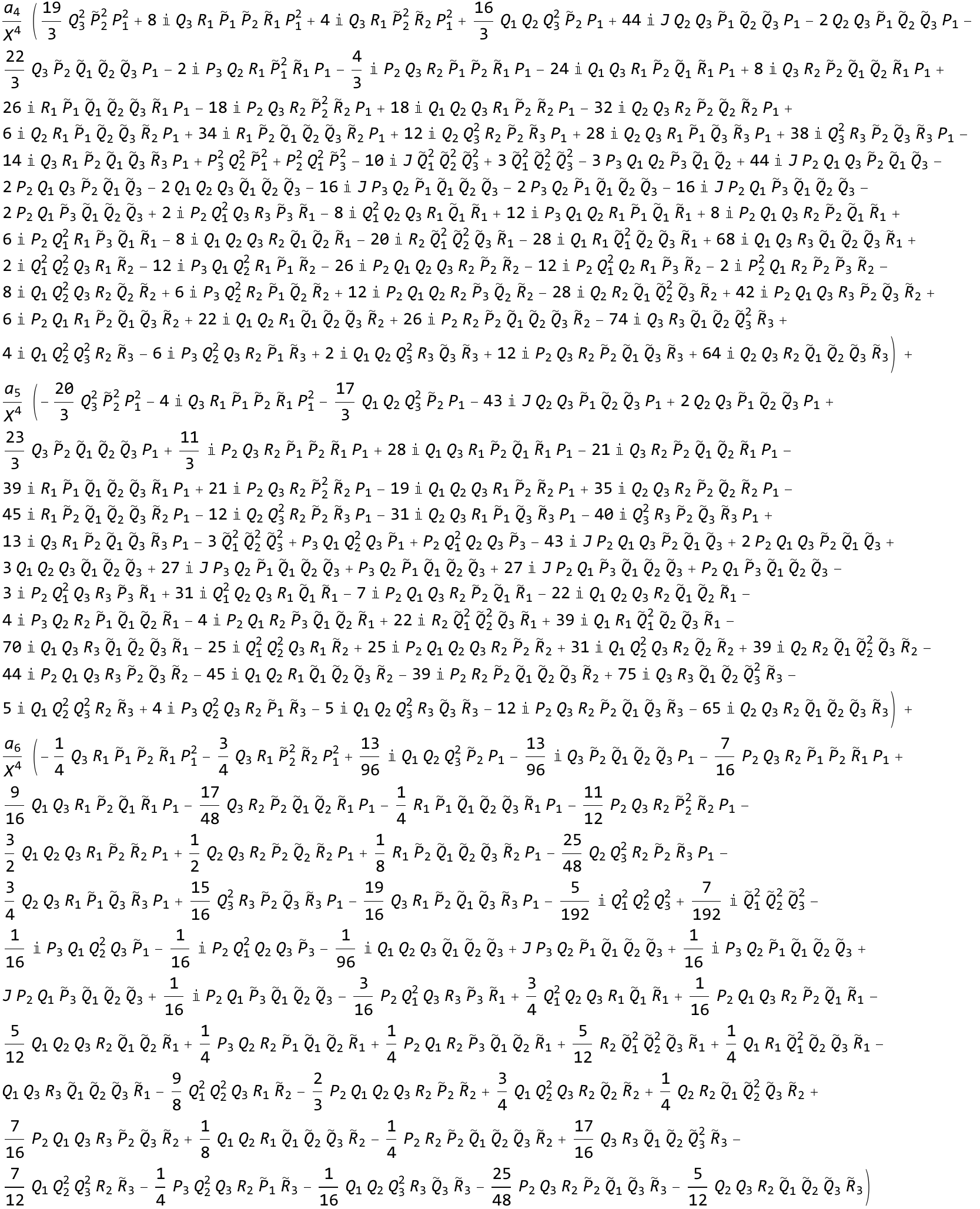} &&
\end{flalign*} 
The 6 conserved structures are linearly independent in superspace, however, after bar-projection the parity-odd structures with coefficients $a_{1}$, $a_{2}$ and $a_{6}$ become linearly dependent and combine to give two independent structures which are proportional to the coefficients $ b_{1} = - \frac{21}{32} a_{1} - \frac{7}{64} a_{2} - \frac{5}{192} a_{6}$, $b_{2} = - \frac{3}{8} a_{1} - \frac{1}{16} a_{2} - \frac{1}{16} a_{6}$. Hence, after bar-projection there are 5 linearly independent structures, which coincide precisely with the parity-even and parity-odd structures in the three-point function $\langle J^{}_{2} J'_{2} J''_{2} \rangle$ presented in \cite{Buchbinder:2023coi}.



\printbibliography[heading=bibintoc,title={References}]



\end{document}